\documentclass[twocolumn,nofootinbib,preprintnumbers,floatfix,superscriptaddress,aps,prd]{revtex4-1}

\usepackage[bookmarks=false]{hyperref}
\usepackage{graphicx}
\usepackage{amsmath}
\usepackage{xspace}
\usepackage[small]{subfigure}

\newcommand{\eq}[1]{Eq.~\eqref{eq:#1}}
\newcommand{\eqs}[2]{Eqs.~\eqref{eq:#1} and \eqref{eq:#2}}
\newcommand{\eqss}[3]{Eqs.~\eqref{eq:#1}, \eqref{eq:#2}, and \eqref{eq:#3}}
\renewcommand{\sec}[1]{Sec.~\ref{sec:#1}}

\newcommand{\secss}[3]{Secs.~\ref{sec:#1}, \ref{sec:#2}, and \ref{sec:#3}}
\newcommand{\subsec}[1]{Sec.~\ref{subsec:#1}}

\newcommand{\fig}[1]{Fig.~\ref{fig:#1}}

\newcommand{\app}[1]{Appendix~\ref{app:#1}}

\newcommand{\abs}[1]{\lvert#1\rvert}
\newcommand{\ord}[1]{\mathcal{O}(#1)}
\newcommand{\vev}[1]{\langle #1 \rangle}

\newcommand{\mae}[3]{\langle#1\rvert#2\rvert#3\rangle}
\newcommand{\Mae}[3]{\bigl\langle#1\bigr\rvert#2\bigr\rvert#3\bigr\rangle}
\newcommand{\MAe}[3]{\Bigl\langle#1\Bigr\rvert#2\Bigr\rvert#3\Bigr\rangle}
\newcommand{\bra}[1]{\langle#1\rvert}
\newcommand{\ket}[1]{\lvert#1\rangle}
\newcommand{\braket}[2]{\langle#1\rvert#2\rangle}
\newcommand{\ang}[1]{\langle #1 \rangle}

\newcommand{\intlim}[3]{\int_{#1}^{#2}\! \df #3 \,}

\newcommand{\df}{\mathrm{d}}
\newcommand{\img}{\mathrm{i}}
\newcommand{\Li}{\mathrm{Li}}

\newcommand{\sdt}{\!\cdot\!}
\newcommand{\tr}{\textrm{tr}}
\newcommand{\fr}{\frac}

\newcommand{\lra}{\leftrightarrow}

\newcommand{\al}{\alpha}
\newcommand{\bt}{\beta}
\newcommand{\ga}{\gamma}
\newcommand{\Ga}{\Gamma}
\newcommand{\de}{\delta}

\newcommand{\eps}{\epsilon}
\newcommand{\ve}{\varepsilon}
\newcommand{\la}{\lambda}

\newcommand{\si}{\sigma}

\newcommand{\w}{\omega}
\newcommand{\balpha}{{\bar \alpha}}
\newcommand{\bbeta}{{\bar \beta}}
\newcommand{\bgamma}{{\bar \gamma}}
\newcommand{\bdelta}{{\bar \delta}}

\newcommand{\cA}{{\mathcal A}}
\newcommand{\cB}{{\mathcal B}}

\newcommand{\cL}{{\mathcal L}}

\newcommand{\cP}{{\mathcal P}}

\newcommand{\bn}{\bar{n}}
\newcommand{\bq}{{\bar{q}}}

\newcommand{\vC}{\vec{C}}

\newcommand{\vT}{\bar{T}}

\newcommand{\hD}{\widehat{D}}
\newcommand{\hatt}{\hat{t}}
\newcommand{\hT}{\widehat{T}}
\newcommand{\hU}{\widehat{U}}
\newcommand{\hX}{\widehat{X}}
\newcommand{\hZ}{\widehat{Z}}
\newcommand{\hga}{\widehat{\gamma}}

\newcommand{\hDe}{\widehat{\Delta}}

\newcommand{\Dslash}{D\!\!\!\!\slash}
\newcommand{\kslash}{k\!\!\!\slash}
\newcommand{\nslash}{n\!\!\!\slash}
\newcommand{\bnslash}{\bar{n}\!\!\!\slash}
\newcommand{\pslash}{p\!\!\!\slash}

\newcommand{\qslash}{q\!\!\!\slash}

\newcommand{\hH}{\widehat{H}}
\newcommand{\hS}{\widehat{S}}
\newcommand{\hV}{\widehat{V}}

\newcommand{\GeV}{\:\mathrm{GeV}}

\newcommand{\nn}{\nonumber}

\newcommand{\lp}{\tilde p}        
\newcommand{\ldel}{\tilde \delta} 

\newcommand{\Ecm}{E_\mathrm{cm}}

\newcommand{\soft}{{s}}

\renewcommand{\P}{\mathrm{P}}       
\newcommand{\C}{\mathrm{C}}       
\newcommand{\bnP}{\overline {\mathcal P}}

\newcommand{\IR}{\mathrm{IR}}
\newcommand{\UV}{\mathrm{UV}}

\newcommand{\hard}{\mathrm{hard}}
\newcommand{\fin}{\mathrm{fin}}
\renewcommand{\div}{\mathrm{div}}
\newcommand{\tree}{\mathrm{tree}}
\newcommand{\cusp}{\mathrm{cusp}}

\newcommand{\id}{\mathbf{1}}

\newcommand{\Op}{\vec{O}}

\newcommand{\zero}{{(0)}}
\newcommand{\one}{{(1)}}
\newcommand{\two}{{(2)}}

\allowdisplaybreaks[4]

\begin{document}


\preprint{\vbox{\hbox{MIT--CTP 4303}\hbox{NIKHEF 2014-022}\hbox{INT-PUB-11-044}\hbox{DESY 15-065}}}

\title{Employing Helicity Amplitudes for Resummation \vspace{0.3cm}}

\author{Ian Moult}
\affiliation{Center for Theoretical Physics, Massachusetts Institute of Technology, Cambridge, MA 02139, USA\vspace{0.5ex}}

\author{Iain W.~Stewart}
\affiliation{Center for Theoretical Physics, Massachusetts Institute of Technology, Cambridge, MA 02139, USA\vspace{0.5ex}}

\author{Frank J.~Tackmann}
\affiliation{Theory Group, Deutsches Elektronen-Synchrotron (DESY), D-22607 Hamburg, Germany\vspace{0.5ex}}

\author{Wouter J.~Waalewijn\vspace{0.5ex}}
\affiliation{Nikhef, Theory Group, Science Park 105, 1098 XG, Amsterdam, The Netherlands\vspace{0.5ex}}
\affiliation{ITFA, University of Amsterdam, Science Park 904, 1018 XE, Amsterdam, The Netherlands\vspace{0.5ex}}

\date{August 10, 2015}

\begin{abstract}

Many state-of-the-art QCD calculations for multileg processes use helicity amplitudes as their fundamental ingredients. We construct a simple and easy-to-use helicity operator basis in soft-collinear effective theory (SCET), for which the hard Wilson coefficients from matching QCD onto SCET are directly given in terms of color-ordered helicity amplitudes. Using this basis allows one to seamlessly combine fixed-order helicity amplitudes at any order they are known with a resummation of higher-order logarithmic corrections. In particular, the virtual loop amplitudes can be employed in factorization theorems to make predictions for exclusive jet cross sections without the use of numerical subtraction schemes to handle real-virtual infrared cancellations. 
We also discuss matching onto SCET in renormalization schemes with helicities in $4$- and $d$-dimensions.
To demonstrate that our helicity operator basis is easy to use, we provide an explicit construction of the operator basis, as well as results for the hard matching coefficients, for $pp\to H + 0,1,2$ jets, $pp\to W/Z/\gamma + 0,1,2$ jets, and $pp\to 2,3$ jets. These operator bases are completely crossing symmetric, so the results can easily be applied to processes with $e^+e^-$ and $e^-p$ collisions.

\end{abstract}

\maketitle

\phantom{x}
\newpage
\phantom{x}
\newpage
\tableofcontents

\newpage
\section{Introduction}
\label{sec:intro}

The production of hadronic jets is one of the most basic processes at particle colliders. Processes including a vector boson ($W$, $Z$, $\gamma$) or Higgs boson together with jets provide probes of the Standard Model (SM), and are also dominant backgrounds for many new-physics searches. Optimizing the precision and discovery potential of these channels requires accurate predictions of the SM backgrounds. Furthermore, the growth of the jet substructure field has sparked a renewed interest in the study of jets themselves, both for an improved understanding of QCD, and for applications to identify boosted heavy objects in and beyond the SM.

Precise predictions for jet production require perturbative calculations including both fixed-order corrections as well as logarithmic resummation. QCD corrections to processes with jets are typically enhanced due to phase space restrictions. Such restrictions often introduce sensitivity to low momentum scales, $p$, of order a few tens of GeV, in addition to the hard scale, $Q$, which is of order the partonic center-of-mass energy. In this case, the perturbative series contains large double logarithms $\alpha_s^n\ln^m(p/Q)$ with $m \leq 2n$. To obtain the best possible perturbative predictions, these logarithms should be resummed to all orders in $\alpha_s$.

There has been tremendous progress in the calculation of fixed-order perturbative amplitudes in QCD using the spinor helicity formalism \cite{DeCausmaecker:1981bg,Berends:1981uq,Gunion:1985vca,Xu:1986xb}, color ordering techniques \cite{Berends:1987me,Mangano:1987xk,Mangano:1988kk,Bern:1990ux} and unitarity based methods \cite{Bern:1994zx,Bern:1994cg}. NLO predictions are now available for a large number of high multiplicity final states, including $pp\rightarrow V+$ up to 5 jets \cite{Giele:1991vf, Arnold:1988dp, Giele:1993dj, Bern:1997sc, Ellis:2008qc, Berger:2009zg, Berger:2009ep, Berger:2010vm, Berger:2010zx, Ita:2011wn, Bern:2013gka}, $pp\rightarrow$ up to 5 jets \cite{Bern:1993mq, Kunszt:1993sd, Kunszt:1994tq, Bern:1994fz, Nagy:2001fj, Bern:2011ep, Badger:2012pf, Badger:2013yda}, and $pp\rightarrow  H+$ up to 3 jets \cite{Campbell:2006xx, Kauffman:1996ix, Schmidt:1997wr, Dixon:2009uk, Badger:2009hw, Campbell:2010cz, vanDeurzen:2013rv, Cullen:2013saa, Campanario:2013fsa}, and there are many efforts~\cite{Binoth:2010ra,AlcarazMaestre:2012vp, Ossola:2007ax, Berger:2008sj,Binoth:2008uq, Mastrolia:2010nb, Badger:2010nx, Fleischer:2010sq, Cullen:2011kv, Hirschi:2011pa, Bevilacqua:2011xh, Cullen:2011ac, Cascioli:2011va, Badger:2012pg, Actis:2012qn, Peraro:2014cba, Alwall:2014hca, Cullen:2014yla} to fully automatize the computation of one-loop corrections to generic helicity amplitudes.

For high-multiplicity jet events, the resummation of large logarithms is typically achieved with parton shower Monte Carlo programs. Here, the hard process enters through tree-level (and also one-loop) matrix elements and the QCD corrections due to final-state and initial-state radiation are described by the parton shower.
The parton shower resums logarithms at the leading logarithmic (LL) accuracy, with some subleading improvements, but it is difficult to reliably assess and systematically improve its logarithmic accuracy. 

The approach we will take in this paper is to match onto soft-collinear effective theory (SCET)~\cite{Bauer:2000ew, Bauer:2000yr, Bauer:2001ct, Bauer:2001yt}, the effective theory describing the soft and collinear limits of QCD. In SCET, the QCD corrections at the hard scale are captured by process-dependent Wilson coefficients. The low-energy QCD dynamics does not depend
on the details of the hard scattering (other than the underlying Born kinematics), similar to the parton shower picture. Resummation in SCET is achieved analytically through renormalization group evolution (RGE) in the effective theory, allowing one to systematically improve the logarithmic accuracy and assess the associated perturbative uncertainties.
For example, for dijet event shape variables in $e^+e^-$ collisions, SCET has enabled resummation to N$^3$LL accuracy and global fits for $\alpha_s(m_Z)$~\cite{Becher:2008cf, Chien:2010kc, Abbate:2010xh, Abbate:2012jh, Hoang:2014wka, Hoang:2015hka}. The analytic higher-order resummation can also be used to improve the Monte Carlo parton-shower description~\cite{Alioli:2012fc, Alioli:2013hqa, Alioli:2015toa}.
Furthermore, SCET allows for the direct calculation of exclusive jet cross sections, eliminating the need for numerical subtraction schemes for real emissions up to power corrections.

An important prerequisite for employing SCET is to obtain the hard matching coefficients, which are extracted from the fixed-order QCD amplitudes. The matching for $V+2$ parton and $H+2$ parton processes is well known from the QCD quark and gluon form factors, and is known to three loops~\cite{Baikov:2009bg, Gehrmann:2010ue, Abbate:2010xh}. The matching for $V+3$ partons~\cite{Becher:2009th, Becher:2011fc, Becher:2012xr, Becher:2013vva}, and $H+3$ partons~\cite{Liu:2012sz, Jouttenus:2013hs, Huang:2014mca, Becher:2014tsa},  has been performed at both NLO and NNLO. Partonic processes with four external quarks have been studied in SCET in Refs.~\cite{Chiu:2008vv, Zhu:2010mr, Kelley:2010qs, Wang:2010ue, Ahrens:2010zv, Ahrens:2011mw, Li:2014ula, Liu:2014oog}, and the matching for all massless $2\to2$ processes has been obtained at NLO in Ref.~\cite{Kelley:2010fn} and recently at NNLO in Ref.~\cite{Broggio:2014hoa}.

For high-multiplicity processes, the usual approach to constructing an operator basis with explicit Lorentz indices and gamma matrices is laborious. In this paper, we introduce a convenient formalism, based on helicity operators, which allows for a seamless matching for higher multiplicity processes onto SCET. A first look at the formalism discussed here was already given in Ref.~\cite{Stewart:2012yh}.  Indeed, results for helicity amplitudes are already employed in the SCET matching calculations mentioned above, though without the construction of corresponding SCET operators.

In the spinor helicity formalism, the individual helicity amplitudes (i.e. the amplitudes for given fixed external helicities) are calculated, as opposed to calculating the amplitude for arbitrary external spins in one step and then summing over all spins at the end. One advantage is that the individual helicity amplitudes typically yield more compact expressions. And since they correspond to distinct external states, they can be squared and summed at the end. Helicity amplitudes remove the large redundancies in the usual description of (external) gauge fields, allowing for much simplified calculations particularly for amplitudes with many external gluons.

As we will see, this helicity-based approach is also advantageous in SCET.
In SCET, as we will review in \subsec{scet}, collinear fields carry label directions corresponding to the directions of jets in the process, which provide natural lightlike vectors with which to define fields of definite helicity. As we will demonstrate, the construction of an appropriate operator basis becomes simple when using operators built out of fields with definite helicity. Furthermore, using such a helicity operator basis greatly facilitates the matching of QCD onto SCET, because one can directly utilize the known QCD helicity amplitudes for the matching. Together, this substantially simplifies the study of high-multiplicity jet processes with SCET.

\subsection{Overview}

Consider a process with $N$ final-state jets and $L$ leptons, photons, or other nonstrongly interacting particles, with the  underlying hard Born process
\begin{equation} \label{eq:interaction}
\kappa_a (q_a)\, \kappa_b(q_b) \to \kappa_1(q_1) \dotsb \kappa_{N+L}(q_{N+L})
\,,\end{equation}
where $\kappa_{a,b}$ denote the colliding partons, and $\kappa_i$ denote the outgoing quarks, gluons, leptons, and other particles with momenta $q_i$. The incoming partons are along the beam directions, $q_{a,b}^\mu = x_{a,b} P_{a,b}^\mu$, where $x_{a,b}$ are the momentum fractions and $P^\mu_{a,b}$ the (anti)proton momenta.  For definiteness, we consider two colliding partons, but our discussion of the matching will be completely crossing symmetric, so it applies equally well to $ep$ and $ee$ collisions.

In SCET, the active-parton exclusive jet cross section corresponding to \eq{interaction} can be proven to factorize for a variety of jet resolution variables.\footnote{Here active parton refers to initial-state quarks or gluons. Proofs of factorization with initial-state hadrons must also account for effects due to Glaubers~\cite{Collins:1988ig}, which may or may not cancel, and whose relevance depends on the observable in question~\cite{Gaunt:2014ska,Zeng:2015iba}.} The factorized expression for the exclusive jet cross section can be written schematically in the form
\begin{align} \label{eq:sigma}
\df\sigma &=
\int\!\df x_a\, \df x_b\, \df \Phi_{N+L}(q_a \!+ q_b; q_1, \ldots)\, M(\{q_i\})
\\\nn &\quad \times
\sum_{\kappa} \tr\,\bigl[ \hH_{\kappa}(\{q_i\}) \hS_\kappa \bigr] \otimes
\Bigl[ B_{\kappa_a} B_{\kappa_b} \prod_J J_{\kappa_J} \Bigr]
+ \dotsb
\,.\end{align}
Here, $\df \Phi_{N+L}(q_a\!+ q_b; q_1, \ldots)$ denotes the Lorentz-invariant phase space for the Born process in \eq{interaction}, and $M(\{q_i\})$ denotes the measurement made on the hard momenta of the jets (which in the factorization are approximated by the Born momenta $q_i$). The dependence on the underlying hard interaction is encoded in the hard function $\hH_{\kappa}(\{q_i\})$, where $\{q_i\} \equiv \{q_1, \ldots, q_{N+L}\}$, the sum over $\kappa \equiv \{\kappa_a, \kappa_b, \ldots \kappa_{N+L}\}$ is over all relevant partonic processes, and the trace is over color. Any dependence probing softer momenta, such as measuring jet masses or low $p_T$s, as well as the choice of jet algorithm, will affect the precise form of the factorization, but not the hard function $\hH_\kappa$. This dependence enters through the definition of the soft function $\hS_\kappa$ (describing soft radiation), jet functions $J_{\kappa_J}$ (describing energetic final-state radiation in the jets) and the beam functions $B_i$ (describing energetic initial-state radiation along the beam direction). More precisely, the beam function is given by $B_i = \sum_{i'} {\cal I}_{i i'} \otimes f_{i'}$ with $f_i$ the parton distributions of the incoming protons, and ${\cal I}_{i i'}$ a perturbatively calculable matching coefficient depending on the measurement definition~\cite{Stewart:2009yx}. The ellipses at the end of \eq{sigma} denote power-suppressed corrections. All functions in the factorized cross section depend only on the physics at a single scale. This allows one to evaluate all functions at their own natural scale, and then evolve them to a common scale using their RGE. This procedure resums the large logarithms of scale ratios appearing in the cross section to all orders in perturbation theory.

The explicit form of the factorization theorem in \eq{sigma}, including field-theoretic definitions for the jet, beam, and soft functions is known for a number of exclusive jet cross sections and measurements of interest. For example, factorization theorems exist for the $N$-jet cross section defined using $N$-jettiness~\cite{Stewart:2009yx, Stewart:2010tn, Stewart:2010pd, Berger:2010xi, Jouttenus:2011wh, Jouttenus:2013hs, Kang:2013nha, Kang:2013lga, Stewart:2015waa}. These have also been utilized to include higher-order resummation in Monte Carlo programs~\cite{Alioli:2012fc, Alioli:2013hqa, Alioli:2015toa}, and are the basis of the $N$-jettiness subtraction method for fixed-order calculations~\cite{Boughezal:2015dva, Gaunt:2015pea}. In addition, there has been a focus on color-singlet production at small $q_T$~\cite{Catani:2000vq, Becher:2010tm, GarciaEchevarria:2011rb, Chiu:2012ir, Catani:2013tia}, as well as the factorization of processes defined with jet algorithms~\cite{Delenda:2006nf, Bauer:2008jx, Ellis:2010rwa, Walsh:2011fz, Kelley:2012zs, Banfi:2012yh, Becher:2012qa, Tackmann:2012bt, Banfi:2012jm, Liu:2012sz, Becher:2013xia, Stewart:2013faa, Shao:2013uba, Li:2014ria, Jaiswal:2014yba, Gangal:2014qda, Becher:2014aya}, jet shape variables~\cite{Hornig:2009vb, Ellis:2009wj, Bauer:2011uc, Feige:2012vc, Chien:2012ur, Larkoski:2014tva, Larkoski:2014pca, Chien:2014nsa, Procura:2014cba, Becher:2015gsa, Larkoski:2015zka, Larkoski:2015kga}, or fragmentation properties~\cite{Procura:2009vm, Liu:2010ng, Jain:2011xz, Procura:2011aq, Krohn:2012fg, Waalewijn:2012sv, Chang:2013rca, Bauer:2013bza, Ritzmann:2014mka} for identified jets. The same hard functions also appear in threshold resummation factorization formulas, which are often used to obtain an approximate higher order result for inclusive cross sections.

The focus of our paper is the hard function $\hH_{\kappa}(\{q_i\})$ in \eq{sigma}, which contains the process-dependent underlying hard interaction of \eq{interaction}, but is independent of the particular measurement. In SCET, the dependence on the hard interaction is encoded in the Wilson coefficients, $\vC$, of a basis of operators built out of SCET fields. The Wilson coefficients can be calculated through a matching calculation from QCD onto the effective theory. The hard function appearing in the factorization theorem is then given by
\begin{align}
\hH_\kappa(\{q_i\})
&= \sum_{\{\la_i\}}
  \vC_{\la_1\cdot\cdot(\cdot\cdot\la_n)}(\{q_i\})\,
  \vC_{\la_1 \cdot\cdot(\cdot\cdot\la_n)}^\dagger (\{q_i\})
\,.\end{align}
Here, the $\{ \lambda_i \}$ denote helicity labels and the sum runs over all relevant helicity configurations. The $\vC$ are vectors in color space, and the hard function is therefore a matrix in color space.

For processes of higher multiplicities, the construction of a complete basis of SCET operators, and the subsequent matching calculation, becomes laborious due to the proliferation of Lorentz and color structures, similar to the case of high-multiplicity fixed-order calculations using standard Feynman diagrams. The use of SCET helicity fields introduced in this paper, combined with analogous color management techniques as used in the calculation of amplitudes, makes the construction of an operator basis extremely simple, even in the case of high-multiplicity processes. Furthermore, with this basis choice, the SCET Wilson coefficients are precisely given by the IR-finite parts of the color-ordered QCD helicity amplitudes, rendering the matching procedure almost trivial. Combining the results for the hard function with known results for the soft, jet, and beam functions, then allows for the resummation of jet observables in higher multiplicity processes, which are ubiquitous at the LHC.

The remainder of this paper is organized as follows.
In \subsec{helicity}, we review the notation for the spinor-helicity formalism. Additional useful helicity and color identities can be found in \app{useful}. We provide a brief summary of SCET in \subsec{scet}. In \sec{basis}, we introduce SCET helicity fields and operators, and describe the construction of the helicity and color basis, as well as its symmetry properties. In \sec{matching}, we discuss the matching from QCD onto the SCET helicity operators, including a discussion of the dependence on the regularization and renormalization scheme. We then demonstrate the matching explicitly for $H+0,1,2$ jets in \sec{higgs}, $V+0,1,2$ jets in \sec{vec}, and $pp\to2,3$ jets in \sec{pp}. Explicit results for the required helicity amplitudes are collected in the appendices. In \sec{running}, we discuss the general renormalization group evolution of the hard coefficients, which involves mixing between different color structures, to all orders. We give explicit results for the anomalous dimensions for up to $4$ colored particles plus an arbitrary number of uncolored particles. We conclude in \sec{conclusions}.

\section{Notation}
\label{sec:helicity}

\subsection{Helicity Formalism}
\label{subsec:helicity}

We will use the standard notation for the spinor algebra (for a review see for example Refs.~\cite{Dixon:1996wi, Dixon:2013uaa}).
Consider the four-component spinor $u(p)$ of a massless Dirac particle with momentum $p$, satisfying the massless Dirac equation,
\begin{equation} \label{eq:Dirac}
\pslash\, u(p)=0
\,, \qquad
p^2 = 0
\,.\end{equation}
The charge conjugate (antiparticle) spinor $v(p)$ also satisfies \eq{Dirac}, and we can choose a representation such that $v(p) = u(p)$. The spinors and conjugate spinors for the two helicity states are denoted by
\begin{align} \label{eq:braket_def}
\ket{p\pm} &= \frac{1 \pm \ga_5}{2}\, u(p)
\,,\nn\\
\bra{p\pm} &= \mathrm{sgn}(p^0)\, \bar{u}(p)\,\frac{1 \mp \ga_5}{2}
\,.\end{align}
For massless particles chirality and helicity agree while for antiparticles they are opposite, so $\ket{p+} = u_+(p) = v_-(p)$ corresponds to positive (negative) helicity for particles (antiparticles). The spinors $\ket{p\pm}$ are defined by \eqs{Dirac}{braket_def} for both physical ($p^0 >0$) and unphysical ($p^0<0$) momenta. Their explicit expression, including our overall phase convention, is given in \app{helicity}.

The spinor products are denoted by
\begin{equation}
\langle p q \rangle = \braket{p-}{q+}
\,,\qquad
[p q] = \braket{p+}{q-}
\,.\end{equation}
They satisfy
\begin{align}
\langle pq \rangle = - \langle qp \rangle
\,,\quad
[pq] = - [qp]
\,,\quad
\langle pq \rangle [qp] = 2p\cdot q
\,.\end{align}
Additional relations are collected in \app{helicity}. The minus sign for $p^0 < 0$ in \eq{braket_def} is included so the spinor relations are invariant under inverting the signs of momenta, $p\to -p$, when crossing particles between the initial and final state, e.g. $\langle (-p)q \rangle [q(-p)] = 2(-p) \cdot q$.

If there are several momenta $p_i$, it is common to abbreviate
\begin{equation}
\ket{p_i\pm} = \ket{i\pm}
\,,\qquad
\langle p_i p_j \rangle = \langle ij \rangle
\,,\qquad
[p_i p_j] = [ij]
\,.\end{equation}

The polarization vectors of an outgoing gluon with momentum $p$ are given in the helicity formalism by
\begin{equation}
 \ve_+^\mu(p,k) = \frac{\mae{p+}{\ga^\mu}{k+}}{\sqrt{2} \langle kp \rangle}
\,,\quad
 \ve_-^\mu(p,k) = - \frac{\mae{p-}{\ga^\mu}{k-}}{\sqrt{2} [kp]}
\,,\end{equation}
where $k$ is an arbitrary reference vector with $k^2=0$, which fixes the gauge of the external gluons.
Using the relations in \app{helicity}, it is easy to check that
\begin{align}
p\cdot \ve_\pm(p,k) &= k\cdot \ve_\pm(p,k) = 0
\,,\nn\\
\ve_\pm(p,k) \cdot \ve_\pm(p,k) &= 0
\,,\nn\\
\ve_\pm(p,k) \cdot \ve_\mp(p,k) &= -1
\,,\nn\\
\ve_\pm^*(p,k) &= \ve_\mp(p,k)
\,,\end{align}
as is required for physical polarization vectors. With $p^\mu = E (1,0,0,1)$, the choice $k^\mu = E (1,0,0,-1)$ yields the conventional 
\begin{equation}
\ve_\pm^\mu(p, k) =
\frac{1}{\sqrt{2}}\, (0,1,\mp\img,0)\,.
\end{equation}
\subsection{SCET}
\label{subsec:scet}

Soft-collinear effective theory is an effective field theory of QCD that
describes the interactions of collinear and soft particles~\cite{Bauer:2000ew,
  Bauer:2000yr, Bauer:2001ct, Bauer:2001yt} in the presence of a hard
interaction.\footnote{Throughout this paper, we will for simplicity use the notation of SCET$_{\rm I}$. The theory SCET$_{\rm II}$ \cite{Bauer:2002aj} is required for a certain class of observables, for example $p_T$-dependent measurements or vetoes. The helicity operator formalism presented here applies identically to constructing SCET$_{\rm II}$ operators. The collinear operators and matching coefficients are the same for both cases. } Collinear particles are characterized by having large energy and
small invariant mass. To separate the large and small momentum components, it is
convenient to use light-cone coordinates. We define two light-cone vectors
\begin{equation}
n^\mu = (1, \vec{n})
\,,\qquad
\bn^\mu = (1, -\vec{n})
\,,\end{equation}
with $\vec{n}$ a unit three-vector, which satisfy $n^2 = \bn^2 = 0$ and  $n\cdot\bn = 2$.
Any four-momentum $p$ can be decomposed as
\begin{equation} \label{eq:lightcone_dec}
p^\mu = \bn\sdt p\,\frac{n^\mu}{2} + n\sdt p\,\frac{\bn^\mu}{2} + p^\mu_{n\perp}
\,.\end{equation}
An ``$n$-collinear'' particle has momentum $p$ close to the $\vec{n}$ direction,
so that $p$ scales as $(n\!\cdot\! p, \bn \!\cdot\! p, p_{n\perp}) \sim
\bn\!\cdot\! p$ $\,(\la^2,1,\la)$, with $\la \ll 1$ a small parameter. For example, for
a jet of collinear particles in the $\vec{n}$ direction with total momentum 
$p_J$, $\bn \sdt p_J \simeq 2E_J$ corresponds to the large energy of the jet, 
while $n \sdt p_J \simeq m_J^2/E_J \ll E_J$, where $m_J$ is the jet mass, so $\la^2 \simeq m_J^2/E_J^2 \ll 1$.

To construct the fields of the effective theory, the momentum of $n$-collinear particles is written as
\begin{equation} \label{eq:label_dec}
p^\mu = \lp^\mu + k^\mu = \bn\sdt\lp\, \frac{n^\mu}{2} + \lp_{n\perp}^\mu + k^\mu\,,
\,\end{equation}
where $\bn\cdot\lp \sim Q$ and $\lp_{n\perp} \sim \la Q$ are the large momentum
components, while $k\sim \la^2 Q$ is a small residual momentum. Here, $Q$ is the scale of the hard interaction,
and the effective theory expansion is in powers of $\la$.

The SCET fields for $n$-collinear quarks and gluons, $\xi_{n,\lp}(x)$ and
$A_{n,\lp}(x)$, are labeled by the collinear direction $n$ and their large
momentum $\lp$. They are written in position space with respect to the residual
momentum and in momentum space with respect to the large momentum components.
Derivatives acting on the fields pick out the residual momentum dependence,
$\img\partial^\mu \sim k \sim \la^2 Q$. The large label momentum is obtained
from the label momentum operator $\cP_n^\mu$, e.g. $\cP_n^\mu\, \xi_{n,\lp} =
\lp^\mu\, \xi_{n,\lp}$. If there are several fields, $\cP_n$ returns the sum of
the label momenta of all $n$-collinear fields. For convenience, we define
$\bnP_n = \bn\cdot\cP_n$, which picks out the large momentum component.  Frequently,
we will only keep the label $n$ denoting the collinear direction, while the
momentum labels are summed over (subject to momentum conservation) and are
suppressed in our notation.

Collinear operators are constructed out of products of fields and Wilson lines
that are invariant under collinear gauge
transformations~\cite{Bauer:2000yr,Bauer:2001ct}.  The smallest building blocks
are collinearly gauge-invariant quark and gluon fields, defined as
\begin{align} \label{eq:chiB}
\chi_{n,\w}(x) &= \Bigl[\delta(\w - \bnP_n)\, W_n^\dagger(x)\, \xi_n(x) \Bigr]
\,,\nn\\
\cB_{n,\w\perp}^\mu(x)
&= \frac{1}{g}\Bigl[\delta(\w + \bnP_n)\, W_n^\dagger(x)\,\img D_{n\perp}^\mu W_n(x)\Bigr]
\,.\end{align}
With this definition of $\chi_{n,\w}$, we have $\w > 0$ for an incoming quark and $\w < 0$ for an outgoing antiquark. For $\cB_{n,\w\perp}$, $\w > 0$ ($\w < 0$) corresponds to an outgoing (incoming) gluon.
In \eq{chiB}
\begin{equation}
\img D_{n\perp}^\mu = \cP^\mu_{n\perp} + g A^\mu_{n\perp}\,,
\end{equation}
is the collinear covariant derivative and
\begin{equation} \label{eq:Wn}
W_n(x) = \biggl[\sum_\text{perms} \exp\Bigl(-\frac{g}{\bnP_n}\,\bn\sdt A_n(x)\Bigr)\biggr]
\end{equation}
is a Wilson line of $n$-collinear gluons in label momentum space. 
The label operators $\bnP_n$ in \eqs{chiB}{Wn} only act inside the square brackets. 
$W_n(x)$ sums up arbitrary emissions of $n$-collinear gluons from an $ \bar n$-collinear quark
or gluon, which are $\ord{1}$ in the power counting. Since $W_n(x)$ is
localized with respect to the residual position $x$, we can treat
$\chi_{n,\w}(x)$ and $\cB_{n,\w}^\mu(x)$ like local quark and gluon fields. For later use, we give the expansion of the collinear gluon field
\begin{align}\label{eq:gluon_expand}
\cB^\mu_{n,\perp}=A^\mu_{n\perp}-\frac{p^\mu_\perp}{\bar n\cdot p}\bar n \cdot A_{n,p} +\cdots.
\end{align}
Here the ellipses denote terms in the expansion with more than 2 collinear gluon fields, which are not required for our matching calculations.

In our case the effective theory contains several collinear sectors, $n_1, n_2,
\ldots$~\cite{Bauer:2002nz}, where the collinear fields for a given sector
$n_i^\mu=(1,\vec n_i)$ describe a jet in the direction $\vec n_i$, and we also
define $\bar n_i^\mu=(1,-\vec n_i)$.  A fixed-order QCD amplitude with $N$
colored legs is then matched onto operators in SCET with $N$ different collinear
fields.  The different collinear directions have to be well separated, which
means
\begin{equation} \label{eq:nijsep}
  n_i\sdt n_j \gg \la^2 \qquad\text{for}\qquad i\neq j
\,.\end{equation}
The infrared singularities associated with collinear or soft limits of legs in
QCD are entirely described by the Lagrangian and dynamics of SCET itself, so the
QCD amplitudes are only used to describe the hard kinematics away from infrared
singular limits.

Two different $n_i$ and $n_i'$ with $n_i\cdot n_i' \sim \lambda^2$ both
describe the same jet and corresponding collinear physics. Thus, each collinear
sector can be labeled by any member of a set of equivalent vectors, $\{n_i\}$, which are related by reparametrization invariance~\cite{Manohar:2002fd}.
The simplest way to perform the matching is to choose $n_i$ such that the large
label momentum is given by
\begin{equation} \label{eq:pdefault}
\lp_i^\mu = \w_i\,\frac{n_i^\mu}{2} \,,
\end{equation}
with $\lp_{n_i \perp}^\mu =0$.

In general, operators will have sums over distinct equivalence classes, $\{n_i\}$, and matrix elements select a representative vector to describe particles in a particular collinear direction. For many leading power applications there is only a single collinear field in each sector, and we may simply set the large label momentum of that building block field to that of the external parton using the following simple relation,  
\begin{equation}
\int\!\df\lp\, \ldel(\lp - p)\,f(\lp) = f\Bigl(\bn_i\cdot p\,\frac{n_i}{2}\Bigr)
\,,\end{equation}
where $p$ is collinear with the $i$'th jet.
Here the tildes on the integration measure and delta function ensure that the integration over equivalence classes is properly implemented.\footnote{The precise definition of this delta function and measure are
\begin{align} \label{eq:labelsums}
\ldel(\lp_i - p) &\equiv \delta_{\{n_i\},p}\,\delta(\w_i - \bn_i\cdot p)
\,,\nn\\
\int\!\df\lp &\equiv \sum_{\{n_i\}} \int\!\df\w_i
\,,\end{align}
where 
\begin{equation}
\delta_{\{n_i\},p} =
\begin{cases}
   1 &\quad n_i\cdot p = \ord{\lambda^2}
   \,,\\
   0 &\quad \text{otherwise}
\,.\end{cases}
\end{equation}
The Kronecker delta is nonzero if the collinear momentum $p$ is in the $\{ n_i\}$
equivalence class, i.e. $p$ is close enough to be considered as collinear with
the $i$th jet.  The sum in the second line of \eq{labelsums} runs over the different
equivalence classes. }
Because of this, at leading power, the issue of equivalence classes can largely be ignored. 

Particles that exchange large momentum of $\ord{Q}$ between different jets are
off shell by $\ord{n_i\cdot n_j \,Q^2}$. They are integrated out by matching
QCD onto SCET.  Before and after the hard interaction the jets described by the
different collinear sectors evolve independently from each other, with only soft
radiation between the jets.  The corresponding soft degrees of freedom are
described in the effective theory by soft quark and gluon fields, $q_\soft(x)$
and $A_\soft(x)$, which only have residual soft momentum dependence
$\img\partial^\mu \sim \la^2Q$.  They couple to the collinear sectors via the
soft covariant derivative
\begin{equation}
\img D_\soft^\mu = \img \partial^\mu + g A_\soft^\mu\,,
\end{equation}
acting on the collinear fields. At leading power in $\la$, $n$-collinear
particles only couple to the $n\sdt A_\soft$ component of soft gluons, so the
leading-power $n$-collinear Lagrangian only depends on $n\sdt D_\soft$. For example, for
$n$-collinear quarks~\cite{Bauer:2000yr, Bauer:2001ct}
\begin{equation} \label{eq:L_n}
\cL_n = \bar{\xi}_n \Bigl(\img n\sdt D_\soft + g\,n\sdt A_n 
 + \img\Dslash_{n\perp} W_n \frac{1}{\bnP_n}\, W_n^\dagger\,\img\Dslash_{n\perp}
 \Bigr)\frac{\bnslash}{2} \xi_n
\,.\end{equation}
The leading-power $n$-collinear Lagrangian for gluons is given in
Ref.~\cite{Bauer:2001yt}.

\section{SCET Operator Basis}
\label{sec:basis}

In this section, we describe in detail how to construct a basis of helicity and color operators in SCET, which greatly simplifies the construction of a complete operator basis and also facilitates the matching process. Usually, a basis of SCET operators obeying the symmetries of the problem is constructed from the fields $\chi_{n,\omega}$, $\cB^\mu_{n,\omega \perp}$, as well as Lorentz and color structures. This process becomes quite laborious due to the large number of structures which appear for higher multiplicity processes, and the reduction to a minimal basis of operators quickly becomes nontrivial. Instead, we work with a basis of operators with definite helicity structure constructed from scalar SCET building blocks, which, as we will show, has several advantages. First, this simplifies the construction of the operator basis, because each independent helicity configuration gives rise to an independent
helicity operator. In this way, we automatically obtain the minimal number of
independent operators as far as their Lorentz structure is concerned.  Second, operators with distinct helicity structures do not mix under renormalization group evolution, as will be discussed in detail in \sec{running}.  The reason is that
distinct jets can only exchange soft gluons in SCET, which at leading order in
the power counting means they can transfer color but not spin [see \eq{L_n}]. Therefore, the only nontrivial aspect of the operator basis is the color degrees of freedom. The different color structures mix under renormalization group evolution, but their mixing only depends on the color representations and not on the specific helicity configuration.

\subsection{Helicity Fields}
\label{subsec:fields}

We start by defining quark and gluon fields of definite helicity, out of which we
can build operators with a definite helicity structure. To simplify our
discussion we will take all momenta and polarization vectors as outgoing, and
label all fields and operators by their outgoing helicity and momenta. Crossing symmetry, and crossing relations are discussed in \subsec{crossing}.

We define a gluon field of definite helicity%
\footnote{The label $\pm$ on $\cB_\pm$ refers to helicity and should not be confused with light-cone components.}
\begin{equation} \label{eq:cBpm_def}
\cB^a_{i\pm} = -\ve_{\mp\mu}(n_i, \bn_i)\,\cB^{a\mu}_{n_i,\w_i\perp_i}
\,,\end{equation}
where $a$ is an adjoint color index. For $n_i^\mu=(1,0,0,1)$, 
we have
\begin{equation}
\ve_\pm^\mu(n_i, \bn_i) = \frac{1}{\sqrt{2}}\, (0,1,\mp\img,0)
\,,\end{equation}
in which case
\begin{equation}
\cB^a_{i\pm} = \frac{1}{\sqrt{2}} \bigl(\cB^{a,1}_{n_i,\w_i\,\perp_i}
   \pm \img \cB^{a,2}_{n_i,\w_i\,
  \perp_i} \bigr)
\,.\end{equation}

For an external gluon with outgoing polarization vector $\ve(p,k)$ and
outgoing momentum $p$ in the $n_i$-collinear direction, the contraction with
the field $\cB_{i\pm}^a$ contributes 
\begin{equation}\label{eq:pol_A}
 -\ve_{\mp\mu}(n_i, \bn_i)\Bigl[\ve_{\perp_i}^\mu(p,k)
 - \frac{p_{\perp_i}^\mu}{\bn_i\cdot p}\, \bn_i\cdot \ve(p,k)\Bigr]
\,,\end{equation}
where we have used the expansion of the collinear gluon field given in \eq{gluon_expand}.
Since $\ve_{\mp}(n_i,\bn_i)$ is perpendicular to both $n_i$ and $\bn_i$, we
can drop the $\perp_i$ labels in brackets. A convenient choice for the reference
vector is to take $k = \bn_i$, for which the second term in brackets vanishes. Equation~\eqref{eq:pol_A} then becomes
\begin{equation}
-\ve_\mp(n_i, \bn_i)\cdot\ve(p,\bn_i)
\,,\end{equation}
which is equal to 0 or 1 depending on the helicity of $\ve(p,\bn_i)$.
Adopting this choice, the tree-level Feynman rules for
an outgoing gluon with polarization $\pm$ (so $\ve =\ve_\pm$), momentum
$p$ (with $p^0 > 0$), and color $a$ are
\begin{align}
 \Mae{g_\pm^a(p)}{\cB_{i\pm}^b}{0} &= \delta^{a b}\, \ldel(\lp_i - p)
\,,\nn\\
 \Mae{g_\mp^a(p)}{\cB_{i\pm}^b}{0} &= 0
\,.\end{align}
Note that $\cB_{i\pm}^b=\cB_{i\pm}^b(0)$, so we do not get a phase from the
residual momentum.  Similarly, for an incoming gluon with incoming polarization
$\mp$ ($\ve = \ve_\mp$, so $\ve^* = \ve_\pm$), incoming momentum $-p$
(with $p^0 < 0$), and color $a$, we have
\begin{align}
 \Mae{0}{\cB_{i\pm}^b}{g_\mp^a(-p)} &= \delta^{ab}\, \ldel(\lp_i - p)
\,,\nn\\
 \Mae{0}{\cB_{i\pm}^b}{g_\pm^a(-p)} &= 0
\,.\end{align}

We define quark fields with definite helicity\footnote{Technically speaking chirality, although we work in a limit where all external quarks can be treated as massless.} as
\begin{equation} \label{eq:chipm_def}
\chi^\alpha_{i\pm} = \frac{1\pm\gamma_5}{2}\,\chi_{n_i,-\w_i}^\alpha
\,,\qquad
\bar\chi^{\bar \alpha}_{i\pm} = \bar\chi_{n_i,\w_i}^{\bar \alpha}\,\frac{1\mp\gamma_5}{2}
\,,\end{equation}
where $\alpha$ and $\bar \alpha$ are fundamental and antifundamental color indices respectively.

For external quarks with
$n_i$-collinear momentum $p$, the fields contribute factors of the form
\begin{align}
\frac{1\pm\gamma_5}{2}\,\frac{\nslash_i\bnslash_i}{4}\,u(p)
&= \frac{\nslash_i\bnslash_i}{4}\,\ket{p\pm}
= \ket{p\pm}_{n_i}
\,,\end{align}
where in the last equality, we have defined a shorthand notation $\ket{p\pm}_{n_i}$ for the SCET projected spinor. The spinor $\ket{p\pm}_{n_i}$ is proportional to $\ket{n\pm}$; see \eq{ketn}.

The tree-level Feynman rules for incoming ($p^0 < 0$)
and outgoing ($p^0 > 0$) quarks with helicity $+/-$ and color $\alpha$ are then given by
\begin{align} \label{eq:chipm_q}
\Mae{0}{\chi^\beta_{i+}}{q_+^\balpha(-p)}
&= \delta^{\beta\balpha}\,\ldel(\lp_i - p)\, \ket{(-p_i)+}_{n_i}
\,,\nn\\
\Mae{0}{\chi^\beta_{i-}}{q_-^\balpha(-p)}
&= \delta^{\beta\balpha}\,\ldel(\lp_i - p)\, \ket{(-p_i)-}_{n_i}
\,,\nn\\
\Mae{q_+^\alpha(p)}{\bar\chi^\bbeta_{i+}}{0}
&= \delta^{\alpha\bbeta}\,\ldel(\lp_i - p)\, {}_{n_i\!}\bra{p_i+}
\,,\nn\\
\Mae{q_-^\alpha(p)}{\bar\chi^\bbeta_{i-}}{0}
&= \delta^{\alpha\bbeta}\,\ldel(\lp_i - p)\, {}_{n_i\!}\bra{p_i-}
\,,\end{align}
and similarly for antiquarks
\begin{align} \label{eq:chipm_qbar}
\Mae{0}{\bar\chi^\bbeta_{i+}}{\bar{q}_-^\alpha(-p)}
&= \delta^{\alpha\bbeta}\,\ldel(\lp_i - p)\,{}_{n_i\!}\bra{(-p_i)+}
\,, \nn \\
\Mae{0}{\bar\chi^\bbeta_{i-}}{\bar{q}_+^\alpha(-p)}
&= \delta^{\alpha\bbeta}\,\ldel(\lp_i - p)\,{}_{n_i\!}\bra{(-p_i)-}
\,, \nn \\
\Mae{\bar{q}_-^\balpha(p)}{\chi^\beta_{i+}}{0}
&= \delta^{\beta\balpha}\,\ldel(\lp_i - p)\, \ket{p_i+}_{n_i}
\,, \nn \\
\Mae{\bar{q}_+^\balpha(p)}{\chi^\beta_{i-}}{0}
&= \delta^{\beta\balpha}\,\ldel(\lp_i - p)\, \ket{p_i-}_{n_i}
\,.\end{align}
The corresponding Feynman rules with the helicity of the external (anti)quark
flipped vanish.

To avoid the explicit spinors in \eqs{chipm_q}{chipm_qbar}, and exploit the fact
that fermions come in pairs, we also define fermionic vector currents of definite helicity
\begin{align} \label{eq:jpm_def}
J_{ij+}^{\balpha\beta}
&=  \frac{\sqrt{2}\, \ve_-^\mu(n_i, n_j)}{\sqrt{\phantom{2}\!\!\omega_i \,  \omega_j }}\, \frac{\bar{\chi}^\balpha_{i+}\, \gamma_\mu \chi^\beta_{j+}}{\langle n_i n_j\rangle}\,, \nn \\
J_{ij-}^{\balpha\beta}
&= -\, \frac{ \sqrt{2}\, \ve_+^\mu(n_i, n_j)}{\sqrt{\phantom{2}\!\! \omega_i \,  \omega_j }}\, \frac{\bar{\chi}^\balpha_{i-}\, \gamma_\mu \chi^\beta_{j-}}{[n_i n_j]}
\,,\end{align}
where $\omega_i=\bn_i\cdot \tilde p_i$ from \eq{pdefault}, as well as a scalar current
\begin{align} \label{eq:jS_def}
J_{ij0}^{\balpha\beta}&=\frac{2}{\sqrt{\phantom{2}\!\! \omega_i \,  \omega_j}}\frac{\bar \chi^\balpha_{i+}\chi^\beta_{j-}}{ [n_i n_j]   }\,, \nonumber \\
(J^\dagger)_{ij0}^{\balpha\beta}&=  \frac{2}{\sqrt{\phantom{2}\!\! \omega_i \,  \omega_j}}  \frac{\bar \chi^\balpha_{i-}\chi^\beta_{j+}} { \langle n_i  n_j \rangle }
\,.\end{align}
In \eqs{jpm_def}{jS_def} the flavor labels of the quarks have not been made explicit, but in general the two quark fields in a current can have different flavors (for example in $W$ production). Since we are using a basis of physical polarization states it is not necessary to introduce more complicated Dirac structures. For example, pseudovector and pseudoscalar currents, which are usually introduced using $\gamma^5$, are incorporated through the relative coefficients of operators involving $J_+$, $J_-$ or $J_0$, $J_0^\dagger$. As we shall see, this greatly simplifies the construction of the operator basis in the effective theory.

At leading power, there is a single collinear field in each collinear sector, so we can choose $n_i^\mu=p_i^\mu/p_i^0$ to represent the equivalence class $\{n_i\}$, so that $p_i^\mu = \frac12 \bn\cdot p_i\, n_i^\mu$ which gives
\begin{equation}
\ket{p\pm}_{n_i} = \ket{p\pm} = \Bigl\lvert\bn_i\cdot p\, \frac{n_i}{2}\pm\Bigr\rangle
  = \sqrt{\frac{\bn_i\cdot p}{2}}\: \ket{n_i\pm} \,.
\end{equation}
Since we always work at leading power in this paper, we will always make this choice to simplify the matching.  With this choice, the tree-level Feynman rules for the fermion currents are
\begin{align}\label{eq:feyn_tree}
&\Mae{q_+^{\alpha_1}(p_1)\,\bar{q}_-^{\balpha_2}(p_2)}{J_{12+}^{\bbeta_1\beta_2}}{0}
\,\nn\\ & \quad
= \delta^{\alpha_1\bbeta_1}\,\delta^{\beta_2\balpha_2}\, \ldel(\lp_1 - p_1)\,\ldel(\lp_2 - p_2)
\,, \nn \\
&\Mae{q_-^{\alpha_1}(p_1)\,\bar{q}_+^{\balpha_2}(p_2)}{J_{12-}^{\bbeta_1\beta_2}}{0}
\,\nn\\ & \quad
= \delta^{\alpha_1\bbeta_1}\,\delta^{\beta_2\balpha_2}\, \ldel(\lp_1 - p_1)\,\ldel(\lp_2 - p_2)
\,, \nn \\
&\Mae{q_+^{\alpha_1}(p_1)\,\bar{q}_+^{\balpha_2}(p_2)}{J_{12\,0}^{\bbeta_1\beta_2}}{0}
\,\nn\\ & \quad
= \delta^{\alpha_1\bbeta_1}\,\delta^{\beta_2\balpha_2}\, \ldel(\lp_1 - p_1)\,\ldel(\lp_2 - p_2)
\,, \nn \\
&\Mae{q_-^{\alpha_1}(p_1)\,\bar{q}_-^{\balpha_2}(p_2)}{(J^\dagger)_{12\,0}^{\bbeta_1\beta_2}}{0}
\,\nn\\ & \quad
= \delta^{\alpha_1\bbeta_1}\,\delta^{\beta_2\balpha_2}\, \ldel(\lp_1 - p_1)\,\ldel(\lp_2 - p_2)
\,.\end{align}
The simplicity of these Feynman rules arises due to the unconventional normalization of the operators in \eqs{jpm_def}{jS_def}. This normalization has been chosen to simplify the matching of QCD amplitudes onto SCET operators, as will be seen in \sec{matching}.

We will also make use of leptonic versions of the above currents. These are defined analogously,
\begin{align} \label{eq:jpm_lep_def}
J_{ij+}&=  \frac{\sqrt{2}\, \ve_-^\mu(n_i, n_j)}{\sqrt{\phantom{2}\!\!\omega_i \,  \omega_j }}\,  \frac{\bar{\ell}_{i+}\, \gamma_\mu \ell_{j+}}{\langle n_i n_j \rangle}\,, \nonumber \\
J_{ij-}&=-\, \frac{\sqrt{2}\, \ve_+^\mu(n_i, n_j)}{\sqrt{\phantom{2}\!\! \omega_i \,  \omega_j}}\, \frac{\bar{\ell}_{i\pm}\, \gamma_\mu \ell_{j\pm}}{[n_i n_j]}
\,.\end{align}
Unlike the collinear quark field $\chi$, the leptonic field $\ell$ does not carry color and so does not contain a strong-interaction Wilson line.

All couplings in the SM, except to the Higgs boson, preserve chirality. This limits the need for the scalar current, especially when considering only massless external quarks. In the SM the scalar current can arise through explicit couplings to the Higgs, in which case, even though we still treat the external quarks as massless, the Wilson coefficient for the scalar operator will contain the quark Yukawa coupling. This is relevant for example for $H b\bar b$ processes. The scalar current can also arise through off-diagonal CKM-matrix elements connecting two massless external quarks through a massive quark appearing in a loop. This can occur in multiple vector boson production, or from electroweak loop corrections, neither of which will be discussed in this paper. When constructing an operator basis in \subsec{helicityops}, we ignore the scalar current, as it is not relevant for the examples that we will treat in this paper. However, it should be clear that the construction of the basis in \subsec{helicityops} can be trivially generalized to incorporate the scalar current if needed.

\subsection{Helicity Operator Basis}
\label{subsec:helicityops}

Using the definitions for the gluon and quark helicity fields in \eqs{cBpm_def}{jpm_def}, we can construct operators for a given number of external partons with definite helicities and color. (As discussed at the end of the previous section, for the processes we consider in this paper we do not require the scalar current $J_S$.) In the general case with CKM-matrix elements, we must allow for the two quark flavors within a single current to be different. The situation is simplified in QCD processes, where one can restrict to currents carrying a single flavor label.

For an external state with $n$ particles of definite helicities $\pm$, colors $a_i$, $\alpha_i$, $\bar \alpha_i$, and flavors $f$, $f'$, ..., a complete basis of operators is given by
\begin{align} \label{eq:Opm_gen}
&O_{\pm\pm\dotsb(\pm\dotsb;\dotsb\pm)}^{a_1 a_2 \dotsb \balpha_{i-1} \alpha_i \dotsb \balpha_{n-1} \alpha_n}
(\lp_1, \lp_2, \ldots, \lp_{i-1}, \lp_i, \ldots, \lp_{n-1}, \lp_n)
\nn\\ & \qquad
= S\, \cB_{1\pm}^{a_1}\,\cB_{2\pm}^{a_2}\dotsb J_{f\,i-1,i\pm}^{\balpha_{i-1}\alpha_i}
\dotsb J_{f'\,n-1,n\pm}^{\balpha_{n-1}\alpha_n}
\,.\end{align}
For example, $f=q$ indicates that both quark fields in the current have flavor $q$. When it is necessary to distinguish different flavors with the same current, for example when we consider processes involving W bosons in \sec{vec}, we use a label $f=\bar u d$ such as $J_{\bar u d 12-}$. For simplicity, we will also often suppress the dependence of the operator on the label momenta $\lp_i$. For the operator subscripts, we always put the helicity labels of the gluons first and those of the quark currents in brackets, with the labels for quark currents with different flavor labels $f$ and $f'$ separated by a semicolon, as in \eq{Opm_gen}. The $\pm$ helicity labels of the individual gluon fields and quark currents can all vary independently. Operators with nonzero matching coefficients are restricted to the color-conserving subspace. We will discuss the construction of the color basis in \subsec{color}.

The symmetry factor $S$ in \eq{Opm_gen} is included to simplify the matching. It is given by
\begin{equation} \label{eq:Opm_S}
S = \frac{1}{\prod\limits_i n_{i}^+!\, n_{i}^-!}
\,,\end{equation}
where $n_i^\pm$ denotes the number of fields of type $i=g,u,\bar u, d, \bar d, \dots$ with helicity $\pm$. We also use
\begin{equation}
n = \sum_i (n_i^+ + n_i^-)\,,
\end{equation}
to denote the total number of fields in the operator. Each ${\cal B}_i$ counts as one field, and each $J$ has two fields.

For each set of external particles of definite helicities, colors, and flavors, there is only one independent operator, since the physical external states have been completely specified. All Feynman diagrams contributing to this specific external state will be included in the Wilson coefficient of that specific operator. For the case of pure QCD, quarks always appear in pairs of the same flavor and same chirality, and therefore can be assembled into quark currents labeled by a single flavor. In this case, to keep track of the minimal number of independent operators, we can simply order the helicity labels, and only consider operators of the form
\begin{align} \label{eq:op_struct}
O_{+\cdot\cdot(\cdot\cdot-)}
&= O_{\underbrace{+\dotsb+}\,\underbrace{-\dotsb-}\,(\underbrace{+\dotsb+}\,\underbrace{-\dotsb-})}
\,,
\\\nn &\qquad\quad
n_g^+ \qquad\, n_g^- \,\,\qquad n_q^+ \,\qquad n_q^-
\end{align}
and analogously for any additional quark currents with different quark flavors.\footnote{In the general case with off-diagonal CKM-matrix elements, there is some more freedom in the choice of the operator basis, because quarks of the same flavor do not necessarily appear in pairs. However, it is still true that only a single operator is needed for a specific external state. For example, for external quarks $u_-, \bar d_+, \bar s_+, c_-$, one could either use the operators $J_{us-}J_{cd-}, \text{or the operators }J_{cs-}J_{ud-}$ (where the color structures have been suppressed). Since different helicity combinations are possible, a single flavor assignment does not suffice to construct a complete helicity basis, and one must sum over a basis of flavor assignments. As an example explicitly demonstrating this, we will consider the case of $pp \to W +$ jets in \sec{vec}.}

With the operator basis constructed, for a given $n$-parton process we can match hard scattering processes in QCD onto the leading-power hard-scattering Lagrangian
\begin{equation} \label{eq:Leff}
\cL_\hard \!=\!\! \int\!\! \prod_{i=1}^n\df \lp_i\,
C_{+\cdot\cdot(\cdot\cdot-)}^{a_1\dotsb \alpha_n}(\lp_1, \ldots, \lp_n)\,
O_{+\cdot\cdot(\cdot\cdot-)}^{a_1 \dotsb \alpha_n}(\lp_1, \ldots, \lp_n)
 ,
\end{equation}
where a sum over all color indices is implicit. Lorentz invariance implies that the Wilson coefficients only depend on Lorentz invariant combinations of the momenta. This hard Lagrangian is used in conjunction with the collinear and soft Lagrangians that describe the dynamics of the soft and collinear modes; see for example \eq{L_n}.

We emphasize that \eq{Leff} provides a complete basis in SCET for well-separated jets and additional nonhadronic particles at leading power.  We will discuss in more detail in \sec{matching} the matching and regularization schemes, and demonstrate that no evanescent operators are generated for this case. At subleading power, the SCET operators would involve additional derivative operators, soft fields, or multiple SCET building blocks from the same collinear sector.

\subsection{Example with a $Z$-Boson Exchange}

It is important to note that all kinematic dependence of the hard process, for example, its angular distributions, is encoded in the Wilson coefficients. Since the Wilson coefficients can (in principle) carry an arbitrary kinematic dependence, our choice of helicity basis imposes no restriction on the possible structure or mediating particles of the hard interaction. For example, the spin of an intermediate particle may modify the angular distribution of the decay products, and hence the Wilson coefficients, but this can always be described by the same basis of helicity operators.

As a simple example to demonstrate this point we consider $e^+e^-\to e^+e^- $ at tree level. This process can proceed through either an off-shell $\gamma$ or $Z$ boson. Because the SM couplings to both of these particles preserve chirality, a basis of operators for this process is given by
\begin{align}
O_{(++)}
&= \frac{1}{4}\, J_{e\,12+}\, J_{e\,34+}
\,, \nn \\
O_{(+-)}
&= J_{e\,12+}\, J_{e\,34-}
\,, \nn \\
O_{(--)}
&= \frac{1}{4}\, J_{e\,12-}\, J_{e\,34-}
\,,\end{align}
where the leptonic current is defined in \eq{jpm_lep_def}. The fact that this is a complete basis relies only on the fact that the couplings preserve chirality, and is independent of e.g.~the possible number of polarizations of the mediating $Z$ or $\gamma$.

We now consider the calculation of the Wilson coefficients for the matching to these operators (the matching procedure is discussed in detail in \sec{matching}). At tree level, the Wilson coefficients are easily calculated, giving
\begin{align} \label{eq:Cexample}
C_{(++)}&=- e^2  \bigl[1 + v_{R}^e v_{R}^e P_Z(s_{12})\bigr]
 \frac{2[13]\ang{24}}{s_{12}} + (1 \lra 3) \,,
 \nn \\
C_{(+-)}&=-e^2  \bigl[1+ v_{R}^e v_{L}^e P_Z(s_{12})\bigr]
\frac{2[14]\ang{23}}{s_{12}}  \,,
\nn \\
C_{(--)}&=-e^2  \bigl[1 + v_{L}^ev_{L}^e P_Z(s_{12})\bigr]
\frac{2[24]\ang{13}}{s_{12}} + (1 \lra 3) \,.
\end{align}
Here $s_{12}=(p_1+p_2)^2$, $P_Z$ is the ratio of the $Z$ and photon propagators,
\begin{equation}
P_{Z}(s) = \frac{s}{s-m_Z^2 + \img \Ga_{Z} m_{Z}}
\,,\end{equation}
and the couplings $v_{L,R}$ to the $Z$ boson are
\begin{align}
 v_L^e = \frac{1 - 2 \sin^2 \theta_W}{\sin(2\theta_W)}
 \,, \quad
 v_R^e= - \frac{2 \sin^2 \theta_W}{\sin(2\theta_W)}
\,.\end{align}
Note that the presence of the spinor factors in \eq{Cexample} occur due to our normalization conventions for the currents. 

Now, consider calculating the scattering amplitude in the effective theory, for example for the case when both electrons have positive helicity. The matrix element in the effective theory gives
\begin{align}
&\Mae{e^-_+(p_1)\,e^+_-(p_2)\, e^-_+(p_3)\,e^+_-(p_4)}{\img \cL_\hard}{0}
\nn\\ 
& 
= \img \MAe{e^-_+(p_1)\,e^+_-(p_2)\, e^-_+(p_3)\,e^+_-(p_4)}{\int\!\prod_{i=1}^n\df \lp_i\,C_{++} O_{++}}{0}
\nn\\ 
& 
= - \img e^2  \big[1 + v_{R}^e v_{R}^e P_Z(s_{12})\big] 
 \frac{2[13]\ang{24}}{s_{12}} + (1 \lra 3) \,,
\end{align}
using the Feynman rules of \eq{feyn_tree}. The effective theory therefore reproduces the full theory scattering amplitude. The same is true of the other helicity configurations, so the familiar angular distributions for $e^+e^-\to e^+e^- $, as well as the different couplings of the $Z$ to left- and right-handed particles, are entirely encoded in the Wilson coefficients.

\subsection{Color Basis} \label{subsec:color}

In addition to working with a basis of operators with definite helicity, we can also choose a color basis that facilitates the matching. When constructing a basis of operators in SCET, we are free to choose an arbitrary color basis. With respect to color, we can think of \eq{Leff} as having a separate Wilson coefficient for each color configuration. For specific processes the color structure of the Wilson coefficients can be further decomposed as
\begin{equation} \label{eq:Cpm_color}
C_{+\cdot\cdot(\cdot\cdot-)}^{a_1\dotsb\alpha_n}
= \sum_k C_{+\cdot\cdot(\cdot\cdot-)}^k T_k^{a_1\dotsb\alpha_n}
\equiv \vT^{ a_1\dotsb\alpha_n} \vC_{+\cdot\cdot(\cdot\cdot-)}
\,.\end{equation}
Here, $\vT^{ a_1\dotsb\alpha_n}$ is a row vector whose entries  $T_k^{a_1\dotsb\alpha_n}$ are suitable color structures that together
provide a complete basis for all allowed color structures, but which do not necessarily all have to be independent. In other words, the elements of $\vT^{ a_1\dotsb\alpha_n}$ span the color-conserving subspace of the full color space spanned by $\{a_1 \dotsb\alpha_n\}$, and $\vC$ is a vector in this subspace. Throughout this paper we will refer to the elements of $\vT^{ a_1\dotsb\alpha_n}$ as a color basis, although they will generically be overcomplete, since this allows for simpler choices of color structures. As discussed below, due to the overcompleteness of the bases, some care will be required for their consistent usage.

Using \eq{Cpm_color}, we can rewrite \eq{Leff} as
\begin{equation} \label{eq:Leff_alt}
\cL_\hard = \!\int\!\prod_{i=1}^n\!\df \lp_i\,
\Op^\dagger_{+\cdot\cdot(\cdot\cdot-)}(\lp_1, \ldots, \lp_n)
\vC_{+\cdot\cdot(\cdot\cdot-)}(\lp_1, \ldots, \lp_n)
,\end{equation}
where $\Op^\dagger$ is a conjugate vector defined by
\begin{equation} \label{eq:Opm_color}
\Op^\dagger_{+\cdot\cdot(\cdot\cdot-)}
= O_{+\cdot\cdot(\cdot\cdot-)}^{a_1\dotsb\alpha_n}\, \vT^{ a_1\dotsb\alpha_n}
\,.\end{equation}
While the form $C_{+\cdot\cdot(\cdot\cdot-)}^{a_1\dotsb \alpha_n}\,
O_{+\cdot\cdot(\cdot\cdot-)}^{a_1 \dotsb \alpha_n}$ in \eq{Leff} is more convenient to discuss the matching and the symmetry properties of operators and Wilson coefficients, the alternative form in \eq{Leff_alt} is more convenient to discuss the mixing of the color structures under renormalization.

For low multiplicities of colored particles it can be convenient to use orthogonal color bases, e.g., the singlet-octet basis for $q\bar q q' \bar q'$ is orthogonal. However, using orthogonal bases becomes increasingly difficult for higher multiplicity processes, and the color bases used for many fixed-order calculations are not orthogonal. (See e.g.\ Refs.~\cite{Keppeler:2012ih, Sjodahl:2015qoa} for a discussion of the use of orthogonal bases for $SU(N)$.) The use of a nonorthogonal color basis implies that when written in component form in a particular basis, the conjugate $\vC^\dagger$ of the vector $\vC$ is not just given by the naive complex conjugate transpose of the components of the vector. Instead, we have
\begin{align}\label{eq:def_dagger}
 \vec C^\dagger = \left[ C^{ a_1\dotsb\alpha_n} \right]^* \vT^{a_1\dotsb\alpha_n}
= \vC^{*T}\, \hT
\,,\end{align}
where
\begin{equation}\label{eq:hatT_def}
\hT = \sum_{a_1,\ldots,\alpha_n} (\vT^{a_1\dotsb\alpha_n})^\dagger \vT^{a_1\dotsb\alpha_n}
\end{equation}
is the matrix of color sums for the chosen basis. If the basis is orthogonal (orthonormal), then $\hT$ is a diagonal matrix (identity matrix). Note that \eq{hatT_def} implies that by definition $\hT^{* T} = \hT$.

Similar to \eq{def_dagger}, for an abstract matrix $\widehat X$ in color space, the components of its Hermitian conjugate $\widehat X^\dagger$ when written in a particular basis are given in terms of the components of $\widehat X$ as
\begin{equation} \label{eq:def_daggermatrix}
\widehat X^\dagger = \hT^{-1}\, \widehat X^{*T}\, \hT
\,.\end{equation}

A proper treatment of the nonorthogonality of the color basis is also important in the factorization theorem of \eq{sigma}. Here, the color indices of the Wilson coefficients are contracted with the soft function as
\begin{align} \label{eq:trHS}
\left[C^{ a_1 \dotsb\alpha_n} \right]^*  S_\kappa^{a_1 \dotsb\alpha_n  b_1\dotsb\beta_n }    C^{b_1\dotsb\beta_n}
&= \vC^\dagger\, \hS_\kappa \vC \nonumber \\
& = \vC^{*T} \hT\, \hS_\kappa \vC
\,.\end{align}
At tree level, the soft function is simply the color-space identity
\begin{align}\label{eq:soft_id}
\hS_\kappa = \id \,,
\end{align}
which follows from its color basis independent definition in terms of Wilson lines [see e.g.\ Ref.~\cite{Jouttenus:2011wh} or \eq{softfunction_def}]. Here we have suppressed the dependence of $\hS$ on soft momenta. The action of the identity on an element of the color space is defined by
\begin{align}
(\id \vT)^{\cdots a_i \cdots \alpha_j \cdots} = \vT^{\cdots a_i \cdots \alpha_j \cdots}\,,
\end{align}
and its matrix representation in any color basis is given by $\id = \text{diag}(1,1,\cdots , 1)$.
In the literature, see e.g.\ Refs.~\cite{Kidonakis:1998nf,Kelley:2010fn, Kelley:2010qs,Broggio:2014hoa,Becher:2015gsa}, often a different convention is used, where the $\hT$ matrix is absorbed into the definition of the soft function. In this convention, the soft function becomes explicitly basis dependent and is not the same as the basis-independent color-space identity.  One should be careful to not identify the two.

As an example to demonstrate our notation for the color basis, consider the process $ggq\bq$. A convenient choice for a complete basis of color structures is
\begin{align} \label{eq:ggqqcol}
\vT^{ ab \alpha\bbeta}
&= \Bigl(
   (T^a T^b)_{\alpha\bbeta}\,,\, (T^b T^a)_{\alpha\bbeta} \,,\, \tr[T^a T^b]\, \delta_{\alpha\bbeta}
   \Bigr)
\nn\\
&\equiv \begin{pmatrix}
   (T^a T^b)_{\alpha\bbeta} \\ (T^b T^a)_{\alpha\bbeta} \\ 
   \tr[T^a T^b]\, \delta_{\alpha\bbeta}
\end{pmatrix}^{\!\!\!T}
.\end{align}
For cases with many color structures we will write $\vT$ as the transpose of a column vector as above. The transpose in this case only refers to the vector itself, not to the individual color structures. The color-sum matrix for this particular basis is
\begin{align}
\widehat T_{ggq\bq}
&= (\vT^{ ab \alpha\bbeta})^\dagger \vT^{ ab \alpha\bbeta}
\nn \\
&= \frac{C_F N}{2}
\begin{pmatrix}
   2C_F & 2C_F - C_A & 2T_F \\
   2C_F - C_A & 2C_F & 2T_F \\
   2T_F & 2T_F & 2T_F N
\end{pmatrix}.
\end{align}
Our conventions for color factors are given in \app{color}.
Explicit expressions for $\widehat T$ for the bases used in this paper are given in \app{treesoft} for up to five partons.

Depending on the application, different choices of color basis can be used.
For example, in fixed-order QCD calculations, color ordering  \cite{Berends:1987me,Mangano:1987xk,Mangano:1988kk,Bern:1990ux} is often used to organize color information and simplify the singularity structure of amplitudes, while the color flow basis \cite{Maltoni:2002mq} is often used to interface with Monte Carlo generators. For a brief review of the color decomposition of QCD amplitudes, see \app{color_decomp}. Choosing a corresponding color basis in SCET has the advantage that the Wilson coefficients are given directly by the finite parts of the color-stripped helicity amplitudes, as defined in \eq{matching_general}, which can be efficiently calculated using unitarity methods. In this paper we will use color bases corresponding to the color decompositions of the QCD amplitudes when giving explicit results for the matching coefficients, although we emphasize that an arbitrary basis can be chosen depending on the application.

Finally, note that the color structures appearing in the decomposition of a QCD amplitude up to a given loop order may not form a complete basis. The color basis in SCET must be complete even if the matching coefficients of some color structures are zero to a given loop order, since all structures can in principle mix under renormalization group evolution, as will be discussed in \sec{running}. In this case, we always choose a complete basis in SCET such that the color structures appearing in the amplitudes to some fixed order are contained as a subset.

\subsection{Parity and Charge Conjugation}
\label{subsec:discrete}

Under charge conjugation, the fields transform as
\begin{align} \label{eq:Cfield}
\C\, \cB^a_{i\pm}\, T^a_{\alpha\bbeta}\,\C &= - \cB^a_{i \pm} T^a_{\beta\balpha}
\,,\nn\\
\C\, J^{\balpha\beta}_{ij\pm}\,\C &= -J^{\bbeta\alpha}_{ji\mp}
\,.\end{align}
The minus sign on the right-hand side of the second equation comes from
anticommutation of the fermion fields.

Under parity, the fields transform as
\begin{align} \label{eq:Pfield}
\P\, \cB^a_{i\pm}(\lp_i, x)\, \P
&= e^{\pm2\img \phi_{n_i}}\cB^a_{i \mp}(\lp_i^\P, x^\P)
\,,\nn\\
\P\, J^{\balpha\beta}_{ij\pm}(\lp_i, \lp_j, x)\, \P
&= e^{\pm \img (\phi_{n_i}-\phi_{n_j})} J^{\balpha\beta}_{ij\mp}(\lp_i^\P, \lp_j^\P, x^\P)
\,,\end{align}
where we have made the dependence on $\lp_i$ and $x$ explicit, and the
parity-transformed vectors are $\lp_i^\P = \w_i\,\bn_i/2$, $x^\P_\mu = x^\mu$. The $\phi_{n_i}$ are real phases, whose exact definition is given in \app{helicity}. The phases appearing in the parity transformation of the helicity operators exactly cancel the phases appearing in the corresponding helicity amplitude under a parity transformation. This overall phase is determined by the little group scaling (see \app{helicity} for a brief review).

Using the transformations of the helicity fields under parity and charge
conjugation in \eqs{Cfield}{Pfield}, it is straightforward to determine how
these discrete symmetries act on the helicity operators. Parity and charge
conjugation invariance of QCD implies that the effective Lagrangian in \eq{Leff}
must also be invariant. (For amplitudes involving electroweak interactions, parity and charge conjugation invariance are explicitly violated. This is treated by extracting parity and charge violating couplings from the operators and amplitudes. See \sec{vec} for a discussion.) This then allows one to derive corresponding relations for the Wilson coefficients.

To illustrate this with a nontrivial example we consider the $ggq\bq$ process. The operators transform under charge conjugation as
\begin{align}
&\C\, O_{\la_1\la_2(\pm)}^{ab\,\balpha\beta}(\lp_1, \lp_2; \lp_3, \lp_4)\, \vT^{ ab\al\bbeta}\,\C
\nn\\ & \qquad
= \C\, S\, \cB_{1\la_1}^a \cB_{2\la_2}^b J_{34\pm}^{\balpha\beta} \, \vT^{ ab\al\bbeta}\,\C
\nn\\ & \qquad
= - O_{\la_1\la_2(\mp)}^{ba\,\balpha\beta}(\lp_1, \lp_2; \lp_4, \lp_3)\, \vT^{ ab\al\bbeta}
\,,\end{align}
where $\lambda_{1,2}$ denote the gluon helicities, and $\vT^{ ab\al\bt}$ is as given in \eq{ggqqcol}. From the invariance of \eq{Leff} we can infer that the Wilson coefficients must satisfy
\begin{equation} 
C_{\la_1\la_2(\pm)}^{ab\,\alpha\bbeta}(\lp_1, \lp_2; \lp_3, \lp_4)
= -C_{\la_1\la_2(\mp)}^{ba\,\alpha\bbeta}(\lp_1, \lp_2; \lp_4, \lp_3)
\,.\end{equation}
In the color basis of \eq{ggqqcol}, we can write this as
\begin{align}  \label{eq:ggqq_charge_basis}
\vC_{\la_1\la_2(\pm)}(\lp_1, \lp_2; \lp_3, \lp_4)
&= \hV \vC_{\la_1\la_2(\mp)}(\lp_1, \lp_2; \lp_4, \lp_3)
\,,\nn \\
\text{with}\qquad
\hV & =
\begin{pmatrix}
  0 & -1 & 0 \\
  -1 & 0 & 0 \\
  0 & 0 & -1
\end{pmatrix}
\,.\end{align}

Now consider the behavior under parity. For concreteness we consider the case of positive helicity gluons. The operators transform as
\begin{align}
&\P\, O_{++(\pm)}^{ab\,\balpha\beta}(\lp_1, \lp_2; \lp_3, \lp_4)\, \P
= \P\, \frac{1}{2} \cB_{1+}^a \cB_{2+}^b J_{34\pm}^{\balpha\beta} \, \P
\\ &
= e^{ \img (2\phi_{n_1}+2\phi_{n_2}\pm(\phi_{n_3}-\phi_{n_4}))} O_{--(\mp)}^{ab\,\balpha\beta}(\lp_1^P, \lp_2^P; \lp_3^P, \lp_4^P)\, 
\,. \nn\end{align}
The invariance of \eq{Leff}
under parity then implies that the Wilson coefficients satisfy
\begin{align} \label{eq:ggqq_parity}
&\vC_{++(\pm)}(\lp_1, \lp_2; \lp_3, \lp_4) \nn\\
&= \vC_{--(\mp)}(\lp_1^P, \lp_2^P; \lp_3^P, \lp_4^P) e^{ -\img (2\phi_{n_1}+2\phi_{n_2}\pm(\phi_{n_3}-\phi_{n_4}))} \nn \\
&= \vC_{--(\mp)}(\lp_1, \lp_2; \lp_3, \lp_4) \Big|_{\langle..\rangle \leftrightarrow [..]}
\,.\end{align}
Here we have introduced the notation $ \langle .. \rangle \leftrightarrow [..]$ to indicate that all angle and square spinors have been switched in the Wilson coefficient. The fact that the phase appearing in the parity transformation of the operator exactly matches the phase arising from evaluating the Wilson coefficient with parity related momenta is guaranteed by little group scaling, and will therefore occur generically. See \eqs{app_parity1}{app_parity2} and the surrounding discussion for a review.

Below we will use charge conjugation to reduce the number of Wilson
coefficients for which we have to carry out the matching explicitly. We
will use parity only when it helps to avoid substantial repetitions in the
matching.

\subsection{Crossing Symmetry}
\label{subsec:crossing}

Our basis is automatically crossing symmetric, since the gluon fields $\cB_{i\pm}$ can absorb or emit a gluon and the quark current $J_{ij\pm}$ can destroy or produce a quark-antiquark pair, or destroy and create a quark or antiquark. We will first illustrate how to use crossing symmetry in an example and then describe how to technically have crossing symmetric Wilson coefficients.

We will again consider the process $ggq\bq$ as an example. Due to our outgoing conventions, the default Wilson coefficient is for the unphysical processes with all outgoing particles:
\begin{equation}
  0 \to g_+^a(p_1) g_-^b(p_2) q_+^\alpha(p_3) \bq_-^\bbeta(p_4) : C_{+-(+)}^{ab\al\bbeta}(\lp_1,\lp_2;\lp_3,\lp_4)
\,,\end{equation}
where we picked one specific helicity configuration for definiteness.
Crossing a particle from the final state to the initial state flips its helicity, changes the sign of its momentum, and changes it to its antiparticle.
In addition we get a minus sign for each crossed fermion, though in practice these can be ignored as they do not modify the cross section.
This allows one to obtain the Wilson coefficient for any crossing. For example, for the following possible crossings, the Wilson coefficients are given by
\begin{align} \label{eq:crossing}
  g_+^a(p_1) g_-^b(p_2) \to q_+^\al(p_3) \bq_-^\bbeta(p_4) & : C_{+-(+)}^{ba\al\bbeta}(-\lp_2,-\lp_1;\lp_3,\lp_4)\,,
  \nn \\
  g_+^a(p_1) q_+^\balpha(p_2) \to g_+^b(p_3) q_+^\bt(p_4) & : -C_{+-(+)}^{ba\bt\balpha}(\lp_3,-\lp_1;\lp_4,-\lp_2)\,,
  \nn \\
  g_+^a(p_1) \bq_-^\al(p_2) \to g_+^b(p_3) \bq_-^\bbeta(p_4) & : -C_{+-(+)}^{ba\al\bbeta}(\lp_3,-\lp_1;-\lp_2,\lp_4)\,,
  \nn \\
  q_+^\balpha(p_1) \bq_-^\beta(p_2) \to g_+^a(p_3) g_-^b(p_4) & : C_{+-(+)}^{ab\bt\balpha}(\lp_3,\lp_4;-\lp_2,-\lp_1)\,.
\end{align}

Since the signs of momenta change when crossing particles between the final and initial state, care is required in taking the proper branch cuts to maintain crossing symmetry for the Wilson coefficients. In terms of the Lorentz invariants 
\begin{align}
  s_{ij}=(p_i+p_j)^2
\end{align}
this amounts to the choice of branch cut defined by $s_{ij} \to s_{ij} + \img 0$. In particular, we write all logarithms as
\begin{equation} \label{eq:Lij_def}
L_{ij} \equiv \ln\Bigl(-\frac{s_{ij}}{\mu^2} -\img0 \Bigr) = \ln\Bigl(\frac{s_{ij}}{\mu^2}\Bigr) - \img\pi \theta(s_{ij})
\,.\end{equation}
For spinors, crossing symmetry is obtained by defining the conjugate spinors $\bra{p\pm}$ as was done in \eq{braket_def}, resulting in the following relation
\begin{equation}
\bra{p\pm} = \mathrm{sgn}(p^0)\, \overline{\ket{p\pm}}
\,.\end{equation}
The additional minus sign for negative $p^0$ is included to use the same branch (of the square root inside the spinors) for both spinors and conjugate spinors, i.e., for $p^0 > 0$ we have
\begin{align}
\ket{(-p)\pm} &= \img\ket{p\pm}\,,
\nn \\
\bra{(-p)\pm} &=
-(-\img) \bra{p\pm} = \img\bra{p\pm}
\,.\end{align}
In this way all spinor identities are automatically valid for both positive and negative momenta, which makes it easy to use crossing symmetry.

\subsection{Hard Function}
\label{subsec:hard}

In the factorized expression for the cross section given in \eq{sigma},
the dependence on the underlying hard Born process appears through the hard function $\hH_\kappa$. In terms of the Wilson coefficients of the operator basis in the effective theory, the hard function for a particular partonic channel $\kappa$ is given by
\begin{align}\label{eq:hard_C}
  \hH_\kappa(\{\lp_i\})
  = \sum_{\{\la_i\}}
  \vC_{\la_1\cdot\cdot(\cdot\cdot\la_n)}(\{\lp_i\})\,
  \vC_{\la_1 \cdot\cdot(\cdot\cdot\la_n)}^\dagger (\{\lp_i\})
 \,, 
\end{align}
where $\{\lp_i\} \equiv \{\lp_1, \lp_2, \ldots\}$. For unpolarized experiments we simply sum over all helicity operators, so $\hH_\kappa(\{\lp_i\})$ with its sum over helicities in \eq{hard_C} appears as a multiplicative factor. It is important to note that the color indices of the Wilson coefficients are not contracted with each other, rather they are contracted with the color indices of the soft function through the trace seen in \eq{sigma}.

As an explicit example to demonstrate the treatment of both color and helicity indices, we consider the contribution of the $ggq\bq$ partonic channel to the $pp \to 2$ jets process. In this case, the Wilson coefficients are given by $\vC_{\la_1 \la_2 (\la_3)}$, where $\lambda_1,\lambda_2$ denote the helicities of the gluons, $\lambda_3$ denotes the helicity of the quark current, and recall that the vector denotes the possible color structures, which were given explicitly for this case in \eq{ggqqcol}. The hard function for this partonic channel is then given by
\begin{align} \label{eq:ggqqhard}
  \hH_{ggq\bq}(\{\lp_i\})
  & = \!\sum_{\{\la_i\}}\!
  \vC_{\la_1 \la_2 (\la_3)}(\{\lp_i\})\,
  \vC_{\la_1 \la_2 (\la_3)}^\dagger (\{\lp_i\})
  \nn \\
  & = 
  \vC_{++ (+)}
  \vC_{++ (+)}^\dagger +
  \vC_{+- (+)}
  \vC_{+- (+)}^\dagger +
  \nn \\ & \quad
  \vC_{-+ (+)}
  \vC_{-+ (+)}^\dagger +
  \vC_{-- (+)}
  \vC_{-- (+)}^\dagger +
  \nn \\ & \quad
  \vC_{++ (-)}
  \vC_{++ (-)}^\dagger +
  \vC_{+- (-)}
  \vC_{+- (-)}^\dagger +
  \nn \\ & \quad
  \vC_{-+ (-)}
  \vC_{-+ (-)}^\dagger +
  \vC_{-- (-)}
  \vC_{-- (-)}^\dagger
 \,.
\end{align}
Here, explicit expressions are only needed for $\vC_{++(+)}, \vC_{+-(+)}$ and $\vC_{--(+)}$. One can obtain $\vC_{-+(+)}$ using Bose symmetry simply by interchanging the gluons,
\begin{align}
\vC_{-+(+)}^{ab\alpha\bbeta}(\lp_1,\lp_2;\lp_3,\lp_4) &= \vC_{+-(+)}^{ba\alpha\bbeta}(\lp_2,\lp_1;\lp_3,\lp_4)
\,,\end{align}
or equivalently,
\begin{align}
\vC_{-+(+)}(\lp_1,\lp_2;\lp_3,\lp_4) &= \hV \vC_{+-(+)}(\lp_2,\lp_1;\lp_3,\lp_4)
\,,\nn \\
\text{with}\qquad
\hV & =
\begin{pmatrix}
  0 & 1 & 0 \\
  1 & 0 & 0 \\
  0 & 0 & 1
\end{pmatrix}
\,.\end{align}
As explained in \subsec{ggqqbarbasis}, the remaining $\vC_{\la_1\la_2(-)}$ can be obtained from the expressions for the other Wilson coefficients by charge conjugation.

In \eq{ggqqhard}, the Wilson coefficients are vectors in the color basis of \eq{ggqqcol} and thus the hard function is a matrix in this basis. As discussed in \subsec{color}, the tree-level soft function is the color-space identity, i.e.,
\begin{equation}
  S_{ggq\bq}^{\zero\,b_1 b_2 \beta_1 \bbeta_2\, a_1 a_2 \alpha_1 \balpha_2} = \de^{b_1 a_1} \de^{b_2 a_2} \de_{\bt_1\alpha_1} \de_{\bbeta_2\balpha_2}
  \equiv \id
\,.\end{equation}
With the color trace in \eq{sigma} this amounts to contracting the color indices of the Wilson coefficients. In the color basis of \eq{ggqqcol}, this simply becomes
\begin{equation}
  \hS_{ggq\bq}^\zero = \id =
\begin{pmatrix}
  1 & 0 & 0 \\
  0 & 1 & 0 \\
  0 & 0 & 1
\end{pmatrix}
.\end{equation}
The tree-level soft function also has dependence on momenta depending on the measurement being made, which are not shown here.

To demonstrate a complete calculation of the cross section using the factorization theorem of \eq{sigma} together with the hard functions computed using the helicity operator formalism, it is instructive to see how the leading-order cross section is reproduced from \eq{sigma}. We consider the simple case of $H+0$ jets in the $m_t \rightarrow \infty$ limit. For this channel, there is a unique color structure $\delta_{a_1a_2}$, and using the results of \subsec{H0jet} and \app{H0amplitudes}, the lowest order Wilson coefficients are given by
\begin{align}
\vC_{++}(\lp_1,\lp_2;\lp_3)&=\delta_{a_1 a_2} \frac{\alpha_s}{3\pi v} \frac{s_{12}}{2}\frac{[12]}{\langle 1 2\rangle},\\
\vC_{--}(\lp_1,\lp_2;\lp_3)&=\delta_{a_1 a_2} \frac{\alpha_s}{3\pi v} \frac{s_{12}}{2}\frac{\langle12\rangle}{[ 1 2]},\\
\vC_{+-}(\lp_1,\lp_2;\lp_3)&=\vC_{-+}(\lp_1,\lp_2;\lp_3)=0\,,
\end{align}
where $v=\left (  \sqrt{2} G_F  \right )^{-1/2}=246$GeV. Note that these are simply the helicity amplitudes for the process, as will be shown more generally in \sec{matching}.
Analytically continuing to physical momenta, squaring, and summing over helicities, the tree-level hard function is given by
\begin{align}
  H_{ggH}^{\zero\, a_1 a_2\, b_1 b_2}(\lp_1,\lp_2;\lp_3)
  &=  \Bigl| \frac{\al_s}{3\pi v} \frac{s_{12}}{2} \Bigr|^2\, 2\, \de_{a_1 a_2} \de_{b_1 b_2} 
  \nn \\ &=
  \frac{\al_s^2 s_{12}^2}{18\pi^2 v^2} \, \de_{a_1 a_2} \de_{b_1 b_2} 
\,.\end{align}
Note that only 2 of the 4 helicity configurations contribute, hence the factor of 2.

The tree-level gluon beam functions are given by the gluon PDFs. Since there are no jets in the final state, there are no jet functions. The tree-level soft function is the identity in color space%
\footnote{Since there is only one color structure, the tree-level soft function is normally defined as
\begin{equation}
   S^\zero_{gg} = \frac{1}{N^2-1}\,\de_{a_1 a_2} \de_{b_1 b_2}\, \de^{b_1 a_1} \de^{b_2 a_2} = 1
\,.\end{equation}
Here we do not absorb numerical prefactors into our soft functions, because this is not useful for processes with more final-state partons.}
\begin{equation}
  S_{gg}^{\zero\, b_1 b_2\, a_1 a_2} = \de^{b_1 a_1} \de^{b_2 a_2}
\,.\end{equation}
The leading-order cross section is then given by
\begin{align} \label{eq:hard_comp_H}
\si &= \frac{1}{2\Ecm^2} \frac{1}{[2(N^2-1)]^2}
 \int\! \frac{\df x_1}{x_1}\, \frac{\df x_2}{x_2}\, f_g(x_1) f_g(x_2)
\nn \\ & \quad\times
  \int\! \frac{\df^4p_3}{(2\pi)^3}\, \theta(p_3^0)\, \de(p_3^2 \!-\! m_H^2)
\nn \\ & \quad\times
  (2\pi)^4 \delta^4\Bigl(x_1 \Ecm \frac{n_1}{2} \!+\! x_2 \Ecm \frac{n_2}{2} \!-\! p_3 \Bigr)
 \nn \\ & \quad  
\times H_{ggH}^{\zero\, a_1 a_2\, b_1 b_2}(\lp_1,\lp_2;\lp_3)\,  S_{gg}^{\zero\, b_1 b_2\, a_1 a_2}
\nn \\
&= \frac{\al_s^2 m_H^2}{576 \pi v^2 \Ecm^2} \int\! \df Y
 f_g\Bigl(\frac{m_H}{\Ecm} e^Y\Bigr) f_g\Bigl(\frac{m_H}{\Ecm} e^{-Y}\Bigr)
\,.\end{align}
The $1/(2\Ecm^2)$ factor is the flux factor and for each of the incoming gluons we get a $1/[2(N^2-1)]$ from averaging over its spin and color. This is followed by integrals over the gluon PDFs, $f_g$, and the Higgs phase space, where we have restricted to the production of an on-shell Higgs. The final expression in \eq{hard_comp_H} agrees with the standard result, where the first factor is the Born cross section.

We now briefly discuss our choice of normalization. The currents in \eq{jpm_def} were normalized such that
the Wilson coefficients are simply given by the finite part of the QCD helicity amplitudes (see \eq{matching_general} and \sec{matching}). 
This is distinct from the normalization typically used for SCET operators, e.g.~$\bar{\chi_i} \ga^\mu \chi_j$, which is chosen to facilitate the matching to QCD operators. We now show that the extra factors in \eq{jpm_def} arrange themselves to produce the standard normalization for the jet function (or beam function). Starting from the current and its conjugate,
\begin{align}\label{eq:fierzeg}
&J_{ij\pm}^{\balpha\beta}
(J_{ij\pm}^{\bgamma\delta})^\dagger
  \\
&=\frac{\sqrt{2}\ve_\mp^\mu(n_i, n_j)}
  {   \sqrt{ \vphantom{2}    \omega_i\, \omega_j }} \frac{\bar{\chi}^\balpha_{i\pm} \gamma_\mu \chi^\beta_{j\pm}}
   {\langle n_i \!\mp | n_j \pm\rangle}
 \frac{\sqrt{2}\ve_\pm^\nu(n_i, n_j)}
   {\left(\sqrt{\vphantom{2} \omega_i \,  \omega_j }\right)^*}
 \frac{\bar{\chi}^\bdelta_{j\pm}\, \gamma_\nu \chi^\gamma_{i\pm}}
 {\langle n_j\!\pm | n_i \mp\rangle}
\nn \\
&= 4\frac{\de^{\ga\balpha}}{N} \frac{\de^{\beta \bdelta}}{N}\,
\frac{\ve_\pm^\nu(n_i, n_j)\,\ve_\mp^\mu(n_i, n_j)}{2 n_i \sdt n_j\, |\omega_i\, \omega_j|}\,
\tr\Big[\ga_\nu \frac{\nslash_i}{4} \ga_\mu \frac{\nslash_j}{4} \Big] 
\nn \\ & \quad \times
\Big(\bar{\chi}_{i\pm} \frac{\bnslash_i}{2} \chi_{i\pm}\Big)
\Big(\bar{\chi}_{j\pm} \frac{\bnslash_j}{2} \chi_{j\pm}\Big)
 + \ldots
\nn \\
&=2\, \de^{\ga\balpha} \de^{\beta \bdelta}\,
\Big(\frac{1}{2N} \frac{1}{|\w_i|} \bar{\chi}_{i\pm} \frac{\bnslash_i}{2} \chi_{i\pm}\Big)
\Big(\frac{1}{2N} \frac{1}{|\w_j|} \bar{\chi}_{j\pm} \frac{\bnslash_j}{2} \chi_{j\pm}\Big)
\,,\nonumber
\end{align}
where we have rearranged the expression in a factorized form using the SCET Fierz formula in spin
\begin{align}
1 \otimes 1=\frac{1}{2} \biggl[ \frac{\bnslash_i}{2} \otimes \frac{\nslash_i}{2}-\frac{\bnslash_i \gamma_5}{2} \otimes \frac{\nslash_i \gamma_5}{2}-\frac{\bnslash_i \gamma_\perp^\mu}{2} \otimes \frac{\nslash_i \gamma_{\perp\mu}}{2} \biggr]
,\end{align}
which applies for the SCET projected spinors. In the last line of \eq{fierzeg} we have dropped the color nonsinglet terms and terms which vanish when averaging over helicities, which are indicated by ellipses. The delta functions in color space highlight that the jet function does not modify the color structure. The factor $1/\w_{i,j}$, which arises from the normalization of the helicity currents, is part of the standard definition of the jet function and ensures that this operator has the correct mass dimension.

\section{Matching and Scheme Dependence}
\label{sec:matching}

In this section, we discuss the matching of QCD onto the SCET helicity operator basis introduced in the previous section. We start with a discussion of the matching for generic helicity operators in \subsec{genmatch}. In \subsec{renscheme} we discuss in detail the subject of renormalization schemes, and the issue of converting between regularization/renormalization schemes commonly used in spinor-helicity calculations, and those used in SCET. We also demonstrate that evanescent operators are not generated in our basis.

\subsection{Generic Matching} \label{subsec:genmatch}

In this paper, we work to leading order in the power counting, which means we only require operators that contain exactly one field per collinear sector. That is, different $n_i$ in \eq{Opm_gen} are implicitly restricted to belong to different equivalence classes,  $\{n_i\} \neq \{n_j\}$ for $i\neq j$. Operators with more than one field per collinear direction are power-suppressed compared to the respective leading-order operators that have the same set of collinear directions and the minimal number of fields.

At leading order, the Wilson coefficients can thus be determined by computing matrix elements of \eq{Leff}, with all external particles assigned well-separated momenta, so that they belong to separate collinear sectors. The only helicity operator that contributes in this case is the one that matches the set of external helicities, picking out the corresponding Wilson coefficient. Since we only have one external particle per collinear sector, we can simply choose $n_i = p_i/p_i^0$ in the matching calculation to represent the equivalence class $\{n_i\}$.

To compute the matrix element of $\cL_\hard$, we first note that the helicity operators are symmetric (modulo minus signs from fermion anticommutation) under simultaneously interchanging the label momenta and indices of identical fields, and the same is thus also true for their Wilson coefficients. For example, at tree level
\begin{align}
&\Mae{g_+^{a_1}(p_1)\, g_+^{a_2}(p_2)}{O_{++}^{b_1 b_2}}{0}^{\tree}
\nn\\ & \quad
= \frac{1}{2}\bigl[ \delta^{a_1 b_1}\, \delta^{a_2 b_2} \ldel(\lp_1 - p_1)\,\ldel(\lp_2 - p_2)
\nn\\ &\qquad
+ \delta^{a_1 b_2}\, \delta^{a_2 b_1} \ldel(\lp_1 - p_2)\,\ldel(\lp_2 - p_1) \bigr]
\end{align}
so the tree-level matrix element of $\cL_\hard$ gives
\begin{align}
&\Mae{g_+^{a_1}(p_1)\, g_+^{a_2}(p_2)}{\cL_\hard}{0}^{\tree}
\\\nn &\quad
= \frac{1}{2} \bigl[C_{++}^{a_1 a_2}(\lp_1, \lp_2) + C_{++}^{a_2 a_1}(\lp_2, \lp_1)\bigr]
= C_{++}^{a_1 a_2}(\lp_1, \lp_2)
\,.\end{align}
By choosing $n_i = p_i/p_i^0$, the label momenta $\lp_i$ on the right-hand side simply become $\lp_i \equiv \bn\cdot p_i\, n_i/2 = p_i$.

Taking into account the symmetry factor in \eq{Opm_S}, one can easily see that this result generalizes to more than two gluons or quark currents with the same helicity. In the case of identical fermions, the various terms in the operator matrix element have relative minus signs due to fermion anticommutation which precisely match the (anti)symmetry properties of the Wilson coefficients. Hence, the tree-level matrix element of $\cL_\hard$ is equal to the Wilson coefficient that corresponds to the configuration of external particles,
\begin{align} \label{eq:Leff_me}
&\Mae{g_1g_2\dotsb q_{n-1}\bar{q}_n}{\cL_\hard}{0}^{\tree}
\nn\\ & \qquad
= C_{+\cdot\cdot(\cdot\cdot-)}^{a_1 a_2\dotsb\alpha_{n-1}\balpha_n}(\lp_1,\lp_2,\ldots,\lp_{n-1},\lp_n)
\,.\end{align}
Here and below, $g_i \equiv g_\pm^{a_i}(p_i)$ stands for a gluon with helicity $\pm$, momentum $p_i$, color $a_i$, and analogously for (anti)quarks. From \eq{Leff_me} we obtain the generic tree-level matching equation
\begin{equation} \label{eq:matching_LO}
C_{+\cdot\cdot(\cdot\cdot-)}^{a_1\dotsb\balpha_n}(\lp_1,\ldots,\lp_n)
= -\img \cA^\tree(g_1 \dotsb \bar{q}_n)
\,,\end{equation}
where $\cA^\tree$ denotes the tree-level QCD helicity amplitude. Intuitively, since all external particles are energetic and well separated, we are away from any soft or collinear limits and so all propagators in the QCD tree-level diagram are far off shell and can be shrunk to a point. Hence, the tree-level diagram simply becomes the Wilson coefficient in SCET.

The above discussion can be extended to higher orders in perturbation theory. In pure dimensional regularization (where $\epsilon$ is used to simultaneously regulate UV and IR divergences) all bare loop graphs in SCET are scaleless and vanish. Here the UV and IR divergences precisely cancel each other, and the bare matrix elements are given by their tree-level expressions, \eq{Leff_me}. Including the counterterm $\delta_O(\eps_\mathrm{UV})$ due to operator renormalization removes the UV divergences and leaves the IR divergences. Schematically, the renormalized loop amplitude computed in SCET using $\cL_\hard$ is
\begin{equation}
\cA_\mathrm{SCET}
= \int\! (\vev{\vec O^\dagger}^\tree 
   \!+\! \vev{\vec O^\dagger}^\mathrm{loop})\, \img \vec C
= \bigl[1 + \delta_O(\eps_\IR) \bigr] \img \vec C 
,
\end{equation}
where we used that the loop contribution is a pure counterterm and thus proportional to the tree-level expression. In general, the counterterm $\delta_O$ is a matrix in color space, as we will see explicitly in \sec{running} and \app{IRdiv}. By construction, the $1/\epsilon$ IR divergences in the effective theory, $C\,\delta_O(\eps_\IR)$, have to exactly match those of the full theory. Therefore, beyond tree level the matching coefficients in $\overline{\mathrm{MS}}$ are given by the infrared-finite part of the renormalized full-theory amplitude, ${\cal A}_{\rm ren}$, computed in pure dimensional regularization. The IR-finite part is obtained by multiplying ${\cal A}_{\rm ren}$ by SCET $\overline{\rm MS}$ renormalization factors, which cancel the full theory $1/\epsilon_\IR$ poles.  Decomposing the renormalized QCD amplitude in a color basis so that ${\cal A}_{\rm ren}^{a_1 \dotsb\alpha_n}= \vT^{ a_1\dotsb\bar \alpha_n} \vec {\cal A}_{\rm ren}(g_1\dotsb \bar q_n)$, the all-orders form of \eq{matching_LO} becomes
\begin{align}  \label{eq:matching_general}
C_{+\cdot\cdot(\cdot\cdot-)}^{a_1\dotsb\balpha_n}(\lp_1,\ldots,\lp_n)
&= -\img \cA_\fin(g_1\dotsb \bar{q}_n) 
 \\
&\equiv \frac{-\img \,
  \vT^{ a_1\dotsb\bar\alpha_n}   \widehat Z_C^{-1} \vec {\cal A}_{\rm ren}(g_1\dotsb \bar{q}_n)}{ Z_\xi^{n_q/2} Z_A^{n_g/2} }
  \nn
\,.\end{align}
The SCET renormalization factors $\widehat Z_C$, $Z_\xi$, and $Z_A$ are discussed in \subsec{running_general}. At one-loop order this corresponds to taking $(-\img {\cal A}_{\rm ren}^{a_1 \dotsb\bar\alpha_n})$ and simply dropping the $1/\epsilon_\IR$ terms. In \subsec{renscheme} we will discuss in more detail the use of different renormalization schemes to compute $\vec {\cal A}_{\rm ren}(g_1\dotsb \bar{q}_n)$.

If the same color decomposition is used for the QCD amplitude as for the Wilson coefficients in \eq{Cpm_color}, we can immediately read off the coefficients $\vC$ in this color basis from \eq{matching_general}. As an example, consider for simplicity the leading color $n$ gluon amplitude, which has the color decomposition (see \app{color_decomp})
\begin{align}
\cA_n(g_{1}\dotsb g_{n})
&= \img g_s^{n-2} \!\!\sum_{\si \in S_n/Z_n}\!\! \tr[T^{a_{\si(1)}} \dotsb T^{a_{\si(n)}}]
\nn\\ & \quad\times
\sum_i g_s^i\, A_n^{(i)}(\si(1),\ldots,\si(n))
\,,\end{align}
where the first sum runs over all permutations $\sigma$ of $n$ objects ($S_n$) excluding cyclic permutations ($Z_n$). The $A_n^{(i)}$ are the color-ordered or partial amplitudes at $i$ loops. Each is separately gauge invariant and only depends on the external momenta and helicities $(p_i\pm) \equiv (i^\pm)$. If we choose
\begin{equation}
T_k^{a_1\dotsb a_n} = \tr[T^{a_{\si_k(1)}} \dotsb T^{a_{\si_k(n)}}]\,,
\end{equation}
as the color basis in \eq{Cpm_color}, where $\sigma_k$ is the $k$th permutation in $S_n/Z_n$, then the Wilson coefficients in this color basis are given directly by
\begin{align}
C^k_{\lambda_1\cdots\lambda_2}&(\lp_1,\ldots,\lp_n) =\nonumber \\
& g_s^{n-2} \sum_i g_s^i\,  A_{n,\fin}^{(i)}(\si_k(1^{\lambda_1}),\ldots,\si_k(n^{\lambda_2}))
\,,\end{align}
where the subscript ``$\fin$'' denotes the IR-finite part of the helicity amplitude, as defined in \eq{matching_general}. This is easily extended beyond leading color, given a valid choice of subleading color basis. Our basis therefore achieves seamless matching from QCD helicity amplitudes onto SCET operators.

\subsection{Renormalization Schemes}
\label{subsec:renscheme}

In this section we discuss in more detail the issue of renormalization/regularization schemes in QCD and in SCET. In particular, the construction of a basis of helicity operators discussed in \sec{basis} relied heavily on massless quarks and gluons having two helicity states, which is a feature specific to 4 dimensions. We clarify this issue here and discuss the conversion between various schemes.

In dimensional regularization, divergences are regularized by analytically continuing the particle momenta to $d$ dimensions. In a general scheme, the helicities of quarks and gluons live in $d^g_s$, $d^q_s$ dimensional spaces respectively. We shall here restrict ourselves to schemes where quarks have two helicities, but $d^g_s$ is analytically continued. This is true of most commonly used regularization schemes, but is not necessary \cite{Catani:1996pk}. Different schemes within dimensional regularization differ in their treatment of $d^g_s$ for internal (unobserved) and external (observed) particles. In the conventional dimensional regularization (CDR), 't Hooft-Veltman (HV)~\cite{tHooft:1972fi}, and four-dimensional helicity (FDH)~\cite{Bern:1991aq, Bern:2002zk} schemes the internal/external polarizations are treated in $d/d$ (CDR), $d/4$ (HV), $4/4$ (FDH) dimensions. 

For helicity-based computations, the FDH scheme has the advantage of having all helicities defined in 4 dimensions, where the spinor-helicity formalism applies, as well as preserving supersymmetry. Indeed, most of the recent one-loop computations of helicity amplitudes utilize on-shell methods and therefore employ the FDH scheme. However, most existing calculations of SCET matrix elements (jet, beam, and soft functions) use $d$-dimensional internal gluons, corresponding to the CDR/HV schemes.\footnote{Recently while this paper was being finalized, a calculation of the inclusive jet and soft functions in both FDH and dimensional reduction (DRED)~\cite{Siegel:1979wq} appeared in Ref.~\cite{Broggio:2015dga}. The conclusions of this section agree with their study of the regularization scheme dependence of QCD amplitudes.} As we will discuss below, CDR and HV are identical for matching onto SCET.

Although the FDH scheme is convenient for helicity amplitude computations, it leads to subtleties beyond NLO~\cite{Kilgore:2011ta,Boughezal:2011br}. As explained in Ref.~\cite{Boughezal:2011br}, this discrepancy arises due to the different number of dimensions for the momenta in the loop integral and the spin space, leading to components of the gluon field whose couplings to quarks are not protected by gauge invariance and require separate renormalization. Nevertheless, it has been shown that FDH is a consistent regularization scheme to NNLO \cite{Broggio:2015dga}. The presence of these extra degrees of freedom in the FDH scheme is quite inconvenient in the formal construction of SCET, especially when working to subleading power. Because of this fact, and because most SCET calculations are performed in CDR/HV, our discussion of SCET schemes will focus on regularization schemes where the dimension of the gluon field and the momentum space are analytically continued in the same manner. We will also discuss how full-theory helicity amplitudes in the FDH scheme are converted to CDR/HV for the purposes of matching to SCET.

We will now describe how helicity amplitudes in the FDH scheme can be converted to CDR/HV.
To get a finite correction from the $\ord{\eps}$ part of the gluon polarization requires a factor from either ultraviolet (UV) or infrared (IR) $1/\eps$ divergences. Although the regularization of UV and IR divergences is coupled in pure dimensional regularization schemes by use of a common $\epsilon$, they can in principle be separately regulated, and we discuss their role in the scheme conversion separately below.

When matching to SCET, the UV regulators in the full and effective theory need not be equal. Indeed, the effective theory does not reproduce the UV of the full theory. In massless QCD, scheme dependence due to the UV divergences only affects the coupling constant through virtual (internal) gluons. Therefore, the CDR and HV schemes have the same standard $\overline{\text{MS}}$ coupling, $\alpha_s(\mu)$, while FDH has a different coupling, $\alpha_s^\mathrm{FDH}(\mu)$. The conversion between these couplings is achieved by a perturbatively calculable shift, known to two loops~\cite{Altarelli:1980fi, Kunszt:1993sd, Bern:2002zk}
\begin{align} \label{eq:DR_HV_UV}
 \al_s^\mathrm{FDH}(\mu) &= \al_s(\mu) \biggl[1 + \frac{C_A}{3} \frac{\al_s(\mu)}{4\pi}
 \nn \\ & \quad
 + \Bigl(\frac{22}{9}\,C_A^2 - 2 C_F T_F n_f\Bigr) \bigg(\frac{\al_s(\mu)}{4\pi}\bigg)^2 \biggr]
\,.\end{align}
This replacement rule for the coupling captures the effect of the scheme choice from UV divergences. One can therefore perform a matching calculation, treating $\alpha_s$ in the full and effective theories as independent parameters that can be defined in different schemes. A conversion between schemes can then be used to ensure that the matching coefficients are written entirely in terms of $\alpha_s$ defined in one scheme, for example using \eq{DR_HV_UV}. The issue of UV regularization is therefore simple to handle in the matching.

The structure of $1/\eps^2$ and $1/\eps$ IR divergences in one-loop QCD amplitudes is well known, and allows one to determine their effect on converting amplitudes from FDH to CDR/HV. For a QCD amplitude involving $n_q$ (anti)quarks and $n_g$ gluons the FDH and HV one-loop amplitudes $\cA^\one$ are related by~\cite{Kunszt:1993sd, Catani:1996pk}
\begin{equation} \label{eq:DR_HV_IR}
 \cA_\mathrm{HV}^\one = \cA_\mathrm{FDH}^\one
 - \frac{\alpha_s}{4\pi} \Bigl(\frac{n_q}{2} C_F + \frac{n_g}{6} C_A\Bigr)  \cA^\zero
\,,\end{equation}
where $\cA^\zero$ denotes the tree-level amplitude, and the precise scheme of the $\alpha_s$ entering here is a two-loop effect. At one loop, the FDH scheme can therefore be consistently used when calculating full-theory helicity amplitudes and results can easily be converted to HV with \eqs{DR_HV_UV}{DR_HV_IR} for use in SCET Wilson coefficients.

We will now compare CDR and HV schemes for SCET calculations and the construction of the operator basis. In the HV scheme, all external polarizations are 4 dimensional, so that one can use a basis of helicity operators, as was constructed in \sec{basis}. However, in CDR external polarizations are $d$ dimensional, with the limit $d\to 4$ taken. In particular, this implies that one must work with $d-2$ gluon polarizations at intermediate steps, potentially allowing for the presence of evanescent operators corresponding to operators involving the additional components of the gluon field, so-called $\epsilon$-helicities. However, we will now argue that there is no real distinction between the two schemes, and that one does not need to consider evanescent operators in SCET at leading power.

First consider the Wilson coefficients and matching. In the case of CDR, the operator basis must be extended to include operators involving the $\epsilon$-helicities. However, their presence does not affect the matching coefficients for operators with physical helicities, since they do not contribute at tree level and all loop corrections are scaleless and vanish.
Additionally, in \sec{running}, we will discuss the fact that the SCET renormalization of the operators is spin independent at leading power, and therefore there is no mixing under renormalization group evolution between the physical and evanescent operators. For the beam and jet functions, azimuthal symmetry implies that the difference between a field with 2 or $2-2\eps$ polarizations is simply an overall factor of $1-\eps$ and thus can be easily taken into account. The independence of the soft function to the differences in the CDR/HV regularization schemes follows from the insensitivity of the soft emission to the polarization of the radiating parton, which is made manifest by the SCET Lagrangian and the fact that the soft function can be written as a matrix element of Wilson lines. Thus there is no difference between CDR and HV and the helicity operator basis suffices.

\section{Higgs + Jets}
\label{sec:higgs}

In this section, we consider the production of an on-shell Higgs + jets. We give the helicity operator basis and matching relations for $H + 0,1,2$ jets, and the corresponding helicity amplitudes are collected in \app{Hamplitudes}.

\subsection{\boldmath $H + 0$ Jets}
\label{subsec:H0jet}

The $ggH$ and $q\bq H$ processes contribute to the $H+0$ jets process. For $q\bq H$, the scalar current in \eq{jS_def} is required, and the helicity operator basis is given by
\begin{align} \label{eq:qqH_basis}
O_1^{\balpha\beta}
&= J_{12\,0}^{\balpha\beta}\, H_3\,,
\nn \\
O_2^{\balpha\beta}
&= (J^\dagger)_{12\,0}^{\balpha\beta}\, H_3
\,,\end{align}
with the unique color structure
\begin{equation} \label{eq:H0qq_color}
\vT^{ \alpha\bbeta} = \begin{pmatrix} \de_{\alpha\bbeta} \end{pmatrix}
\,.\end{equation}
These operators are relevant when considering Higgs decays to massive quarks, for example $H\to \bar b b$. However, we will not consider this case further since for Higgs production the $b\bar bH$ and $t\bar tH$ contributions are much smaller than the dominant gluon-fusion hard scattering process.

For $ggH$, the basis of helicity operators is given by
\begin{align} \label{eq:ggH_basis}
O_{++}^{ab}
&= \frac{1}{2}\, \cB_{1+}^a\, \cB_{2+}^b\,  H_3
\,,\nn\\
O_{--}^{ab}
&= \frac{1}{2}\, \cB_{1-}^a\, \cB_{2-}^b\, H_3
\,.\end{align}
The operator $O_{+-}$ is not allowed by angular momentum conservation. Similar helicity operators, extended to include the decay of the Higgs, were used in Ref.~\cite{Moult:2014pja}.
There is again a unique color structure for this process,
\begin{equation} \label{eq:H0_color}
\vT^{ ab} = \begin{pmatrix} \de^{ab} \end{pmatrix}
\,.\end{equation}
Writing the QCD helicity amplitudes as
\begin{align}
\cA(g_1 g_2 H_3) &= \img \delta^{a_1 a_2}\, A(1, 2; 3_H)
\,,\end{align}
the Wilson coefficients for $ggH$ are given by
\begin{align} \label{eq:ggH_coeffs}
\vC_{++}(\lp_1, \lp_2; \lp_3) &= A_\fin(1^+, 2^+; 3_H)
\,, \nn \\
\vC_{--}(\lp_1, \lp_2; \lp_3) &= A_\fin(1^-, 2^-; 3_H)
\,.\end{align}
The subscript ``$\fin$'' in \eq{ggH_coeffs} denotes the IR-finite part of the helicity amplitudes, as discussed in \sec{matching}.
Note that the two amplitudes appearing in \eq{ggH_coeffs} are related by parity.
The results for the gluon amplitudes up to NNLO are given in \app{H0amplitudes}. They correspond to the usual gluon-fusion process, where the Higgs couples to a (top) quark loop at leading order. The LO amplitude including the dependence on the mass of the quark running in the loop is well known. The NLO amplitudes are also known including the full quark-mass dependence~\cite{Dawson:1990zj, Djouadi:1991tka, Spira:1995rr, Harlander:2005rq, Anastasiou:2006hc}, while the NNLO~\cite{Harlander:2000mg, Harlander:2009bw, Pak:2009bx} and N$^3$LO~\cite{Baikov:2009bg,Gehrmann:2010ue} amplitudes are known in an expansion in $m_H/m_t$. 

\subsection{\boldmath $H + 1$ Jet}

The $gq\bar qH$ and $gggH$ processes contribute to the $H+1$ jet process. For $gq\bar q$, the basis of helicity operators is given by
\begin{align}
O_{+(+)}^{a\, \balpha\beta}
&= \cB_{1+}^a\, J_{23+}^{\balpha\beta}\, H_4
\,,\nn\\
O_{-(+)}^{a\, \balpha\beta}
&= \cB_{1-}^a\, J_{23+}^{\balpha\beta}\, H_4
\,,\nn\\
O_{+(-)}^{a\, \balpha\beta}
&= \cB_{1+}^a\, J_{23-}^{\balpha\beta}\, H_4
\,,\nn\\
O_{-(-)}^{a\, \balpha\beta}
&= \cB_{1-}^a\, J_{23-}^{\balpha\beta}\, H_4
\,.\end{align}
Note that we consider only QCD corrections to the $ggH$ process, so the $q\bar q$ pair is described by $J_{ij\pm}$.
For $ggg$, the helicity operator basis is
\begin{align} \label{eq:H1_basis}
O_{+++}^{abc}
&= \frac{1}{3!}\, \cB_{1+}^a\, \cB_{2+}^b\, \cB_{3+}^c\, H_4
\,,\nn\\
O_{++-}^{abc}
&= \frac{1}{2}\, \cB_{1+}^a\, \cB_{2+}^b\, \cB_{3-}^c\, H_4
\,,\nn\\
O_{--+}^{abc}
&= \frac{1}{2}\, \cB_{1-}^a\, \cB_{2-}^b\, \cB_{3+}^c\, H_4
\,,\nn\\
O_{---}^{abc}
&= \frac{1}{3!}\, \cB_{1-}^a\, \cB_{2-}^b\, \cB_{3-}^c\, H_4
\,.\end{align}
For both cases the color space is one dimensional and we use the respective color structures as basis elements
\begin{equation} \label{eq:H1_color}
\vT^{ a\alpha\bbeta} = \begin{pmatrix} T^a_{\alpha\bbeta} \end{pmatrix}
\,, \qquad
\vT^{ abc} = \begin{pmatrix} \img f^{abc} \end{pmatrix}
\,.\end{equation}

In principle, there could be another independent color structure, $d^{abc}$, for $gggH$.
The $gggH$ operators transform under charge conjugation as
\begin{align}
&\C\, O_{\la_1\la_2\la_3}^{abc}(\lp_1, \lp_2, \lp_3; \lp_4)\, \vT^{ abc} \,\C
\nn\\ & \qquad
= -O_{\la_1\la_2\la_3}^{cba}(\lp_1, \lp_2, \lp_3; \lp_4)\, \vT^{ abc}
\,.\end{align}
Charge conjugation invariance of QCD thus leads to
\begin{equation} \label{eq:H1_charge}
C_{\la_1\la_2\la_3}^{abc}(\lp_1, \lp_2, \lp_3; \lp_4)
= - C_{\la_1\la_2\la_3}^{cba}(\lp_1, \lp_2, \lp_3; \lp_4)
\,,\end{equation}
which implies that the $d^{abc}$ color structure cannot arise to all orders in perturbation theory, so it suffices to consider $\img f^{abc}$ as in \eq{H1_color}. This also means that the $d^{abc}$ color structure cannot be generated by mixing under renormalization group evolution, which will be seen explicitly in \eq{mix23gluons}.

Using \eq{H1_color}, we write the QCD helicity amplitudes as
\begin{align}
\cA(g_1 g_2 g_3 H_4) &= \img\, (\img f^{a_1 a_2 a_3})\, A(1, 2, 3; 4_H)
\,,\nn\\
\cA(g_1 q_{2} \bq_{3} H_4) &= \img\, T^{a_1}_{\al_2 \balpha_3}\, A(1; 2_q, 3_\bq; 4_H)
\,.\end{align}
The Wilson coefficients for $gq\bq H$ are then given by
\begin{align} \label{eq:gqqH_coeffs}
\vC_{+(+)}(\lp_1;\lp_2,\lp_3;\lp_4) &= A_\fin(1^+; 2_q^+, 3_\bq^-; 4_H)
\,, \nn \\
\vC_{-(+)}(\lp_1;\lp_2,\lp_3;\lp_4) &= A_\fin(1^-; 2_q^+, 3_\bq^-; 4_H)
\,, \nn \\[0.5ex]
\vC_{+(-)}(\lp_1; \lp_2, \lp_3; \lp_4) &= \vC_{+(+)}(\lp_1; \lp_3, \lp_2; \lp_4)
\,, \nn \\
\vC_{-(-)}(\lp_1; \lp_2, \lp_3; \lp_4) &= \vC_{-(+)}(\lp_1; \lp_3, \lp_2; \lp_4)
\,,\end{align}
where the last two coefficients follow from charge conjugation invariance.
The Wilson coefficients for $gggH$ are given by
\begin{align} \label{eq:gggH_coeffs}
\vC_{+++}(\lp_1,\lp_2,\lp_3;\lp_4) &= A_\fin(1^+, 2^+, 3^+; 4_H)
\,,\nn\\
\vC_{++-}(\lp_1,\lp_2,\lp_3;\lp_4) &= A_\fin(1^+, 2^+, 3^-; 4_H)
\,,\nn\\[0.5ex]
\vC_{--+}(\lp_1, \lp_2, \lp_3; \lp_4)
&= \vC_{++-}(\lp_1, \lp_2, \lp_3; \lp_4) \Big|_{\langle..\rangle \leftrightarrow [..]}
\,,\nn\\
\vC_{---}(\lp_1, \lp_2, \lp_3; \lp_4)
&= \vC_{+++}(\lp_1, \lp_2, \lp_3; \lp_4) \Big|_{\langle..\rangle \leftrightarrow [..]}
\,,\end{align}
where the last two relations follow from parity invariance. As before, the subscript ``$\fin$'' in \eqs{gqqH_coeffs}{gggH_coeffs} denotes the finite part of the IR divergent amplitudes. 
The NLO helicity amplitudes were calculated in Ref.~\cite{Schmidt:1997wr}, and are given in \app{H1amplitudes}, and the NNLO helicity amplitudes were calculated in Ref.~\cite{Gehrmann:2011aa}. Both calculations were performed in the $m_t\to\infty$ limit.
At NLO, the first corrections in $m_H^2/m_t^2$ were obtained in Ref.~\cite{Neill:2009mz}.

\subsection{\boldmath $H + 2$ Jets}

For $H+2$ jets, the $q\bq\, q'\bq'H$, $q\bq\, q\bq H$, $ggq\bq H$, and $ggggH$ processes contribute, each of which we discuss in turn. Again, we consider only QCD corrections to the $ggH$ process, so $q\bar q$ pairs are described by the helicity currents $J_{ij\pm}$. The LO helicity amplitudes for $H + 2$ jets in the $m_t\to\infty$ limit were calculated in Refs.~\cite{Dawson:1991au, Kauffman:1996ix} and are collected in \app{H2amplitudes} for each channel. The LO amplitudes including the $m_t$ dependence were calculated in \cite{DelDuca:2001fn} (but explicit expressions for $ggggH$ were not given due to their length). The NLO helicity amplitudes were computed in Refs.~\cite{Berger:2006sh, Badger:2007si, Glover:2008ffa, Dixon:2009uk, Badger:2009hw, Badger:2009vh}.

\subsubsection{$q\bq\, q'\bq' H$ and $q\bq\, q\bq H$}

For the case of distinct quark flavors, $q\bq\, q'\bq'H$, the helicity basis consists of four independent operators,
\begin{align} \label{eq:qqQQH_basis}
O_{(+;+)}^{\balpha\bt\bgamma\de}
&= J_{q\, 12+}^{\balpha\bt}\, J_{q'\, 34+}^{\bgamma\de}\, H_5
\,,\nn\\
O_{(+;-)}^{\balpha\bt\bgamma\de}
&= J_{q\, 12+}^{\balpha\bt}\, J_{q'\, 34-}^{\bgamma\de}\, H_5
\,,\nn\\
O_{(-;+)}^{\balpha\bt\bgamma\de}
&= J_{q\, 12-}^{\balpha\bt}\, J_{q'\, 34+}^{\bgamma\de}\, H_5
\,,\nn\\
O_{(-;-)}^{\balpha\bt\bgamma\de}
&= J_{q\, 12-}^{\balpha\bt}\, J_{q'\, 34-}^{\bgamma\de}\, H_5
\,,\end{align}
where the additional labels on the quark currents indicate the quark flavors. For the case of identical quark flavors, $q\bq\,q\bq H$, the basis only has three independent helicity operators,
\begin{align} \label{eq:qqqqH_basis}
O_{(++)}^{\balpha\bt\bgamma\de}
&= \frac{1}{4}\, J_{12+}^{\balpha\bt}\, J_{34+}^{\bgamma\de}\, H_5
\,,\nn\\
O_{(+-)}^{\balpha\bt\bgamma\de}
&= J_{12+}^{\balpha\bt}\, J_{34-}^{\bgamma\de}\, H_5
\,,\nn\\
O_{(--)}^{\balpha\bt\bgamma\de}
&= \frac{1}{4}\, J_{12-}^{\balpha\bt}\, J_{34-}^{\bgamma\de}\, H_5
\,,\end{align}
since both quark currents have the same flavor. In both cases we use the color basis
\begin{equation} \label{eq:qqqqH_color}
\vT^{ \al\bbeta\ga\bdelta} =
2T_F\Bigl(
  \de_{\al\bdelta}\, \de_{\ga\bbeta}\,,\, \delta_{\al\bbeta}\, \de_{\ga\bdelta}
\Bigr)
\,.\end{equation}

The QCD helicity amplitudes for $q\bq\,q'\bq'H$ can be color decomposed in the basis of \eq{qqqqH_color} as
\begin{align} \label{eq:qqQQH_QCD}
 \cA(q_{1} \bq_{2} q_3' \bq_4' H_5)
&= 2 \img T_F  \Bigl[
\de_{\al_1\balpha_4} \de_{\al_3\balpha_2}  A(1_q,2_\bq;3_{q'},4_{\bq'};5_H)
\nn\\ & \quad
+ \frac{1}{N}\,\de_{\al_1 \balpha_2} \de_{\al_3 \balpha_4} B(1_q,2_\bq;3_{q'},4_{\bq'};5_H) \Bigr]
\,,\end{align}
where we have included a factor of $1/N$ for convenience. The amplitude vanishes when the quark and antiquark of the same flavor have the same helicity, in accordance with the fact that the operators of \eq{qqQQH_basis} provide a complete basis of helicity operators.
For identical quark flavors, $q\bq\,q\bq H$, the amplitudes can be obtained from the $q\bq\,q'\bq'H$ amplitudes using the relation
\begin{align}
\cA(q_{1} \bq_{2} q_3 \bq_4 H_5)= \cA(q_{1} \bq_{2} q_3' \bq_4' H_5)-\cA(q_{1} \bq_{4} q_3' \bq_2' H_5)
\,.\end{align}
The Wilson coefficients for $q\bq\,q'\bq'H$ are then given by
\begin{align} \label{eq:qqQQH_coeffs}
\vC_{(+;+)}(\lp_1,\lp_2;\lp_3,\lp_4;\lp_5)
&= \begin{pmatrix}
  A_\fin(1_q^+,2_\bq^-; 3_{q'}^+, 4_{\bq'}^-; 5_H) \\
  \tfrac{1}{N} B_\fin(1_q^+,2_\bq^-; 3_{q'}^+, 4_{\bq'}^-; 5_H)
\end{pmatrix}
,\nn\\
\vC_{(+;-)}(\lp_1,\lp_2;\lp_3,\lp_4;\lp_5)
&= \begin{pmatrix}
  A_\fin(1_q^+,2_\bq^-; 3_{q'}^-, 4_{\bq'}^+; 5_H) \\
 \tfrac{1}{N} B_\fin(1_q^+,2_\bq^-; 3_{q'}^-, 4_{\bq'}^+; 5_H)
\end{pmatrix}
,\nn\\
\vC_{(-;+)}(\lp_1,\lp_2;\lp_3,\lp_4;\lp_5) &= \vC_{(+;-)}(\lp_2,\lp_1;\lp_4,\lp_3;\lp_5)
\,,\nn\\
\vC_{(-;-)}(\lp_1,\lp_2;\lp_3,\lp_4;\lp_5) &= \vC_{(+;+)}(\lp_2,\lp_1;\lp_4,\lp_3;\lp_5)
\,,\end{align}
and for $q\bq\,q\bq H$ they are given in terms of the amplitudes $A_{\rm fin}$ and $B_{\rm fin}$ for $q\bq\,q'\bq'H$ by
\begin{widetext}
\begin{align} \label{eq:qqqqH_coeffs}
\vC_{(++)}(\lp_1,\lp_2;\lp_3,\lp_4;\lp_5)
&= \begin{pmatrix}
  A_\fin(1_q^+,2_\bq^-; 3_{q}^+, 4_{\bq}^-; 5_H)
  -\frac{1}{N} B_\fin(1_q^+, 4_{\bq}^-; 3_{q}^+, 2_{\bq}^-; 5_H) \\
  \frac{1}{N} B_\fin(1_q^+,2_\bq^-; 3_{q}^+, 4_{\bq}^-; 5_H) -A_\fin(1_q^+,4_\bq^-; 3_{q}^+, 2_{\bq}^-; 5_H)
\end{pmatrix}
, \nn \\[1ex]
\vC_{(+-)}(\lp_1,\lp_2;\lp_3,\lp_4;\lp_5)
&= \begin{pmatrix}
  A_\fin(1_q^+,2_\bq^-; 3_{q}^-, 4_{\bq}^+; 5_H) \\
  \tfrac{1}{N} B_\fin(1_q^+,2_\bq^-; 3_{q}^-, 4_{\bq}^+; 5_H)
\end{pmatrix}
,\nn \\
\vC_{(--)}(\lp_1,\lp_2;\lp_3,\lp_4;\lp_5) &= \vC_{(++)}(\lp_2,\lp_1;\lp_4,\lp_3;\lp_5)
\,.\end{align}
\end{widetext}
The relations for $\vC_{(-;\pm)}$ and $\vC_{(--)}$ follow from charge conjugation invariance.
Note that there is no exchange term for $\vC_{(+-)}$, since the amplitude vanishes when the quark and antiquark of the same flavor have the same helicity (both $+$ or both $-$). Also, recall that the symmetry factors of $1/4$ in \eq{qqqqH_basis} already take care of the interchange of identical (anti)quarks, so there are no additional symmetry factors needed for $\vC_{(++)}$. Explicit expressions for the required amplitudes at tree level are given in \app{qqqqH}.

\subsubsection{$gg q\bar q H$}

For $gg q\bar q H$, the helicity basis consists of a total of six independent operators,
\begin{align} \label{eq:ggqqH_basis}
O_{++(+)}^{ab\, \balpha\beta}
&= \frac{1}{2}\, \cB_{1+}^a\, \cB_{2+}^b\, J_{34+}^{\balpha\beta}\, H_5
\,,\nn\\
O_{+-(+)}^{ab\, \balpha\beta}
&= \cB_{1+}^a\, \cB_{2-}^b\, J_{34+}^{\balpha\beta}\, H_5
\,,\nn\\
O_{--(+)}^{ab\, \balpha\beta}
&= \frac{1}{2} \cB_{1-}^a\, \cB_{2-}^b\, J_{34+}^{\balpha\beta}\, H_5
\,,\nn\\
O_{++(-)}^{ab\, \balpha\beta}
&= \frac{1}{2}\, \cB_{1+}^a\, \cB_{2+}^b\, J_{34-}^{\balpha\beta}\, H_5
\,,\nn\\
O_{+-(-)}^{ab\, \balpha\beta}
&= \cB_{1+}^a\, \cB_{2-}^b\, J_{34-}^{\balpha\beta}\, H_5
\,,\nn\\
O_{--(-)}^{ab\, \balpha\beta}
&= \frac{1}{2} \cB_{1-}^a\, \cB_{2-}^b\, J_{34-}^{\balpha\beta}\, H_5
\,.\end{align}
We use the color basis already given in \eq{ggqqcol},
\begin{equation} \label{eq:ggqqH_color}
\vT^{ ab \alpha\bbeta}
= \Bigl(
   (T^a T^b)_{\alpha\bbeta}\,,\, (T^b T^a)_{\alpha\bbeta} \,,\, \tr[T^a T^b]\, \delta_{\alpha\bbeta}
   \Bigr)
\,.\end{equation}

Using \eq{ggqqH_color}, the color decomposition of the QCD helicity amplitudes into partial amplitudes is
\begin{align} \label{eq:ggqqH_QCD}
&\cA\bigl(g_1 g_2\, q_{3} \bq_{4} H_5 \bigr)
\nn\\ & \quad
= \img \sum_{\sigma\in S_2} \bigl[T^{a_{\sigma(1)}} T^{a_{\sigma(2)}}\bigr]_{\alpha_3\balpha_4}
\,A(\sigma(1),\sigma(2); 3_q, 4_\bq; 5_H)
\nn\\ & \qquad
+ \img\, \tr[T^{a_1} T^{a_2}]\,\delta_{\alpha_3\balpha_4}\, B(1,2; 3_q, 4_\bq; 5_H)
\,.\end{align}
The $B$ amplitudes vanish at tree level. From \eq{ggqqH_QCD} we can read off the Wilson coefficients,
\begin{align} \label{eq:ggqqH_coeffs}
\vC_{+-(+)}(\lp_1,\lp_2;\lp_3,\lp_4;\lp_5) &=
\begin{pmatrix}
   A_\fin(1^+,2^-;3_q^+,4_\bq^-; 5_H) \\
   A_\fin(2^-,1^+;3_q^+,4_\bq^-; 5_H) \\
   B_\fin(1^+,2^-;3_q^+,4_\bq^-; 5_H) \\
\end{pmatrix}
,\nn\\
\vC_{++(+)}(\lp_1,\lp_2;\lp_3,\lp_4;\lp_5)
&= \begin{pmatrix}
   A_\fin(1^+,2^+;3_q^+,4_\bq^-; 5_H) \\
   A_\fin(2^+,1^+;3_q^+,4_\bq^-; 5_H) \\
   B_\fin(1^+,2^+;3_q^+,4_\bq^-; 5_H) \\
\end{pmatrix}
,\nn\\
\vC_{--(+)}(\lp_1,\lp_2;\lp_3,\lp_4;\lp_5)
&= \begin{pmatrix}
   A_\fin(1^-,2^-;3_q^+,4_\bq^-; 5_H) \\
   A_\fin(2^-,1^-;3_q^+,4_\bq^-; 5_H) \\
   B_\fin(1^-,2^-;3_q^+,4_\bq^-; 5_H) \\
\end{pmatrix}
.\end{align}
The Wilson coefficients of the last three operators in \eq{ggqqH_basis} are obtained by charge conjugation as discussed in \subsec{discrete}. Under charge conjugation, the operators transform as
\begin{align}
&\C\, O_{\la_1\la_2(\pm)}^{ab\,\balpha\beta}(\lp_1, \lp_2; \lp_3, \lp_4;\lp_5)\, \vT^{ab\,\al\bbeta}\,\C
\nn\\ & \qquad
= - O_{\la_1\la_2(\mp)}^{ba\,\balpha\beta}(\lp_1, \lp_2; \lp_4, \lp_3;\lp_5)\, \vT^{ab\,\al\bbeta}
\,,\end{align}
so charge conjugation invariance of QCD implies
\begin{align}
\vC_{\la_1\la_2(-)}(\lp_1,\lp_2;\lp_3,\lp_4;\lp_5)
&= \hV \vC_{\la_1\la_2(+)}(\lp_1,\lp_2;\lp_4,\lp_3;\lp_5)
\nn\\
\text{with}\qquad
\hV & =
\begin{pmatrix}
  0 & -1 & 0 \\
  -1 & 0 & 0 \\
  0 & 0 & -1
\end{pmatrix}
.\end{align}
Explicit expressions for the required amplitudes at tree level are given in \app{ggqqH}.

\subsubsection{$ggggH$}

For $ggggH$, the helicity basis consists of five independent operators,
\begin{align} \label{eq:ggggH_basis}
O_{++++}^{abcd} &= \frac{1}{4!}\, \cB_{1+}^a \cB_{2+}^b \cB_{3+}^c \cB_{4+}^d\, H_5
\,,\nn\\
O_{+++-}^{abcd} &= \frac{1}{3!}\, \cB_{1+}^a \cB_{2+}^b \cB_{3+}^c \cB_{4-}^d\, H_5
\,,\nn\\
O_{++--}^{abcd} &= \frac{1}{4}\, \cB_{1+}^a \cB_{2+}^b \cB_{3-}^c \cB_{4-}^d\, H_5
\,,\nn\\
O_{---+}^{abcd} &= \frac{1}{3!}\, \cB_{1-}^a \cB_{2-}^b \cB_{3-}^c \cB_{4+}^d\, H_5
\,,\nn\\
O_{----}^{abcd} &= \frac{1}{4!}\, \cB_{1-}^a \cB_{2-}^b \cB_{3-}^c \cB_{4-}^d\, H_5
\,.\end{align}
We use the basis of color structures
\begin{equation} \label{eq:ggggH_color}
\vT^{ abcd} =
\frac{1}{2\cdot 2T_F}\begin{pmatrix}
\tr[abcd] + \tr[dcba] \\ \tr[acdb] + \tr[bdca] \\ \tr[adbc] + \tr[cbda] \\
2\tr[ab]\, \tr[cd] \\ 2\tr[ac]\, \tr[db] \\ 2\tr[ad]\, \tr[bc]
\end{pmatrix}^{\!\!\!T}
,\end{equation}
where we have used the shorthand notation
\begin{equation}
\tr[ab] = \tr[T^a T^b]
\,,\qquad
\tr[abcd] = \tr[T^a T^b T^c T^d]
\,.\end{equation}
Note that the three independent color structures with a minus sign instead of the plus sign in the first three lines in \eq{ggggH_color} can be eliminated using charge conjugation invariance, see \subsec{ggggbasis}.

The color decomposition of the QCD helicity amplitudes into partial amplitudes using the color basis in \eq{ggggH_color} is
\begin{align} \label{eq:ggggH_QCD}
\cA(g_1 g_2 g_3 g_4 H_5)
&=\frac{ \img}{2T_F} \biggl[\sum_{\si \in S_4/Z_4}\! \tr[a_{\si(1)} a_{\si(2)} a_{\si(3)} a_{\si(4)}]
\nn \\ & \qquad \times
A\bigl(\si(1),\si(2),\si(3),\si(4); 5_H\bigr)
\nn\\ &\quad
+ \sum_{\si \in S_4/Z_2^3}\! \tr[a_{\si(1)} a_{\si(2)}] \tr[a_{\si(3)} a_{\si(4)}]
\nn\\ & \qquad \times
B\bigl(\si(1),\si(2),\si(3),\si(4); 5_H\bigr) \biggr]
\,,\end{align}
where the $B$ amplitudes vanish at tree level. From \eq{ggggH_QCD} we obtain the Wilson coefficients,
\begin{align} \label{eq:ggggH_coeffs}
\vC_{++--}(\lp_1,\lp_2,\lp_3,\lp_4;\lp_5)
&= \begin{pmatrix}
  2A_\fin(1^+,2^+,3^-,4^-; 5_H) \\
 2A_\fin(1^+,3^-,4^-,2^+; 5_H) \\
  2A_\fin(1^+,4^-,2^+,3^-; 5_H) \\
  B_\fin(1^+,2^+,3^-,4^-; 5_H) \\
  B_\fin(1^+,3^-,4^-,2^+; 5_H) \\
  B_\fin(1^+,4^-,2^+,3^-; 5_H)
\end{pmatrix}
\!,\nn\\
\vC_{+++-}(\lp_1,\lp_2,\lp_3,\lp_4;\lp_5)
&= \begin{pmatrix}
  2A_\fin(1^+,2^+,3^+,4^-; 5_H) \\
  2A_\fin(1^+,3^+,4^-,2^+; 5_H) \\
  2A_\fin(1^+,4^-,2^+,3^+; 5_H) \\
  B_\fin(1^+,2^+,3^+,4^-; 5_H) \\
  B_\fin(1^+,3^+,4^-,2^+; 5_H) \\
  B_\fin(1^+,4^-,2^+,3^+; 5_H) \\
\end{pmatrix}
\!,\nn\\
\vC_{++++}(\lp_1,\lp_2,\lp_3,\lp_4;\lp_5)
&= \begin{pmatrix}
2A_\fin(1^+,2^+,3^+,4^+; 5_H) \\
  2A_\fin(1^+,3^+,4^+,2^+; 5_H) \\
  2A_\fin(1^+,4^+,2^+,3^+; 5_H) \\
  B_\fin(1^+,2^+,3^+,4^+; 5_H) \\
  B_\fin(1^+,3^+,4^+,2^+; 5_H) \\
  B_\fin(1^+,4^+,2^+,3^+; 5_H) \\
\end{pmatrix}
\!,\nn\\
\vC_{---+}(\lp_1, \ldots; \lp_5)
&= \vC_{+++-}(\lp_1, \ldots; \lp_5) \Big|_{\langle..\rangle \leftrightarrow [..]}
\,,\nn\\
\vC_{----}(\lp_1, \ldots; \lp_5)
&= \vC_{++++}(\lp_1, \ldots; \lp_5) \Big|_{\langle..\rangle \leftrightarrow [..]}
\,.\end{align}
The last two coefficients follow from parity invariance. The factors of two in the first three entries of the coefficients come from combining the two traces in the first three entries in \eq{ggggH_color} using charge conjugation invariance. Because of the cyclic symmetry of the traces, the partial amplitudes are invariant under the corresponding cyclic permutations of their first four arguments, which means that most of the amplitudes in \eq{ggggH_coeffs} are not independent. Explicit expressions for the necessary amplitudes at tree level are given in \app{ggggH}.

\section{Vector Boson + Jets}
\label{sec:vec}

In this section, we give the helicity operator basis and the corresponding matching for the production of a $\gamma$, $Z$, or $W$ vector boson in association with up to two jets. The corresponding helicity amplitudes are collected in \app{Zamplitudes}.

We work at tree level in the electroweak coupling and consider only QCD corrections, so any external $q\bar q$ pairs are described by the helicity vector currents $J_{ij\pm}$ in \eq{jpm_def}. We always include the subsequent leptonic decays $\ga/Z \to \ell \bar \ell$, $W^\pm \to \nu \bar \ell/\ell \bar \nu$. In the following, for $\ga/Z$ processes, $\ell$ stands for any charged lepton or neutrino flavor, and $q$ stands for any quark flavor. For $W$ processes, we use $\ell$ to denote any charged lepton flavor and $\nu$ the corresponding neutrino flavor. Similarly, we use $u$ and $d$ to denote any up-type or down-type quark flavor (i.e. not necessarily first generation quarks only).

The operators in the helicity bases satisfy the transformation properties under C and P as discussed in \subsec{discrete}. However, the weak couplings in the amplitudes explicitly violate C and P. Therefore, to utilize the C and P transformations of the operators and minimize the number of required amplitudes and Wilson coefficients, it is useful to separate the weak couplings from the amplitudes.

We define $P_Z$ and $P_W$ as the ratios of the $Z$ and $W$ propagators to the photon propagator,
\begin{equation}
P_{Z,W}(s) = \frac{s}{s-m_{Z,W}^2 + \img \Ga_{Z,W} m_{Z,W}}
\,.\end{equation}
The left- and right-handed couplings $v_{L,R}$ of a particle to the $Z$ boson are, as usual,
\begin{align}
 v_L^i = \frac{2 T_3^i - 2Q^i \sin^2 \theta_W}{\sin(2\theta_W)}
 \,, \quad
 v_R^i = - \frac{2Q^i \sin^2 \theta_W}{\sin(2\theta_W)}
\,,\end{align}
where $T_3^i$ is the third component of weak isospin, $Q^i$ is the electromagnetic charge in units of $\abs{e}$, and $\theta_W$ is the weak mixing angle.

The $\gamma/Z$ amplitudes can then be decomposed as
\begin{align} \label{eq:Z_expand}
&\cA(\dotsb \ell \bar \ell)
\nn \\ & \quad
= e^2
\biggl\{ \bigl[Q^\ell Q^q + v_{L,R}^\ell v_{L,R}^q P_Z(s_{\ell\bar\ell})\bigr] \cA_q(\dotsb \ell \bar \ell)
\nn \\ & \qquad
+\sum_{i=1}^{n_f} \Bigl[Q^\ell Q^i + v_{L,R}^\ell\frac{v_L^i + v_R^i}{2} P_Z(s_{\ell\bar\ell}) \Bigr] \cA_v(\dotsb \ell \bar\ell)
\nn \\ & \qquad
+ \frac{v_{L,R}^\ell}{\sin(2\theta_W)} P_Z(s_{\ell\bar\ell})\, \cA_{a}(\dotsb \ell \bar\ell) \biggr\}
\,.\end{align}
Here, $\cA_q$ corresponds to the usual contribution where the vector boson couples directly to the external quark line with flavor $q$. (There is one such contribution for each external $q\bar q$ pair, and this contribution is absent for pure gluonic amplitudes like $gggZ$.) For $\cA_v$, the $\ga/Z$ couples to an internal quark loop through a vector current and the sum runs over all considered internal quark flavors. For $\cA_a$, the $Z$ boson couples to an internal quark loop through the axial-vector current. This means that when using parity and charge conjugation we have to include an additional relative minus sign for this contribution. We have also made the assumption in \eq{Z_expand} that all quarks, except for the top, are massless. Since $\cA_a$ vanishes when summed over a massless isodoublet, this has the consequence that only the $b,t$ isodoublet contributes to $\cA_a$, hence the lack of sum over flavors. We have made this simplification following the one-loop calculation of Ref.~\cite{Bern:1997sc}, which calculated the amplitude in an expansion in $1/m_t^2$, assuming all other kinematic invariants to be smaller than the top mass. From the point of view of constructing a basis these assumptions are trivial to relax.

The $W^\mp$ amplitudes can be written as
\begin{align}\label{eq:W_expand}
\cA(\dotsb \ell^- \bar \nu^+)
&= \frac{e^2 V_{ud}}{2\sin^2 \theta_W}\, P_W(s_{\ell\bar\nu})\, \cA_q(\dotsb \ell^- \bar \nu^+)
\,, \nn \\
\cA(\dotsb \nu^- \bar \ell^+)
&= \frac{e^2 V_{ud}^\dagger}{2\sin^2 \theta_W}\, P_W(s_{\nu\bar\ell})\, \cA_q(\dotsb \nu^- \bar \ell^+)
\,,\end{align}
where $V_{ud}$ is the appropriate CKM-matrix element. The $\cA_q$ amplitudes are the same in \eqs{Z_expand}{W_expand}, since all electroweak couplings have been extracted, but we have explicitly included the helicity labels (not to be mistaken as charge labels) to emphasize that these are the only possible helicities. The analogs of $\cA_v$ and $\cA_a$ do not exist for $W$ production.

We note again that \eqs{Z_expand}{W_expand} hold at tree level in the electroweak coupling, which is what we consider in this paper. At this level, the leptons always couple to the vector boson through the currents [see \eq{Fierzetc}]
\begin{equation} \label{eq:lep_reduce}
\mae{p_\ell\pm}{\ga^\mu}{p_{\bar\ell}\pm} = \mae{p_{\bar\ell}\mp}{\ga^\mu}{p_\ell\mp}
\,.\end{equation}
This allows us to obtain the Wilson coefficients for opposite lepton helicities simply by interchanging the lepton momenta.

\subsection{\boldmath $V + 0$ Jets}
\label{subsec:V0Jets}

For $\ga/Z+0$ jets, the partonic process is $q\bar q \ell \bar \ell$, and the basis of helicity operators is
\begin{align} \label{eq:Z0_basis}
O_{(+;\pm)}^{\balpha\bt}
&= J_{q\, 12+}^{\balpha\bt}\, J_{\ell \, 34\pm}
\,,\nn\\
O_{(-;\pm)}^{\balpha\bt}
&= J_{q\, 12-}^{\balpha\bt}\, J_{\ell \, 34\pm}
\,.\end{align}
In principle, the process $gg \ell \bar \ell$ is allowed through the axial anomaly, but its contribution vanishes because in the matching calculation the gluons are taken to be on shell, and we neglect lepton masses.

For $W^\mp+0$ jets, the partonic processes are $u \bar d \ell \bar \nu$ and $d \bar u \nu \bar \ell$, respectively. Since the $W$ only couples to left-handed fields, the helicity basis simplifies to
\begin{align} \label{eq:W0_basis}
O_{(W^-)}^{\balpha\bt}
&= J_{\bar u d\, 12-}^{\balpha\bt}\, J_{\bar\ell \nu\, 34-}
\,,\nn\\
O_{(W^+)}^{\balpha\bt}
&= J_{\bar d u\, 12-}^{\balpha\bt}\, J_{\bar\nu \ell \, 34-}
\,.\end{align}
Here, we have explicitly written out the flavor structure of the currents. However, we use the shorthand subscript $(W^\mp)$ on the operators and Wilson coefficients, since we will not focus any further on the flavor structure. In an explicit calculation, one must of course sum over all relevant flavor combinations.

The unique color structure for $V+0$ jets is
\begin{equation} \label{eq:Z0_color}
\vT^{ \al\bbeta} =  \begin{pmatrix} \de_{\al\bbeta} \end{pmatrix}
,\end{equation}
and extracting it from the amplitudes, we have
\begin{align}
\cA_{q,v,a}(q_{1}\bar q_{2} \ell_{3} \bar \ell_{4})
= \img\, \delta_{\alpha_1 \balpha_2}  \, A_{q,v,a}(1_q,2_{\bar q}; 3_\ell, 4_{\bar \ell})
\,. \end{align}
Here, $A_v$ and $A_a$ first appear at two loops. In addition, $A_a$ is proportional to the top and bottom mass splitting due to isodoublet cancellations. It drops out when both top and bottom are treated as massless (e.g., when the matching scale is much larger than the top mass).

We use the same electroweak decomposition as in \eqs{Z_expand}{W_expand} to write the Wilson coefficients. For $\gamma/Z+0$ jets, we have
\begin{align} \label{eq:Z0_expand_Wilson}
&\vec C_{(\lambda_q;\lambda_\ell)}(\lp_1, \lp_2; \lp_3, \lp_4)
\nn \\ & \quad
= e^2  \,
\biggl\{ \bigl[Q^\ell Q^q + v_{\lambda_\ell}^\ell v_{\lambda_q}^q P_Z(s_{34})\bigr]
\vec C_{q(\lambda_q;\lambda_\ell)}(\dots)
\nn \\ & \qquad
+ \sum_{i=1}^{n_f} \Bigl[Q^\ell Q^i + v_{\lambda_\ell}^\ell\frac{v_L^i + v_R^i}{2} P_Z(s_{34}) \Bigr]
\vec C_{v(\lambda_q;\lambda_\ell)}(\dots)
\nn \\ & \qquad
+ \frac{v_{\lambda_\ell}^\ell}{\sin(2\theta_W)} P_Z(s_{34})\, \vec C_{a(\lambda_q;\lambda_\ell)}(\dots)  \biggr\}
\,,\end{align}
where the weak couplings are determined by the helicity labels of the quark and lepton currents,
\begin{align}
v_{+}^\ell = v_R^\ell
\,,\quad
v_{-}^\ell = v_L^\ell
\,,\qquad
v_{+}^q = v_R^q
\,,\quad
v_{-}^q = v_L^q
\,.\end{align}
For $W + 0$ jets, we simply have
\begin{align}
\vC_{(W^-)}(\lp_1, \lp_2; \lp_3, \lp_4)
&= \frac{e^2 V_{ud}}{2\sin^2 \theta_W}\, P_W(s_{34})\, \vC_{q (-;-)}(\ldots)
\,, \nn \\
\vC_{(W^+)}(\lp_1, \lp_2; \lp_3, \lp_4)
&= \frac{e^2 V_{ud}^{\dagger}}{2\sin^2 \theta_W}\, P_W(s_{34})\, \vC_{q (-;-)}(\ldots)
\,.\end{align}
In all cases, the momentum arguments on the right-hand side are the same as on the left-hand side.
Note that the $\vC_{q(-;-)}$ coefficient is the same in all cases. The Wilson coefficients are given by
\begin{align} \label{eq:qqV_coeffs}
\vC_{x(+;+)}(\lp_1, \lp_2; \lp_3, \lp_4) &= A_{x,\fin}(1_q^+,2_{\bar q}^-;3_\ell^+,4_{\bar \ell}^-)
\,, \nn \\
\vC_{x(+;-)}(\lp_1, \lp_2; \lp_3, \lp_4) &= \vC_{x(+;+)}(\lp_1, \lp_2; \lp_4, \lp_3)
\,, \nn \\
\vC_{q,v(-;\pm)}(\lp_1, \lp_2; \lp_3, \lp_4) &= \vC_{q,v(+;\pm)}(\lp_2, \lp_1; \lp_3, \lp_4)
\,, \nn \\
\vC_{a(-;\pm)}(\lp_1, \lp_2; \lp_3, \lp_4) &= -\vC_{a(+;\pm)}(\lp_2, \lp_1; \lp_3, \lp_4)
\,, \end{align}
where $x = q,v,a$ and as discussed in \sec{matching} the subscript ``$\fin$'' denotes the IR-finite part of the helicity amplitudes.
The second relation follows from \eq{lep_reduce}. The last two relations follow from charge conjugation invariance.
At tree level and one loop only $\vC_q$ receives a nonvanishing contribution. The $A_q$ amplitude is given in \app{V0amplitudes}.

\subsection{\boldmath $V + 1$ Jet}
\label{subsec:V1J}

\subsubsection{$gq\bar qV$}

For $\ga/Z+1$ jet, the partonic process is $g q\bar q \ell \bar\ell$, and the basis of helicity operators is
\begin{align} \label{eq:Z1_basis}
O_{+(+;\pm)}^{a\,\balpha\bt}
&= \cB_{1+}^a\, J_{q\, 23+}^{\balpha\bt}\, J_{\ell \, 45\pm}
\,,\nn\\
O_{+(-;\pm)}^{a\,\balpha\bt}
&= \cB_{1+}^a\, J_{q\, 23-}^{\balpha\bt}\, J_{\ell \, 45\pm}
\,,\nn\\
O_{-(+;\pm)}^{a\,\balpha\bt}
&= \cB_{1-}^a\, J_{q\, 23+}^{\balpha\bt}\, J_{\ell \, 45\pm}
\,,\nn\\
O_{-(-;\pm)}^{a\,\balpha\bt}
&= \cB_{1-}^a\, J_{q\, 23-}^{\balpha\bt}\, J_{\ell \, 45\pm}
\,.\end{align}

For $W^\mp+1$ jet, the partonic processes are $gu \bar d \ell \bar \nu$ and $gd \bar u \nu \bar \ell$, respectively, and the helicity operator basis is
\begin{align} \label{eq:W1_basis}
O_{\pm\, (W^-)}^{a\,\balpha\bt}
&= \cB_{1\pm}^a\, J_{\bar u d\, 23-}^{\balpha\bt}\, J_{\bar\ell \nu\, 45-}
\,,\nn\\
O_{\pm\, (W^+)}^{a\,\balpha\bt}
&= \cB_{1\pm}^a\, J_{\bar d u\, 23-}^{\balpha\bt}\, J_{\bar\nu \ell \, 45-}
\,.\end{align}
The unique color structure for $gq\bar qV$ is
\begin{equation} \label{eq:Z1q_color}
\vT^{ a\, \al\bbeta} = \begin{pmatrix} T^a_{\al\bbeta} \end{pmatrix}
,\end{equation}
and extracting it from each of the amplitudes, we have
\begin{align}
\cA_{x}(g_1 q_{2}\bar q_{3} \ell_{4} \bar \ell_{5})
&=\img\, T^{a_1}_{\alpha_2 \balpha_3} \, A_{x}(1;2_q,3_{\bar q};4_\ell,5_{\bar \ell})
\,,\end{align}
where the subscript $x$ stands for one of $q,v,a$.

As for $V+0$ jets, we write the Wilson coefficients using the electroweak decomposition in \eqs{Z_expand}{W_expand}. For $\gamma/Z+1$ jet, we have
\begin{align}
&\vec C_{\lambda(\lambda_q;\lambda_\ell)}(\lp_1; \lp_2, \lp_3; \lp_4, \lp_5)
\nn \\* & \quad
= e^2  \,
\biggl\{ \bigl[Q^\ell Q^q + v_{\lambda_\ell}^\ell v_{\lambda_q}^q P_Z(s_{45})\bigr]
\vec C_{q\lambda(\lambda_q;\lambda_\ell)}(\ldots)
\nn \\ & \qquad
+ \sum_{i=1}^{n_f} \Bigl[Q^\ell Q^i + v_{\lambda_\ell}^\ell\frac{v_L^i + v_R^i}{2} P_Z(s_{45}) \Bigr]
\vec C_{v\lambda(\lambda_q;\lambda_\ell)}(\ldots)
\nn \\ & \qquad
+ \frac{v_{\lambda_\ell}^\ell}{\sin(2\theta_W)} P_Z(s_{45})\, \vec C_{a\lambda(\lambda_q;\lambda_\ell)}(\ldots) \biggr\}
\,,\end{align}
where the weak couplings are determined by the helicity labels of the quark and lepton currents,
\begin{align}
v_{+}^\ell = v_R^\ell
\,,\quad
v_{-}^\ell = v_L^\ell
\,,\qquad
v_{+}^q = v_R^q
\,,\quad
v_{-}^q = v_L^q
\,.\end{align}
For $W + 1$ jet, we have
\begin{align}
\vC_{\lambda(W^\mp)}(\ldots)
&= \frac{e^2 V_{ud}^{(\dagger)}}{2\sin^2 \theta_W}\, P_W(s_{45})\, \vC_{q \lambda(-;-)}(\ldots)
\,.\end{align}
The Wilson coefficients are given by 
\begin{align} \label{eq:gqqV_coeffs}
\vC_{x+(+;+)}(\lp_1; \lp_2, \lp_3; \lp_4, \lp_5)
&= A_{x,\fin} (1^+;2_q^+,3_{\bar q}^-; 4_\ell^+,5_{\bar \ell}^-)
\,, \nn \\
\vC_{x\lambda(+;-)}(\lp_1; \lp_2, \lp_3; \lp_4, \lp_5)
&= \vC_{x\lambda(+;+)}(\lp_1; \lp_2, \lp_3; \lp_5, \lp_4)
\,, \nn \\
\vC_{q,v\lambda(-;\pm)}(\lp_1; \lp_2, \lp_3; \lp_4, \lp_5)
&= -\vC_{q,v\lambda(+;\pm)}(\lp_1; \lp_3, \lp_2; \lp_4, \lp_5)
\,, \nn \\
\vC_{a\lambda(-;\pm)}(\lp_1; \lp_2, \lp_3; \lp_4, \lp_5)
&= \vC_{a\lambda(+;\pm)}(\lp_1; \lp_3, \lp_2; \lp_4, \lp_5)
\,.\end{align}
The second relation follows from \eq{lep_reduce}, and the last two relations follow from charge conjugation invariance. The Wilson coefficients with a negative helicity gluon follow from parity invariance,
\begin{align}
&\vC_{q,v-(+;\pm)}(\lp_1; \lp_2, \lp_3; \lp_4, \lp_5)
\nn \\ & \qquad
= \vC_{q,v+(-;\mp)}(\lp_1; \lp_2, \lp_3; \lp_4, \lp_5)\Big|_{\langle..\rangle \leftrightarrow [..]}
\,, \nn \\
&\vC_{a-(+;\pm)}(\lp_1; \lp_2, \lp_3; \lp_4, \lp_5)
\nn \\ & \qquad
= - \vC_{a+(-;\mp)}(\lp_1; \lp_2, \lp_3; \lp_4, \lp_5)\Big|_{\langle..\rangle \leftrightarrow [..]}
\,.\end{align}
The helicity amplitudes for $g q \bar q \ell \bar \ell$ were calculated in Ref.~\cite{Giele:1991vf, Arnold:1988dp, Korner:1990sj}. We provide the tree-level and one-loop results in \app{V1amplitudes}. The two-loop amplitudes were computed in Refs.~\cite{Garland:2001tf, Garland:2002ak}.

\subsubsection{$gggV$}

The partonic process $ggg\ell\bar\ell$ first appears at one loop, and thus contributes only at relative $\ord{\alpha_s^2}$ to $\ga/Z+1$ jet. Nevertheless, for the sake of completeness (and curiosity) we briefly discuss it here. The helicity operator basis is
\begin{align}
O_{+++(\pm)}^{abc}
&= \frac{1}{3!}\, \cB_{1+}^a\, \cB_{2+}^b\, \cB_{3+}^c\, J_{\ell \, 45\pm}
\,,\nn\\
O_{++-(\pm)}^{abc}
&= \frac{1}{2}\, \cB_{1+}^a\, \cB_{2+}^b\, \cB_{3-}^c\, J_{\ell \, 45\pm}
\,,\nn\\
O_{+--(\pm)}^{abc}
&= \frac{1}{2}\, \cB_{1+}^a\, \cB_{2-}^b\, \cB_{3-}^c\, J_{\ell \, 45\pm}
\,,\nn\\
O_{---(\pm)}^{abc}
&= \frac{1}{3!}\, \cB_{1-}^a\, \cB_{2-}^b\, \cB_{3-}^c\, J_{\ell \, 45\pm}
\,.\end{align}

The color space is two dimensional. We use the basis
\begin{equation} \label{eq:Z1g_color}
\vT^{abc} =
\Bigl(  \img f^{abc}\,,\, d^{abc} \Bigr)
\,,\end{equation}
in terms of which we can write the $ggg \ell \bar \ell$ amplitudes as
\begin{align}\label{eq:Zggg_amp_color}
\cA_v(g_1 g_2  g_3  \ell_4 \bar \ell_5)
&= \img\, d^{a_1 a_2 a_3} A_v(1,2,3;4_\ell,5_{\bar \ell})
\,, \nn \\
\cA_a(g_1 g_2  g_3  \ell_4 \bar \ell_5)
&= \img\, (\img f^{a_1 a_2 a_3}) A_{a}(1,2,3;4_\ell,5_{\bar \ell})
\,.\end{align}
We will justify shortly that to all orders, only a single color structure appears for each of $\cA_v$, $\cA_a$.
This process can only occur via a closed quark loop, so there is no $\cA_q$ contribution.
The $gggV$ operators transform under charge conjugation as
\begin{align}
&\C\, O_{\la_1\la_2\la_3(\pm)}^{abc}(\lp_1, \lp_2, \lp_3; \lp_4, \lp_5)\, \vT^{ abc} \,\C
\nn\\ & \qquad
= O_{\la_1\la_2\la_3(\mp)}^{cba}(\lp_1, \lp_2, \lp_3; \lp_5, \lp_4)\, \vT^{ abc}
\,.\end{align}
Charge conjugation invariance of QCD thus leads to
\begin{align}
&C_{v\la_1\la_2\la_3(\pm)}^{abc}(\lp_1, \lp_2, \lp_3; \lp_4, \lp_5)
\nn \\ & \quad
= C_{v\la_1\la_2\la_3(\mp)}^{cba}(\lp_1, \lp_2, \lp_3; \lp_5, \lp_4)
\nn \\ & \quad
= C_{v\la_1\la_2\la_3(\pm)}^{cba}(\lp_1, \lp_2, \lp_3; \lp_4, \lp_5)
\,,\end{align}
where we used \eq{lep_reduce} in the last line. This implies that to all orders in the strong coupling, only the fully symmetric color structure $d^{abc}$ can contribute to $\cA_v$ and $\vC_v$. For $\vC_a$ the same relation holds but with an additional minus sign on the right-hand side due to the weak axial-vector coupling in $\cA_a$. This implies that for $\cA_a$ and $\vC_a$ only the fully antisymmetric color structure $\img f^{abc}$ contributes, as given in \eq{Zggg_amp_color}.

We decompose the $ggg\ell \bar \ell$ Wilson coefficients as
\begin{align}
&\vC_{\lambda_1\lambda_2\lambda_3(\lambda_\ell)}
\nn \\ & \quad
= e^2  \,
\biggl\{
\sum_{i=1}^{n_f} \Bigl[Q^\ell Q^i + v_{\lambda_\ell}^\ell\frac{v_L^i + v_R^i}{2} P_Z(s_{45}) \Bigr]
\vC_{v\lambda_1\lambda_2\lambda_3(\lambda_\ell)}
\nn \\ & \qquad
+ \frac{v_{\lambda_\ell}^\ell}{\sin(2\theta_W)} P_Z(s_{45})\, \vC_{a\lambda_1\lambda_2\lambda_3(\lambda_\ell)}
\biggr\}
\,,\end{align}
where
\begin{align}
v_{+}^\ell = v_R^\ell
\,,\qquad
v_{-}^\ell = v_L^\ell
\,,\end{align}
and we have 
\begin{align}
&\vC_{v\lambda_1\lambda_2\lambda_3(+)}(\lp_1, \lp_2, \lp_3; \lp_4, \lp_5)
\nn \\ & \qquad
= \begin{pmatrix} 0 \\ A_{v, \fin} (1^{\lambda_1},2^{\lambda_2}, 3^{\lambda_3}; 4_\ell^+,5_{\bar \ell}^-)    \end{pmatrix}
, \nn \\
&\vC_{a\lambda_1\lambda_2\lambda_3(+)}(\lp_1, \lp_2, \lp_3; \lp_4, \lp_5)
\nn \\ & \qquad
= \begin{pmatrix} A_{a, \fin}(1^{\lambda_1},2^{\lambda_2}, 3^{\lambda_3}; 4_\ell^+,5_{\bar \ell}^-) \\ 0   \end{pmatrix}
\,, \nn \\
&\vC_{v,a\lambda_1\lambda_2\lambda_3(-)}(\lp_1, \lp_2, \lp_3; \lp_4, \lp_5)
\nn \\ & \qquad
= \vC_{v,a\lambda_1\lambda_2\lambda_3(+)}(\lp_1, \lp_2, \lp_3; \lp_5, \lp_4)
\,.\end{align}
For brevity, we have not written out the various gluon helicity combinations.
The one-loop amplitudes for $gggZ$ were calculated in Ref.~\cite{vanderBij:1988ac}, and the two-loop amplitudes were computed in Ref.~\cite{Gehrmann:2013vga}. Since their contribution is very small we do not repeat them here.

\subsection{\boldmath $V + 2$ Jets}
\label{subsec:V2J}

Here we consider the processes $q'\bq' q\bq\, V$, $q\bq\,q\bq\, V$, and $gg\,q\bq\, V$. The $ggggV$ process is allowed as well, but only arises at one loop, so we do not explicitly consider here. It can be treated similarly to $gggV$, but using the $gggg$ color basis analogous to that for $ggggH$ given in \eq{ggggH_color}.

The NLO helicity amplitudes for $V+2$ jets were calculated in Refs.~\cite{Bern:1996ka, Bern:1997sc} assuming that all kinematic scales are smaller than the top mass $m_t$ and including the $1/m_t^2$ corrections. We give the full expressions for the LO results in \app{V2amplitudes}. Since the NLO results are rather long, we do not repeat them, but we show how to convert the results of Refs.~\cite{Bern:1996ka, Bern:1997sc} to our notation.

\subsubsection{$q'\bq' q\bq\, V$ and $q\bq\, q\bq\, V$}

For $q' \bq' q\bq\, \ell \bar \ell$, the helicity operator basis is
\begin{align} \label{eq:Z2_basis_qQ}
O_{(+;+;\pm)}^{\balpha\bt\bgamma\delta}
&= J_{q'\, 12+}^{\balpha\bt}\, J_{q\, 34+}^{\bgamma\delta}\, J_{\ell \, 56\pm}
\,,\nn\\
O_{(+;-;\pm)}^{\balpha\bt\bgamma\delta}
&= J_{q'\, 12+}^{\balpha\bt}\, J_{q\, 34-}^{\bgamma\delta}\, J_{\ell \, 56\pm}
\,,\nn\\
O_{(-;+;\pm)}^{\balpha\bt\bgamma\delta}
&= J_{q'\, 12-}^{\balpha\bt}\, J_{q\, 34+}^{\bgamma\delta}\, J_{\ell \, 56\pm}
\,,\nn\\
O_{(-;-;\pm)}^{\balpha\bt\bgamma\delta}
&= J_{q'\, 12-}^{\balpha\bt}\, J_{q\, 34-}^{\bgamma\delta}\, J_{\ell \, 56\pm}
\,.\end{align}
For identical quark flavors, $q\bq\, q\bq\,\ell \bar \ell$, the basis reduces to
\begin{align}\label{eq:Z2_basis_qq}
O_{(++;\pm)}^{\balpha\bt\bgamma\delta}
&= \frac{1}{4}J_{q\, 12+}^{\balpha\bt}\, J_{q\, 34+}^{\bgamma\delta}\, J_{\ell \, 56\pm}
\,,\nn\\
O_{(+-;\pm)}^{\balpha\bt\bgamma\delta}
&= J_{q\, 12+}^{\balpha\bt}\, J_{q\, 34-}^{\bgamma\delta}\, J_{\ell \, 56\pm}
\,,\nn\\
O_{(--;\pm)}^{\balpha\bt\bgamma\delta}
&= \frac{1}{4}J_{q\, 12-}^{\balpha\bt}\, J_{q\, 34-}^{\bgamma\delta}\, J_{\ell \, 56\pm}
\,.\end{align}
For $W+2$ jets, the corresponding partonic processes are $q \bq\, u \bar d\, \ell \bar \nu$ and $q \bq\, d \bar u\, \nu \bar \ell$,
and the helicity operator basis is
\begin{align} \label{eq:W2_basis_qq}
O_{(\pm; W^-)}^{\balpha\bt\bgamma\delta}
&= J_{q\, 12\pm}^{\balpha\bt}\, J_{\bar u d\, 34-}^{\bgamma\delta}\, J_{\bar\ell \nu\, 56-}
\,,\nn\\
O_{(\pm; W^+)}^{\balpha\bt\bgamma\delta}
&= J_{q\, 12\pm}^{\balpha\bt}\, J_{\bar d u\, 34-}^{\bgamma\delta}\, J_{\bar\nu \ell \, 56-}
\,.\end{align}
We use the color basis
\begin{equation} \label{eq:qqqqV_color}
\vT^{ \al\bbeta\ga\bdelta} =
 2T_F \Bigl(
  \de_{\al\bdelta}\, \de_{\ga\bbeta}\,,\, \delta_{\al\bbeta}\, \de_{\ga\bdelta}
\Bigr)
\,.\end{equation}
For distinct quark flavors, the color decomposition of the amplitudes in this basis is 
\begin{align}\label{eq:qqQQV2Jets}
&\cA_x(q_1' \bq_2' q_3 \bq_4 \ell_5 \bar \ell_6)
\\* & \quad
=2T_F \img\, \de_{\al_1\balpha_4} \de_{\al_3\balpha_2}\, A_x(1_{q'},2_{\bq'};3_q,4_\bq;5_\ell,6_{\bar \ell})
\nn \\ & \qquad
+2T_F \img\,\de_{\al_1 \balpha_2} \de_{\al_3 \balpha_4}\, \frac{1}{N}\, B_x(1_{q'},2_{\bq'};3_q,4_{\bar q};5_\ell,6_{\bar \ell})
\nn\,.\end{align}
For identical quark flavors the amplitudes can be obtained from the distinct flavor amplitudes using
\begin{align}
\cA_x(q_1 \bq_2 q_3 \bq_4 \ell_5 \bar \ell_6)
&= \cA_x(q_1' \bq_2' q_3 \bq_4 \ell_5 \bar \ell_6)
\nn \\ & \quad
- \cA_x(q_1' \bq_4' q_3 \bq_2 \ell_5 \bar \ell_6)
\,,\end{align}
where it is to be understood that the electroweak couplings of $q'$ must also be replaced by those of $q$.

Writing the Wilson coefficients in the decomposition in \eqs{Z_expand}{W_expand}, we have for the $q' \bq' q\bq\, \ell \bar \ell$ channel
\begin{align}
&\vC_{(\lambda_{q'};\lambda_q;\lambda_\ell)}(\lp_1, \lp_2; \lp_3, \lp_4; \lp_5, \lp_6)
\nn \\ & \quad
= e^2  \,
\biggl\{ \bigl[Q^\ell Q^q + v_{\lambda_\ell}^\ell v_{\lambda_q}^q P_Z(s_{56})\bigr]
\nn \\ & \qquad\qquad\times
\vC_{q(\lambda_{q'};\lambda_q;\lambda_\ell)}(\lp_1, \lp_2; \lp_3, \lp_4; \lp_5, \lp_6)
\nn \\ & \qquad
+ \bigl[Q^\ell Q^{q'} + v_{\lambda_\ell}^\ell v_{\lambda_{q'}}^{q'} P_Z(s_{56})\bigr]
\nn \\ & \qquad\quad\times
\vC_{q(\lambda_q;\lambda_{q'};\lambda_\ell)}(\lp_3, \lp_4; \lp_1, \lp_2; \lp_5, \lp_6)
\nn \\ & \qquad
+ \sum_{i=1}^{n_f} \Bigl[Q^\ell Q^i + v_{\lambda_\ell}^\ell\frac{v_L^i + v_R^i}{2} P_Z(s_{56}) \Bigr]
\vC_{v(\lambda_{q'};\lambda_q;\lambda_\ell)}(\ldots)
\nn \\ & \qquad
+ \frac{v_{\lambda_\ell}^\ell}{\sin(2\theta_W)} P_Z(s_{56})\, \vC_{a(\lambda_{q'};\lambda_q;\lambda_\ell)}(\ldots)
\biggr\}
\,,\end{align}
with the weak couplings
\begin{align}
v_{+}^\ell = v_R^\ell
\,,\quad
v_{-}^\ell = v_L^\ell
\,,\qquad
v_{+}^q = v_R^q
\,,\quad
v_{-}^q = v_L^q
\,.\end{align}
The same decomposition is used for the case of identical flavors, $q\bq\,q\bq\,\ell\bar\ell$. 
For the $W^\mp$ channels, $q \bq\, u \bar d\, \ell \bar \nu$ and $q \bq\, d \bar u\, \nu \bar \ell$, we have
\begin{align} \label{eq:W2_expand}
\vC_{(\lambda_q;W^\mp)}(\dots)
&= \frac{e^2 V_{ud}^{(\dagger)}}{2\sin^2 \theta_W}\, P_W(s_{56})\, \vC_{q (\lambda_q;-;-)}(\ldots)
\,.\end{align}

The coefficients for $q'\bq' q\bq\,V$ are given by
\begin{widetext}
\begin{align} \label{eq:qqQQV_coeffs}
\vC_{x(+;+;+)}(\lp_1,\lp_2;\lp_3,\lp_4;\lp_5, \lp_6)
&= \begin{pmatrix}
  A_{x,\fin}(1_{q'}^+, 2_{\bq'}^-; 3_q^+, 4_\bq^-; 5_\ell^+, 6_{\bar \ell}^-) \\
  \tfrac{1}{N} B_{x,\fin}(1_{q'}^+, 2_{\bq'}^-; 3_q^+, 4_\bq^-; 5_\ell^+, 6_{\bar \ell}^-)
\end{pmatrix}
,\nn\\[1ex]
\vC_{x(+;-;+)}(\lp_1,\lp_2;\lp_3,\lp_4;\lp_5, \lp_6)
&= \begin{pmatrix}
  A_{x,\fin}(1_{q'}^+, 2_{\bq'}^-; 3_q^-, 4_\bq^+; 5_\ell^+, 6_{\bar \ell}^-) \\
   \tfrac{1}{N} B_{x,\fin}(1_{q'}^+, 2_{\bq'}^-; 3_q^-, 4_\bq^+; 5_\ell^+, 6_{\bar \ell}^-)
\end{pmatrix}
,\nn\\
\vC_{x(+;\pm;-)}(\lp_1,\lp_2;\lp_3,\lp_4;\lp_5,\lp_6) &= \vC_{x(+;\pm;+)}(\lp_1,\lp_2;\lp_3,\lp_4;\lp_6,\lp_5)
\,,\nn\\
\vC_{q,v(-;+;\pm)}(\lp_1,\lp_2;\lp_3,\lp_4;\lp_5,\lp_6) &= -\vC_{q,v(+;-;\pm)}(\lp_2,\lp_1;\lp_4,\lp_3;\lp_5,\lp_6)
\,,\nn\\
\vC_{a(-;+;\pm)}(\lp_1,\lp_2;\lp_3,\lp_4;\lp_5,\lp_6) &= \vC_{a(+;-;\pm)}(\lp_2,\lp_1;\lp_4,\lp_3;\lp_5,\lp_6)
\,,\nn\\
\vC_{q,v(-;-;\pm)}(\lp_1,\lp_2;\lp_3,\lp_4;\lp_5,\lp_6) &= -\vC_{q,v(+;+;\pm)}(\lp_2,\lp_1;\lp_4,\lp_3;\lp_5,\lp_6)
\,,\nn\\
\vC_{a(-;-;\pm)}(\lp_1,\lp_2;\lp_3,\lp_4;\lp_5,\lp_6) &= \vC_{a(+;+;\pm)}(\lp_2,\lp_1;\lp_4,\lp_3;\lp_5,\lp_6)
\,,\end{align}
and for $q\bq\,q\bq\, V$ they are given in terms of the amplitudes $A_{x,{\rm fin}}$ and $B_{x,{\rm fin}}$ for $q'\bq' q\bq\,V$ by
\begin{align} \label{eq:qqqqV_coeffs}
\vC_{x(++;+)}(\lp_1,\lp_2;\lp_3,\lp_4;\lp_5,\lp_6)
&= \begin{pmatrix}
  A_{x,\fin}(1_{q}^+,2_{\bq}^-; 3_q^+, 4_\bq^-; 5_\ell^+, 6_{\bar \ell}^-)
  -\frac{1}{N} B_{x,\fin}(1_{q}^+, 4_{\bq}^-; 3_q^+,2_\bq^-  ; 5_\ell^+, 6_{\bar \ell}^-) \\
  \frac{1}{N} B_{x,\fin}(1_{q}^+, 2_{\bq}^-; 3_q^+, 4_\bq^-; 5_\ell^+, 6_{\bar \ell}^-)
  - A_{x,\fin}(1_{q}^+, 4_{\bq}^-; 3_q^+,2_\bq^-  ; 5_\ell^+, 6_{\bar \ell}^-)
\end{pmatrix}
, \nn \\[1ex]
\vC_{x(+-;+)}(\lp_1,\lp_2;\lp_3,\lp_4;\lp_5,\lp_6)
&= \begin{pmatrix}
  A_{x,\fin}(1_{q}^+,2_{\bq}^-; 3_q^-, 4_\bq^+; 5_\ell^+, 6_{\bar \ell}^-) \\
  \tfrac{1}{N} B_{x,\fin}(1_{q}^+,2_{\bq}^-; 3_q^-, 4_\bq^+; 5_\ell^+, 6_{\bar \ell}^-)
\end{pmatrix}
,\nn \\
\vC_{x(+\pm;-)}(\lp_1,\lp_2;\lp_3,\lp_4;\lp_5,\lp_6) &= \vC_{x(+\pm;+)}(\lp_1,\lp_2;\lp_3,\lp_4;\lp_6,\lp_5)
\,, \nn \\
\vC_{q,v(--;\pm)}(\lp_1,\lp_2;\lp_3,\lp_4;\lp_5,\lp_6) &= -\vC_{q,v(++;\pm)}(\lp_2,\lp_1;\lp_4,\lp_3;\lp_5,\lp_6)
\,, \nn \\
\vC_{a(--;\pm)}(\lp_1,\lp_2;\lp_3,\lp_4;\lp_5,\lp_6) &= \vC_{a(++;\pm)}(\lp_2,\lp_1;\lp_4,\lp_3;\lp_5,\lp_6)
\,.\end{align}
\end{widetext}
The various relations for the coefficients with flipped helicities follow from \eq{lep_reduce} and charge conjugation invariance. The tree-level helicity amplitudes are given in \app{V2qqqq}.

\subsubsection{$gg\, q\bq\, V$}

For $gg\, q\bq\, \ell \bar \ell$, the helicity operator basis consists of $12$ independent operators,
\begin{align}\label{eq:Z2_basis_g}
O_{++(+;\pm)}^{ab\, \balpha\bt}
&= \frac{1}{2}\, \cB_{1+}^a\, \cB_{2+}^b\, J_{q\, 34+}^{\balpha\bt}\, J_{\ell \, 56\pm}
\,,\nn\\
O_{++(-;\pm)}^{ab\, \balpha\bt}
&= \frac{1}{2}\, \cB_{1+}^a\, \cB_{2+}^b\, J_{q\, 34-}^{\balpha\bt}\, J_{\ell \, 56\pm}
\,,\nn\\
O_{+-(+;\pm)}^{ab\, \balpha\bt}
&= \cB_{1+}^a\, \cB_{2-}^b\, J_{q\, 34+}^{\balpha\bt}\, J_{\ell \, 56\pm}
\,,\nn\\
O_{+-(-;\pm)}^{ab\, \balpha\bt}
&= \cB_{1+}^a\, \cB_{2-}^b\, J_{q\, 34-}^{\balpha\bt}\, J_{\ell \, 56\pm}
\,,\nn\\
O_{--(+;\pm)}^{ab\, \balpha\bt}
&= \frac{1}{2}\, \cB_{1-}^a\, \cB_{2-}^b\, J_{q\, 34+}^{\balpha\bt}\, J_{\ell \, 56\pm}
\,,\nn\\
O_{--(-;\pm)}^{ab\, \balpha\bt}
&= \frac{1}{2}\, \cB_{1-}^a\, \cB_{2-}^b\, J_{q\, 34-}^{\balpha\bt}\, J_{\ell \, 56\pm}
\,.\end{align}
For $W^\mp$, the corresponding partonic processes are $gg\, u \bar d\, \ell \bar \nu$ and $gg\, d \bar u\, \nu \bar \ell$,
and the helicity operator basis reduces to six independent operators,
\begin{align} \label{eq:W2_basis_g}
O_{++\, (W^-)}^{ab\,\balpha\bt}
&= \frac{1}{2}\, \cB_{1+}^a\, \cB_{2+}^b\, J_{\bar u d\, 34-}^{\balpha\bt}\, J_{\bar\ell \nu\, 56-}
\,,\nn\\
O_{+-\, (W^-)}^{ab\,\balpha\bt}
&= \cB_{1+}^a\, \cB_{2-}^b\, J_{\bar u d\, 34-}^{\balpha\bt}\, J_{\bar\ell \nu\, 56-}
\,,\nn\\
O_{--\, (W^-)}^{ab\,\balpha\bt}
&= \frac{1}{2}\, \cB_{1-}^a\, \cB_{2-}^b\, J_{\bar u d\, 34-}^{\balpha\bt}\, J_{\bar\ell \nu\, 56-}
\,,\nn\\
O_{++\, (W^+)}^{ab\,\balpha\bt}
&= \frac{1}{2}\, \cB_{1+}^a\, \cB_{2+}^b\, J_{\bar d u\, 34-}^{\balpha\bt}\, J_{\bar\nu \ell \, 56-}
\,,\nn\\
O_{+-\, (W^+)}^{ab\,\balpha\bt}
&= \cB_{1+}^a\, \cB_{2-}^b\, J_{\bar d u\, 34-}^{\balpha\bt}\, J_{\bar\nu \ell \, 56-}
\,,\nn\\
O_{--\, (W^+)}^{ab\,\balpha\bt}
&= \frac{1}{2}\, \cB_{1-}^a\, \cB_{2-}^b\, J_{\bar d u\, 34-}^{\balpha\bt}\, J_{\bar\nu \ell \, 56-}
\,.\end{align}

We use the color basis
\begin{equation} \label{eq:ggqqV_color}
\vT^{ ab \alpha\bbeta}
= \Bigl(
   (T^a T^b)_{\alpha\bbeta}\,,\, (T^b T^a)_{\alpha\bbeta} \,,\, \tr[T^a T^b]\, \delta_{\alpha\bbeta}
   \Bigr)
\,,\end{equation}
and the amplitudes are color-decomposed as
\begin{align}
&\cA_x(g_1g_2 q_{3}\bar q_{4} \ell_{5} \bar \ell_{6})
\nn \\ & \quad
= \img \sum_{\sigma\in S_2} \bigl[T^{a_{\sigma(1)}} T^{a_{\sigma(2)}}\bigr]_{\alpha_3\balpha_4}
A_x(\sigma(1),\sigma(2);3_q,4_{\bar q};5_\ell,6_{\bar \ell})
\nn \\ & \qquad
+ \img\, \tr[T^{a_1} T^{a_2}]\,\delta_{\alpha_3\balpha_4}\, B_x(1,2;3_q,4_{\bar q};5_\ell,6_{\bar \ell})
\,.\end{align}

Writing the Wilson coefficients in the decomposition in \eqs{Z_expand}{W_expand}, we have for the $gg\, q\bq\, \ell \bar \ell$ channel
\begin{align}
&\vC_{\lambda_1\lambda_2(\lambda_q;\lambda_\ell)}(\lp_1, \lp_2; \lp_3, \lp_4; \lp_5, \lp_6)
\nn \\ & \quad
= e^2  \,
\biggl\{ \bigl[Q^\ell Q^q + v_{\lambda_\ell}^\ell v_{\lambda_q}^q P_Z(s_{56})\bigr]
\vC_{q \lambda_1\lambda_2(\lambda_q;\lambda_\ell)}(\ldots)
\nn \\ & \qquad
+ \sum_{i=1}^{n_f} \Bigl[Q^\ell Q^i + v_{\lambda_\ell}^\ell\frac{v_L^i + v_R^i}{2} P_Z(s_{56}) \Bigr]
\vC_{v\lambda_1\lambda_2(\lambda_q;\lambda_\ell)}(\ldots)
\nn \\ & \qquad
+ \frac{v_{\lambda_\ell}^\ell}{\sin(2\theta_W)} P_Z(s_{56})\, \vC_{a\lambda_1\lambda_2(\lambda_q;\lambda_\ell)}(\ldots)
\biggr\}
\,,\end{align}
with the weak couplings
\begin{align}
v_{+}^\ell = v_R^\ell
\,,\quad
v_{-}^\ell = v_L^\ell
\,,\qquad
v_{+}^q = v_R^q
\,,\quad
v_{-}^q = v_L^q
\,.\end{align}
For the $W^\mp$ channels $gg\, u \bar d\, \ell \bar \nu$ and $gg\, d \bar u\, \nu \bar \ell$, we have
\begin{align}
\vC_{\lambda_1\lambda_2(W^\mp)}(\dots)
&= \frac{e^2 V_{ud}^{(\dagger)}}{2\sin^2 \theta_W}\, P_W(s_{56})\, \vC_{q\lambda_1\lambda_2(-;-)}(\ldots)
\,.\end{align}

The coefficients for $gg\,q\bq\,V$ are then given by
\begin{align} \label{eq:ggqqV_coeffs}
&\vC_{x\la_1\la_2(+;+)}(\lp_1,\lp_2;\lp_3,\lp_4;\lp_5,\lp_6)
\nn \\ & \qquad
=
\begin{pmatrix}
   A_{x,\fin}(1^{\la_1},2^{\la_2};3_q^+,4_\bq^-; 5_\ell^+, 6_{\bar\ell}^-) \\
   A_{x,\fin}(2^{\la_2},1^{\la_1};3_q^+,4_\bq^-; 5_\ell^+, 6_{\bar\ell}^-) \\
   B_{x,\fin}(1^{\la_1},2^{\la_2};3_q^+,4_\bq^-; 5_\ell^+, 6_{\bar\ell}^-) \\
\end{pmatrix}
,\nn\\
&\vC_{x\la_1\la_2(+;-)}(\lp_1,\lp_2;\lp_3,\lp_4;\lp_5,\lp_6)
\nn \\ & \qquad
= \vC_{x\la_1\la_2(+;+)}(\lp_1,\lp_2;\lp_3,\lp_4;\lp_6,\lp_5)
\,.\end{align}
The remaining Wilson coefficients are obtained by charge conjugation invariance as follows,
\begin{align}
&\vC_{q,v\la_1\la_2(-;\pm)}(\lp_1,\lp_2;\lp_3,\lp_4;\lp_5,\lp_6)
\nn \\ & \qquad
= \hV \vC_{q,v\la_1\la_2(+;\pm)}(\lp_1,\lp_2;\lp_4,\lp_3;\lp_5,\lp_6)
\,,\nn\\
&\vC_{a\la_1\la_2(-;\pm)}(\lp_1,\lp_2;\lp_3,\lp_4;\lp_5,\lp_6)
\nn \\ & \qquad
= - \hV \vC_{a\la_1\la_2(+;\pm)}(\lp_1,\lp_2;\lp_4,\lp_3;\lp_5,\lp_6)\,,
\nn\\
&\text{with}\qquad
\hV =
\begin{pmatrix}
  0 & 1 & 0 \\
  1 & 0 & 0 \\
  0 & 0 & 1
\end{pmatrix}
.\end{align}
The tree-level helicity amplitudes are given in \app{V2ggqq}.

\section{\boldmath $pp \to$ Jets}
\label{sec:pp}

In this section, we give the operator basis and matching relations for $pp \to$ $2, 3$ jets. We consider only the QCD contributions, so that quarks only appear in same-flavor quark-antiquark pairs with the same chirality, and so are described by the currents $J_{ij\pm}$. The helicity amplitudes for each channel are given in \app{helicityamplitudes}.

\subsection{\boldmath $pp \to 2$ Jets}
\label{subsec:pp2jets}

For $pp\to 2$ jets, the partonic channels $q\bq\, q'\bq'$, $q\bq\, q\bq$, $q\bar q gg$, and $gggg$ contribute. We will discuss each in turn. The one-loop helicity amplitudes for all partonic channels were first calculated in Ref.~\cite{Kunszt:1993sd}. The tree-level and one-loop results are given in \app{pp2jets_app}. The two-loop amplitudes have also been calculated, and can be found in Refs.~\cite{Bern:2003ck, Glover:2003cm} for $q\bar q gg$, Refs.~\cite{Anastasiou:2000kg, Anastasiou:2000ue, Glover:2004si, Freitas:2004tk} for $q\bq\, q'\bq'$, $q\bq\, q\bq$ and in Refs.~\cite{Glover:2001af, Bern:2002tk} for $gggg$.

\subsubsection{$q\bq\, q'\bq'$ and $q\bq\, q\bq$}
\label{subsec:qqbarqqbarbasis}

In the case of distinct quark flavors, $q\bq\, q'\bq'$, the helicity basis consists of four independent operators,
\begin{align} \label{eq:qqQQ_basis}
O_{(+;+)}^{\balpha\bt\bgamma\delta}
&= J_{q\, 12+}^{\balpha\bt}\, J_{q'\, 34+}^{\bgamma\delta}
\,,\nn\\
O_{(+;-)}^{\balpha\bt\bgamma\delta}
&= J_{q\, 12+}^{\balpha\bt}\, J_{q'\, 34-}^{\bgamma\delta}
\,,\nn\\
O_{(-;+)}^{\balpha\bt\bgamma\delta}
&= J_{q\, 12-}^{\balpha\bt}\, J_{q'\, 34+}^{\bgamma\delta}
\,,\nn\\
O_{(-;-)}^{\balpha\bt\bgamma\delta}
&= J_{q\, 12-}^{\balpha\bt}\, J_{q'\, 34-}^{\bgamma\delta}
\,.\end{align}
For identical quark flavors, $q\bq\,q\bq$, the helicity basis only has three independent operators,
\begin{align} \label{eq:qqqq_basis}
O_{(++)}^{\balpha\bt\bgamma\delta}
&= \frac{1}{4}\, J_{12+}^{\balpha\bt}\, J_{34+}^{\bgamma\delta}
\,, \nn \\
O_{(+-)}^{\balpha\bt\bgamma\delta}
&= J_{12+}^{\balpha\bt}\, J_{34-}^{\bgamma\delta}
\,, \nn \\
O_{(--)}^{\balpha\bt\bgamma\delta}
&= \frac{1}{4}\, J_{12-}^{\balpha\bt}\, J_{34-}^{\bgamma\delta}
\,.\end{align}
Here we have not made the flavor label explicit, since both quark currents have the same flavor. In both cases we use the color basis
\begin{equation} \label{eq:qqqq_color}
\vT^{ \al\bbeta\ga\bdelta} =
2T_F \Bigl(
  \de_{\al\bdelta}\, \de_{\ga\bbeta}\,,\, \delta_{\al\bbeta}\, \de_{\ga\bdelta}
\Bigr)
\,.\end{equation}

The QCD helicity amplitudes for $q\bq\,q'\bq'$ can be color-decomposed in the basis of \eq{qqqq_color} as
\begin{align} \label{eq:qqQQ_QCD}
\cA(q_{1} \bq_{2} q_3' \bq_4')
&= 2T_F \img \Bigl[
\de_{\al_1\balpha_4} \de_{\al_3\balpha_2}  A(1_q,2_\bq;3_{q'},4_{\bq'}
\\\nn & \quad
+ \frac{1}{N}\,\de_{\al_1 \balpha_2} \de_{\al_3 \balpha_4} B(1_q,2_\bq;3_{q'},4_{\bq'}) \Bigr]
\,,\end{align}
where we have included a factor of $1/N$ for convenience. The amplitude vanishes in the case that the quark and antiquark of the same flavor have the same helicity. This is equivalent to the fact that the operators of \eq{qqQQ_basis} provide a complete basis of helicity operators. For identical quark flavors, the QCD amplitudes can be written in terms of the amplitudes for the distinct flavor case as
\begin{align} 
\cA(q_{1} \bq_{2} q_3 \bq_4) = \cA(q_{1} \bq_{2} q_3' \bq_4')-\cA(q_{1} \bq_4' q_3' \bq_2)
\,.\end{align}
The Wilson coefficients for $q\bq\,q'\bq'$ are then given by
\begin{align} \label{eq:qqQQ_coeffs}
\vC_{(+;+)}(\lp_1,\lp_2;\lp_3,\lp_4)
&= \begin{pmatrix}
  A_\fin(1_q^+,2_\bq^-; 3_{q'}^+, 4_{\bq'}^-) \\
\tfrac{1}{N} B_\fin(1_q^+,2_\bq^-; 3_{q'}^+, 4_{\bq'}^-)
\end{pmatrix}
,\nn\\[1ex]
\vC_{(+;-)}(\lp_1,\lp_2;\lp_3,\lp_4)
&= \begin{pmatrix}
  A_\fin(1_q^+,2_\bq^-; 3_{q'}^-, 4_{\bq'}^+) \\
  \tfrac{1}{N} B_\fin(1_q^+,2_\bq^-; 3_{q'}^-, 4_{\bq'}^+)
\end{pmatrix}
,\nn\\
\vC_{(-;+)}(\lp_1,\lp_2;\lp_3,\lp_4) &= \vC_{(+;-)}(\lp_2,\lp_1;\lp_4,\lp_3)
\,,\nn\\
\vC_{(-;-)}(\lp_1,\lp_2;\lp_3,\lp_4) &= \vC_{(+;+)}(\lp_2,\lp_1;\lp_4,\lp_3)
\,,\end{align}
and for $q\bq\,q\bq$ they are given in terms of the amplitudes $A_{{\rm fin}}$ and $B_{{\rm fin}}$ for $q\bq\,q'\bq'$ by
\begin{widetext}
\begin{align} \label{eq:qqqq_coeffs}
\vC_{(++)}(\lp_1,\lp_2;\lp_3,\lp_4)
&= \begin{pmatrix}
  A_\fin(1_q^+,2_\bq^-; 3_{q}^+, 4_{\bq}^-)
  -\frac{1}{N} B_\fin(1_q^+, 4_{\bq}^-; 3_{q}^+, 2_\bq^-) \\
  \frac{1}{N} B_\fin(1_q^+,2_\bq^-; 3_{q}^+, 4_{\bq}^-)-A_\fin(1_q^+,4_{\bq}^-; 3_{q}^+, 2_\bq^-)
\end{pmatrix}
, \nn \\[1ex]
\vC_{(+-)}(\lp_1,\lp_2;\lp_3,\lp_4)
&= \begin{pmatrix}
  A_\fin(1_q^+,2_\bq^-; 3_{q}^-, 4_{\bq}^+) \\
  \tfrac{1}{N} B_\fin(1_q^+,2_\bq^-; 3_{q}^-, 4_{\bq}^+)
\end{pmatrix}
,\nn \\
\vC_{(--)}(\lp_1,\lp_2;\lp_3,\lp_4) &= \vC_{(++)}(\lp_2,\lp_1;\lp_4,\lp_3)
\,.\end{align}
\end{widetext}
The relations for $\vC_{(-;\pm)}$ and $\vC_{(--)}$ follow from charge conjugation invariance.
The Wilson coefficient $\vC_{(+-)}$ is equal to $\vC_{(+;-)}$, since the amplitude vanishes when the quark and antiquark of the same flavor have the same helicity (both $+$ or both $-$), so there is no exchange term. The subscript ``$\fin$'' in \eqs{qqQQ_coeffs}{qqqq_coeffs} denotes the IR-finite part of the helicity amplitudes as discussed in \sec{matching}, see \eq{matching_general}. Recall that the symmetry factors of $1/4$ in \eq{qqqq_basis} already take care of the interchange of identical (anti)quarks, so there are no additional symmetry factors needed for $\vC_{(++)}$. Explicit expressions for all required partial amplitudes at tree level and one loop are given in \app{qqqqamplitudes}.

\subsubsection{$gg q\bar q$}
\label{subsec:ggqqbarbasis}

For $gg q\bar q$, the helicity basis has a total of six independent operators,
\begin{align} \label{eq:ggqq_basis}
O_{++(+)}^{ab\, \balpha\beta}
&= \frac{1}{2}\, \cB_{1+}^a\, \cB_{2+}^b\, J_{34+}^{\balpha\beta}
\,,\nn\\
O_{+-(+)}^{ab\, \balpha\beta}
&= \cB_{1+}^a\, \cB_{2-}^b\, J_{34+}^{\balpha\beta}
\,,\nn\\
O_{--(+)}^{ab\, \balpha\beta}
&= \frac{1}{2} \cB_{1-}^a\, \cB_{2-}^b\, J_{34+}^{\balpha\beta}
\,,\nn\\
O_{++(-)}^{ab\, \balpha\beta}
&= \frac{1}{2}\, \cB_{1+}^a\, \cB_{2+}^b\, J_{34-}^{\balpha\beta}
\,,\nn\\
O_{+-(-)}^{ab\, \balpha\beta}
&= \cB_{1+}^a\, \cB_{2-}^b\, J_{34-}^{\balpha\beta}
\,,\nn\\
O_{--(-)}^{ab\, \balpha\beta}
&= \frac{1}{2} \cB_{1-}^a\, \cB_{2-}^b\, J_{34-}^{\balpha\beta}
\,.\end{align}
Note that the use of a helicity basis has made it easy to count the number of required operators.~\footnote{This should be contrasted with the more complicated basis given in Eq.~(126) of Ref.~\cite{Marcantonini:2008qn} which is built from fields $\chi_{n_i}$ and $\cB_{n_i}^{\perp\mu}$ and standard Dirac structures. It can be reduced to a minimal basis using identities such as $O_2 = - O_1$, $O_8 = O_7 +4 t O_3 - 4 t O_4$ and $O_6=O_5-2O_1+{\cal O}(\epsilon)$ where $t =-\w_1 \w_3 n_1 \sdt n_3/2$, and then can be related to the basis used here.}
For the color structure, we use the basis
\begin{equation} \label{eq:ggqq_color}
\vT^{ ab \alpha\bbeta}
= \Bigl(
   (T^a T^b)_{\alpha\bbeta}\,,\, (T^b T^a)_{\alpha\bbeta} \,,\, \tr[T^a T^b]\, \delta_{\alpha\bbeta}
   \Bigr)
\,.\end{equation}

The color decomposition of the QCD helicity amplitudes into partial amplitudes using the color basis of \eq{ggqq_color} is
\begin{align} \label{eq:ggqq_QCD}
&\cA\bigl(g_1 g_2\, q_{3} \bq_{4} \bigr)
\nn\\ & \quad
= \img \sum_{\sigma\in S_2} \bigl[T^{a_{\sigma(1)}} T^{a_{\sigma(2)}}\bigr]_{\alpha_3\balpha_4}
\,A(\sigma(1),\sigma(2); 3_q, 4_\bq)
\nn\\ & \qquad
+ \img\, \tr[T^{a_1} T^{a_2}]\,\delta_{\alpha_3\balpha_4}\, B(1,2; 3_q, 4_\bq)
\,,\end{align}
from which we can read off the Wilson coefficients,
\begin{align} \label{eq:ggqq_coeffs}
\vC_{+-(+)}(\lp_1,\lp_2;\lp_3,\lp_4) &=
\begin{pmatrix}
   A_\fin(1^+,2^-;3_q^+,4_\bq^-) \\
   A_\fin(2^-,1^+;3_q^+,4_\bq^-) \\
   B_\fin(1^+,2^-;3_q^+,4_\bq^-) \\
\end{pmatrix}
,\nn\\
\vC_{++(+)}(\lp_1,\lp_2;\lp_3,\lp_4)
&= \begin{pmatrix}
   A_\fin(1^+,2^+;3_q^+,4_\bq^-) \\
   A_\fin(2^+,1^+;3_q^+,4_\bq^-) \\
   B_\fin(1^+,2^+;3_q^+,4_\bq^-) \\
\end{pmatrix}
,\nn\\
\vC_{--(+)}(\lp_1,\lp_2;\lp_3,\lp_4)
&= \begin{pmatrix}
   A_\fin(1^-,2^-;3_q^+,4_\bq^-) \\
   A_\fin(2^-,1^-;3_q^+,4_\bq^-) \\
   B_\fin(1^-,2^-;3_q^+,4_\bq^-) \\
\end{pmatrix}
.\end{align}
The remaining coefficients follow from charge conjugation as discussed in \subsec{discrete},
\begin{align}\label{eq:ggqq_charge_matrix}
\vC_{\la_1\la_2(-)}(\lp_1,\lp_2;\lp_3,\lp_4)
&= \hV \vC_{\la_1\la_2(+)}(\lp_1,\lp_2;\lp_4,\lp_3)\,,
\nn\\
\text{with}\qquad
\hV & =
\begin{pmatrix}
  0 & -1 & 0 \\
  -1 & 0 & 0 \\
  0 & 0 & -1
\end{pmatrix}
\,.\end{align}
At tree level, the partial amplitudes are well known, and only the first two entries in $\vC_{+-(\pm)}$ are nonzero. Explicit expressions for all amplitudes at tree level and one loop are given in \app{ggqqamplitudes}.

\subsubsection{$gggg$}
\label{subsec:ggggbasis}

For $gggg$, the helicity basis has five independent operators,
\begin{align} \label{eq:gggg_basis}
O_{++++}^{abcd} &= \frac{1}{4!}\, \cB_{1+}^a \cB_{2+}^b \cB_{3+}^c \cB_{4+}^d
\,,\nn\\
O_{+++-}^{abcd} &= \frac{1}{3!}\, \cB_{1+}^a \cB_{2+}^b \cB_{3+}^c \cB_{4-}^d
\,,\nn\\
O_{++--}^{abcd} &= \frac{1}{4}\, \cB_{1+}^a \cB_{2+}^b \cB_{3-}^c \cB_{4-}^d
\,,\nn\\
O_{---+}^{abcd} &= \frac{1}{3!}\, \cB_{1-}^a \cB_{2-}^b \cB_{3-}^c \cB_{4+}^d
\,,\nn\\
O_{----}^{abcd} &= \frac{1}{4!}\, \cB_{1-}^a \cB_{2-}^b \cB_{3-}^c \cB_{4-}^d
\,.\end{align}
We use the color basis
\begin{equation} \label{eq:gggg_color}
\vT^{abcd} =
\frac{1}{2\cdot 2T_F}\begin{pmatrix}
\tr[abcd] + \tr[dcba] \\ \tr[acdb] + \tr[bdca] \\ \tr[adbc] + \tr[cbda] \\
2\tr[ab] \tr[cd] \\ 2\tr[ac] \tr[db] \\ 2\tr[ad] \tr[bc]
\end{pmatrix}^T
\,,\end{equation}
where we have used the shorthand notation
\begin{equation}
\tr[ab] = \tr[T^a T^b]
\,,\qquad
\tr[abcd] = \tr[T^a T^b T^c T^d]
\,.\end{equation}
Under charge conjugation, the operators transform as
\begin{equation}
\C\, O_{\la_1\la_2\la_3\la_4}^{abcd}\,\vT^{abcd}\, \C = O_{\la_1\la_2\la_3\la_4}^{dcba}\vT^{abcd}
\,.\end{equation}
Thus, charge conjugation invariance of QCD leads to
\begin{equation} \label{eq:gggg_charge}
C_{\la_1\la_2\la_3\la_4}^{abcd} = C_{\la_1\la_2\la_3\la_4}^{dcba}
\,.\end{equation}
In principle, there are three more color structures with a minus sign instead of the plus sign in the first three lines in \eq{gggg_color}. Since charge conjugation is a symmetry of QCD, \eq{gggg_charge} holds to all orders, so these additional color structures cannot contribute. In particular, the color structures in \eq{gggg_color} cannot mix into these additional structures at any order. Hence, it is sufficient to consider the reduced basis in \eq{gggg_color} instead of the 9 different color structures, which were used for example in Ref.~\cite{Kelley:2010fn}. Note that for $N=3$ it is possible to further reduce the color basis by one using the relation
\begin{align}
&\tr[abcd+dcba] + \tr[acdb+bdca] + \tr[adbc+cbda]
\nn\\ & \qquad
= \tr[ab]\tr[cd] + \tr[ac]\tr[db] + \tr[ad]\tr[bc]
\,.\end{align}
We refrain from doing so, since it makes the structure of the anomalous dimension matrix less visible, and because there are no such relations for $N>3$.

The color decomposition of the QCD amplitude into partial amplitudes using the color basis in \eq{gggg_color} is 
\begin{align} \label{eq:gggg_QCD}
\cA(g_1 g_2 g_3 g_4)
&= \frac{\img}{2T_F} \biggl[\sum_{\si \in S_4/Z_4}\! \tr[a_{\si(1)} a_{\si(2)} a_{\si(3)} a_{\si(4)}]
 \nn \\ 
& \qquad \times
A\bigl(\si(1),\si(2),\si(3),\si(4)\bigr)
\nn\\ 
&\quad
+ \sum_{\si \in S_4/Z_2^3}\! \tr[a_{\si(1)} a_{\si(2)}] \tr[a_{\si(3)} a_{\si(4)}]
\nn\\ 
& \qquad \times
B\bigl(\si(1),\si(2),\si(3),\si(4)\bigr) \biggr]
\,,
\end{align}
from which we obtain the Wilson coefficients
\begin{align} \label{eq:gggg_coeffs}
\vC_{++--}(\lp_1,\lp_2,\lp_3,\lp_4)
&= \begin{pmatrix}
  2A_\fin(1^+,2^+,3^-,4^-) \\
  2A_\fin(1^+,3^-,4^-,2^+) \\
  2A_\fin(1^+,4^-,2^+,3^-) \\
  B_\fin(1^+,2^+,3^-,4^-) \\
  B_\fin(1^+,3^-,4^-,2^+) \\
  B_\fin(1^+,4^-,2^+,3^-)
\end{pmatrix}
,\nn\\
\vC_{+++-}(\lp_1,\lp_2,\lp_3,\lp_4)
&= \begin{pmatrix}
  2A_\fin(1^+,2^+,3^+,4^-) \\
  2A_\fin(1^+,3^+,4^-,2^+) \\
  2A_\fin(1^+,4^-,2^+,3^+) \\
  B_\fin(1^+,2^+,3^+,4^-) \\
  B_\fin(1^+,3^+,4^-,2^+) \\
  B_\fin(1^+,4^-,2^+,3^+) \\
\end{pmatrix}
,\nn\\
\vC_{++++}(\lp_1,\lp_2,\lp_3,\lp_4)
&= \begin{pmatrix}
  2A_\fin(1^+,2^+,3^+,4^+) \\
  2A_\fin(1^+,3^+,4^+,2^+) \\
  2A_\fin(1^+,4^+,2^+,3^+) \\
  B_\fin(1^+,2^+,3^+,4^+) \\
  B_\fin(1^+,3^+,4^+,2^+) \\
  B_\fin(1^+,4^+,2^+,3^+) \\
\end{pmatrix}
,\nn\\
\vC_{---+}(\lp_1, \lp_2, \lp_3, \lp_4)
&= \vC_{+++-}(\lp_1, \lp_2, \lp_3, \lp_4)\Big|_{\langle..\rangle \leftrightarrow [..]}
\,,\nn\\
\vC_{----}(\lp_1, \lp_2, \lp_3, \lp_4)
&= \vC_{++++}(\lp_1, \lp_2, \lp_3, \lp_4)\Big|_{\langle..\rangle \leftrightarrow [..]}
\,.\end{align}
The last two coefficients follow from parity invariance. 
The factors of two in the first three entries of the coefficients come from combining the two color structures in the first three entries in \eq{gggg_color} using charge conjugation invariance in~\eq{gggg_charge}. 

The tree-level amplitudes are well known. At tree level, only the $A$ amplitudes with two positive and two negative helicity gluons are nonzero. Because the $A$ amplitudes correspond to a single-trace color structure, which possesses a cyclic symmetry, the corresponding partial amplitudes are invariant under the corresponding cyclic permutations of their arguments. Explicit expressions for the required amplitudes at tree level and one loop are given in \app{ggggamplitudes}.

\subsection{\boldmath $pp \to 3$ Jets}

The four partonic channels $g\,q\bq\, q'\bq'$, $g\, q\bq\, q\bq$, $ggg\, q\bar q$, and $ggggg$ contribute to $pp\to 3$ jets, which we discuss in turn. The one-loop partial amplitudes for the different partonic channels were calculated in Refs.~\cite{Kunszt:1994tq, Bern:1994fz, Bern:1993mq}. Tree-level results for the helicity amplitudes for each partonic process are given in \app{pp3jets_app}.

\subsubsection{$g\, q\bq\, q'\bq'$ and $g\, q\bq\, q\bq$}
\label{subsec:gqqbarqqbarbasis}

For the case of distinct quark flavors, $g\, q\bq\, q'\bq'$, the helicity basis consists of eight independent operators,
\begin{align} \label{eq:gqqQQ_basis}
O_{\pm(+;+)}^{a\,\balpha\bt\bgamma\delta}
&=  \cB_{1\pm}^a\, J_{q\, 23+}^{\balpha\bt}\, J_{q'\, 45+}^{\bgamma\delta}
\,,\nn\\
O_{\pm(+;-)}^{a\,\balpha\bt\bgamma\delta}
&=  \cB_{1\pm}^a\, J_{q\, 23+}^{\balpha\bt}\, J_{q'\, 45-}^{\bgamma\delta}
\,,\nn\\
O_{\pm(-;+)}^{a\,\balpha\bt\bgamma\delta}
&=  \cB_{1\pm}^a\, J_{q\, 23-}^{\balpha\bt}\, J_{q'\, 45+}^{\bgamma\delta}
\,,\nn\\
O_{\pm(-;-)}^{a\,\balpha\bt\bgamma\delta}
&=  \cB_{1\pm}^a\, J_{q\, 23-}^{\balpha\bt}\, J_{q'\, 45-}^{\bgamma\delta}
\,.\end{align}
For identical quark flavors, $g\, q\bq\,q\bq$, the basis reduces to six independent helicity operators,
\begin{align} \label{eq:gqqqq_basis}
O_{\pm(++)}^{a\,\balpha\bt\bgamma\delta}
&= \frac{1}{4}\, \cB_{1\pm}^a\, J_{23+}^{\balpha\bt}\, J_{45+}^{\bgamma\delta}
\,,\nn\\
O_{\pm(+-)}^{a\,\balpha\bt\bgamma\delta}
&= \cB_{1\pm}^a\, J_{23+}^{\balpha\bt}\, J_{45-}^{\bgamma\delta}
\,,\nn\\
O_{\pm(--)}^{a\,\balpha\bt\bgamma\delta}
&= \frac{1}{4}\, \cB_{1\pm}^a\, J_{23-}^{\balpha\bt}\, J_{45-}^{\bgamma\delta}
\,.\end{align}
In both cases we use the color basis
\begin{equation} \label{eq:gqqqq_color}
\vT^{a\,\al\bbeta\ga\bdelta} =
2T_F\Bigl(
  T^a_{\al\bdelta}\, \de_{\ga\bbeta},\, T^a_{\ga\bbeta} \, \de_{\al\bdelta},\,
  T^a_{\al\bbeta}\, \de_{\ga\bdelta},\, T^a_{\ga\bdelta} \, \de_{\al\bbeta}
\Bigr)
.\end{equation}

The QCD helicity amplitudes for $g\,q\bq\,q'\bq'$ can be color decomposed into partial amplitudes in the color basis of \eq{gqqqq_color} as
\begin{widetext}
\begin{align} \label{eq:gqqQQ_QCD}
\cA(g_{1} q_{2} \bq_{3} q_4' \bq_5')
&=  2T_F\img \Bigl[T^{a_1}_{\al_2 \balpha_5} \de_{\al_4\balpha_3} A(1;2_q,3_\bq;4_{q'},5_{\bq'}) +
T^{a_1}_{\al_4 \balpha_3} \de_{\al_2 \balpha_5} A(1;4_{q'},5_{\bq'};2_q,3_\bq)
\nn \\ &\quad
+\frac{1}{N} T^{a_1}_{\al_2 \balpha_3} \de_{\al_4\balpha_5} B(1;2_q,3_\bq;4_{q'},5_{\bq'})
+\frac{1}{N} T^{a_1}_{\al_4 \balpha_5} \de_{\al_2 \balpha_3} B(1;4_{q'},5_{\bq'};2_q,3_\bq) \Bigr]
\,,\end{align}
where we have used the symmetry $q\bar q \leftrightarrow q'\bar q'$, and inserted the factors of $1/N$ for later convenience.  The amplitude vanishes when the quark and antiquark of the same flavor have the same helicity (both $+$ or both $-$), in accordance with the fact that the operators of \eq{gqqQQ_basis} provide a complete basis of helicity operators. For identical quark flavors, the amplitudes can be written in terms of the amplitudes for the distinct flavor case as
\begin{align}
\cA(g_{1} q_{2} \bq_{3} q_4 \bq_5)
&=\cA(g_{1} q_{2} \bq_{3} q_4' \bq_5')-\cA(g_{1} q_{2} \bq_5' q_4' \bq_{3})\,.
\end{align}
The Wilson coefficients for $g\,q\bq\,q'\bq'$ are then given by
\begin{align} \label{eq:gqqQQ_coeffs}
\vC_{+(+;+)}(\lp_1;\lp_2,\lp_3;\lp_4,\lp_5)
&= \begin{pmatrix}
  A_\fin(1^+;2_q^+,3_\bq^-;4_{q'}^+,5_{\bq'}^-) \\
  A_\fin(1^+;4_{q'}^+,5_{\bq'}^-;2_q^+,3_\bq^-) \\
 \frac{1}{N} B_\fin(1^+;2_q^+,3_\bq^-;4_{q'}^+,5_{\bq'}^-)  \\
  \frac{1}{N} B_\fin(1^+;4_{q'}^+,5_{\bq'}^-;2_q^+,3_\bq^-)
\end{pmatrix}
,\nn\\
\vC_{+(+;-)}(\lp_1;\lp_2,\lp_3;\lp_4,\lp_5)
&= \begin{pmatrix}
  A_\fin(1^+;2_q^+,3_\bq^-;4_{q'}^-,5_{\bq'}^+) \\
  A_\fin(1^+;4_{q'}^-,5_{\bq'}^+;2_q^+,3_\bq^-) \\
 \frac{1}{N} B_\fin(1^+;2_q^+,3_\bq^-;4_{q'}^-,5_{\bq'}^+) \\
 \frac{1}{N} B_\fin(1^+;4_{q'}^-,5_{\bq'}^+;2_q^+,3_\bq^-)
\end{pmatrix}
\,,\end{align}
and for $g\,q\bq\,q\bq$ they are given in terms of the amplitudes $A_\fin$ and $B_\fin$ for $g\,q\bq\,q'\bq'$ by
\begin{align} \label{eq:gqqqq_coeffs}
\vC_{+(++)}(\lp_1;\lp_2,\lp_3;\lp_4,\lp_5)
&= \begin{pmatrix}
  A_\fin(1^+;2_q^+,3_\bq^-;4_{q}^+,5_{\bq}^-)  -\frac{1}{N} B_\fin(1^+;2_q^+,5_{\bq}^-;4_{q}^+,3_\bq^-) \\
  A_\fin(1^+;4_{q}^+,5_{\bq}^-;2_q^+,3_\bq^-)  -\frac{1}{N} B_\fin(1^+;4_{q}^+,3_\bq^-;2_{q}^+,5_{\bq}^-)\\
  \frac{1}{N} B_\fin(1^+;2_q^+,3_\bq^-;4_{q}^+,5_{\bq}^-) -A_\fin(1^+;2_q^+,5_{\bq}^-;4_{q}^+,3_\bq^-)  \\
  \frac{1}{N} B_\fin(1^+;4_{q}^+,5_{\bq}^-;2_q^+,3_\bq^-) - A_\fin(1^+;4_{q}^+,3_\bq^-;2_q^+,5_{\bq}^-)
\end{pmatrix}
,\nn\\
\vC_{+(+-)}(\lp_1;\lp_2,\lp_3;\lp_4,\lp_5)
&= \begin{pmatrix}
  A_\fin(1^+;2_q^+,3_\bq^-;4_{q}^-,5_{\bq}^+) \\
  A_\fin(1^+;4_{q}^-,5_{\bq}^+;2_q^+,3_\bq^-) \\
\frac{1}{N}  B_\fin(1^+;2_q^+,3_\bq^-;4_{q}^-,5_{\bq}^+) \\
\frac{1}{N}  B_\fin(1^+;4_{q}^-,5_{\bq}^+;2_q^+,3_\bq^-)
\end{pmatrix}
.\end{align}
\end{widetext}
Charge conjugation invariance of QCD relates the Wilson coefficients,
\begin{align} \label{eq:gqqqq_charge}
\vC_{\la(-;\pm)}(\lp_1;\lp_2,\lp_3;\lp_4,\lp_5) &= \hV \vC_{\la(+;\mp)}(\lp_1;\lp_3,\lp_2;\lp_5,\lp_4)\,,
\nn\\
\vC_{\la(--)}(\lp_1;\lp_2,\lp_3;\lp_4,\lp_5) &= \hV \vC_{\la(++)}(\lp_1;\lp_3,\lp_2;\lp_5,\lp_4)\,,
\end{align}
with
\begin{align}
\hV &=
\begin{pmatrix}
  0 & -1 & 0 & 0 \\
  -1 & 0 & 0 & 0 \\
  0 & 0 & -1 & 0 \\
  0 & 0 & 0 & -1
\end{pmatrix}
.\end{align}
The remaining Wilson coefficients for a negative helicity gluon follow from parity invariance,
\begin{align}
&\vC_{-(+;\pm)}(\lp_1; \lp_2, \lp_3; \lp_4, \lp_5)
\nn \\ & \qquad
=\vC_{+(-;\mp)}(\lp_1; \lp_2, \lp_3; \lp_4, \lp_5)\Big|_{\langle..\rangle \leftrightarrow [..]}
\,, \nn \\
&\vC_{-(++)}(\lp_1; \lp_2, \lp_3; \lp_4, \lp_5)
\nn \\ & \qquad
=\vC_{+(--)}(\lp_1; \lp_2, \lp_3; \lp_4, \lp_5)\Big|_{\langle..\rangle \leftrightarrow [..]}
\,.\end{align}
Explicit expressions for all required partial amplitudes at tree level are given in \app{gqqqqamplitudes}.

\subsubsection{$ggg\, q\bar q$}
\label{subsec:gggqqbarbasis}

For $ggg\, q\bar q$, we have a basis of eight independent helicity operators,
\begin{align} \label{eq:gggqq_basis}
O_{+++(\pm)}^{abc\, \balpha\beta}
&= \frac{1}{3!}\, \cB_{1+}^a\, \cB_{2+}^b\, \cB_{3+}^c\, J_{45\pm}^{\balpha\beta}
\,,\nn\\
O_{++-(\pm)}^{abc\, \balpha\beta}
&= \frac{1}{2}\, \cB_{1+}^a\, \cB_{2+}^b\, \cB_{3-}^c\, J_{45\pm}^{\balpha\beta}
\,,\nn\\
O_{--+(\pm)}^{abc\, \balpha\beta}
&= \frac{1}{2}\, \cB_{1-}^a\, \cB_{2-}^b\, \cB_{3+}^c\, J_{45\pm}^{\balpha\beta}
\,,\nn\\
O_{---(\pm)}^{abc\, \balpha\beta}
&= \frac{1}{3!} \cB_{1-}^a\, \cB_{2-}^b\, \cB_{3-}^c\, J_{45\pm}^{\balpha\beta}
\,,\end{align}
and we use the color basis
\begin{equation} \label{eq:gggqq_color}
\vT^{abc\,\alpha\bbeta} =
\begin{pmatrix}
  [ T^a T^b T^c]_{\alpha\bbeta} \\  [T^b T^c T^a]_{\alpha\bbeta} \\ [T^c T^a T^b]_{\alpha\bbeta} \\ [T^c T^b T^a]_{\alpha\bbeta} \\ [T^a T^c T^b]_{\alpha\bbeta} \\ [T^b T^a T^c]_{\alpha\bbeta} \\
   \tr[T^c T^a] T^b_{\alpha\bbeta} \\  \tr[T^a T^b] T^c_{\alpha\bbeta} \\ \tr[T^b T^c] T^a_{\alpha\bbeta} \\
   \tr[T^a T^b T^c] \delta_{\alpha\bbeta} \\ \tr[T^c T^b T^a] \delta_{\alpha\bbeta}
\end{pmatrix}^T
\,.\end{equation}

The color decomposition of the QCD helicity amplitudes into partial amplitudes using \eq{gggqq_color} is
\begin{align} \label{eq:gggqq_QCD}
\cA\bigl(g_1 g_2 g_3\, q_{4} \bq_{5} \bigr)
&= \img \!\sum_{\sigma\in S_3}\! \bigl[T^{a_{\sigma(1)}} T^{a_{\sigma(2)}} T^{a_{\sigma(3)}}\bigr]_{\alpha_4\balpha_5}
\nn \\ & \qquad\times
A(\sigma(1),\sigma(2),\sigma(3); 4_q, 5_\bq)
\nn\\ & \quad
+ \img \!\sum_{\sigma\in S_3/Z_2}\!\! \tr\bigl[T^{a_{\sigma(1)}} T^{a_{\sigma(2)}}\bigr] T^{a_{\sigma(3)}}_{\alpha_4\balpha_5}
\nn \\ & \qquad\times
B(\sigma(1),\sigma(2),\sigma(3); 4_q, 5_\bq)
\nn \\& \quad
+ \img \!\sum_{\sigma\in S_3/Z_3}\!\! \tr\bigl[T^{a_{\sigma(1)}} T^{a_{\sigma(2)}} T^{a_{\sigma(3)}}\bigr] \de_{\alpha_4\balpha_5}
\nn \\ & \qquad\times
C(\sigma(1),\sigma(2),\sigma(3); 4_q, 5_\bq)
\,,\end{align}
from which we can read off the Wilson coefficients, 
\begin{align} \label{eq:gggqq_coeffs}
\vC_{++\mp(+)}(\lp_1, \dots; \lp_4, \lp_5)
&= \begin{pmatrix}
   A_\fin(1^+,2^+,3^\mp;4_q^+,5_\bq^-) \\
    A_\fin(2^+,3^\mp,1^+;4_q^+,5_\bq^-) \\
    A_\fin(3^\mp,1^+,2^+;4_q^+,5_\bq^-) \\
    A_\fin(3^\mp,2^+,1^+;4_q^+,5_\bq^-) \\
    A_\fin(1^+,3^\mp,2^+;4_q^+,5_\bq^-) \\
    A_\fin(2^+,1^+,3^\mp;4_q^+,5_\bq^-) \\
    B_\fin(3^\mp,1^+,2^+;4_q^+,5_\bq^-) \\
    B_\fin(1^+,2^+,3^\mp;4_q^+,5_\bq^-) \\
    B_\fin(2^+,3^\mp,1^+;4_q^+,5_\bq^-) \\
    C_\fin(1^+,2^+,3^\mp;4_q^+,5_\bq^-) \\
    C_\fin(3^\mp,2^+,1^+;4_q^+,5_\bq^-)
\end{pmatrix}
.\end{align}
Charge conjugation invariance of QCD relates the coefficients with opposite quark helicities,
\begin{align} \label{eq:gggqq_charge}
&\vC_{\la_1\la_2 \la_3(-)}(\lp_1,\lp_2,\lp_3 ; \lp_4, \lp_5)
= \hV \vC_{\la_1\la_2 \la_3(+)}(\lp_1, \lp_2,\lp_3; \lp_5, \lp_4)\,,
\nn \\
&\text{with}\qquad
\hV =
\begin{pmatrix}
  0_{3\times 3} & 1_{3\times3} &  &  \\
  1_{3\times 3} & 0_{3\times3} &  &  \\
   &  & 1_{3\times3} &  &  \\
   &  &  & 0 & 1 \\
   &  &  & 1 & 0
\end{pmatrix}
,\end{align}
where $1_{n\times n}$ denotes the $n$-dimensional identity matrix and the empty entries are all zero.
The remaining coefficients follow from parity invariance
\begin{align}\label{eq:gggqq_parity}
&\vC_{--+(\pm)}(\lp_1,  \lp_2, \lp_3; \lp_4, \lp_5)
\nn \\& \qquad
=\vC_{++-(\mp)}(\lp_1,  \lp_2,\lp_3; \lp_4, \lp_5)\Big|_{\langle..\rangle \leftrightarrow [..]}
\,,\nn \\
&\vC_{---(\pm)}(\lp_1, \lp_2, \lp_3; \lp_4, \lp_5)
\nn \\ & \qquad
=\vC_{+++(\mp)}(\lp_1, \lp_2,\lp_3; \lp_4, \lp_5)\Big|_{\langle..\rangle \leftrightarrow [..]}
\,.\end{align}
At tree level, the partial amplitudes are well known, and only the $A$ amplitudes are nonzero. Furthermore, the partial amplitudes with all negative or all positive helicity gluons vanish.  Combining the charge and parity relations of \eqs{gggqq_charge}{gggqq_parity}, there are only three independent amplitudes at tree level, which we take to be $A(1^+,2^+,3^-;4_q^+,5_{\bq}^-)$, $A(2^+,3^-,1^+;4_q^+,5_{\bq}^-)$, and $A(3^-,1^+,2^+;4_q^+,5_{\bq}^-)$. These amplitudes are given in \app{gggqqamplitudes}.

\subsubsection{$ggggg$}
\label{subsec:gggggbasis}

For $ggggg$, the basis consists of six independent helicity operators,
\begin{align} \label{eq:ggggg_basis}
O_{+++++}^{abcde} &= \frac{1}{5!}\, \cB_{1+}^a \cB_{2+}^b \cB_{3+}^c \cB_{4+}^d \cB_{5+}^e
\,,\nn\\
O_{++++-}^{abcde} &= \frac{1}{4!}\, \cB_{1+}^a \cB_{2+}^b \cB_{3+}^c \cB_{4+}^d \cB_{5-}^e
\,,\nn\\
O_{+++--}^{abcde} &= \frac{1}{2\cdot 3!}\, \cB_{1+}^a \cB_{2+}^b \cB_{3+}^c \cB_{4-}^d \cB_{5-}^e
\,,\nn\\
O_{---++}^{abcde} &= \frac{1}{2 \cdot 3!}\, \cB_{1-}^a \cB_{2-}^b \cB_{3-}^c \cB_{4+}^d \cB_{5+}^e
\,,\nn\\
O_{----+}^{abcde} &= \frac{1}{4!}\, \cB_{1-}^a \cB_{2-}^b \cB_{3-}^c \cB_{4-}^d \cB_{5+}^e
\,,\nn\\
O_{-----}^{abcde} &= \frac{1}{5!}\, \cB_{1-}^a \cB_{2-}^b \cB_{3-}^c \cB_{4-}^d \cB_{5-}^e
\,.\end{align}
As before, we only need one operator for each number of positive and negative helicities.
We use the color basis
\begin{equation} \label{eq:ggggg_color}
\vT^{ abcde} =
\frac{1}{2\cdot 2T_F}\begin{pmatrix}
\tr[abcde]-\tr[edcba] \\
\tr[acdeb]-\tr[bedca] \\
\tr[acbed]-\tr[debca] \\
\tr[abced]-\tr[decba] \\
\tr[abdec]-\tr[cedba] \\
\tr[acbde]-\tr[edbca] \\
\tr[adceb]-\tr[becda] \\
\tr[adcbe]-\tr[ebcda] \\
\tr[aebdc]-\tr[cdbea] \\
\tr[abdce]-\tr[ecdba] \\
\tr[aecbd]-\tr[dbcea] \\
\tr[acebd]-\tr[dbeca] \\
(\tr[ced]-\tr[dec] )\tr[ab] \\ (\tr[abe]-\tr[eba])\tr[cd] \\ (\tr[acd]-\tr[dca])\tr[be] \\ (\tr[bec]-\tr[ceb])\tr[ad] \\ (\tr[adb]-\tr[bda])\tr[ce] \\ (\tr[ace]-\tr[eca])\tr[bd] \\ (\tr[bdc]-\tr[cdb])\tr[ae] \\ (\tr[aed]-\tr[dea])\tr[bc] \\ (\tr[acb]-\tr[bca])\tr[de] \\ (\tr[bed]-\tr[deb])\tr[ac]
\end{pmatrix}^T
,\end{equation}
where we have used the shorthand notation
\begin{equation}
\tr[ab \cdots cd] = \tr[T^a T^b \cdots T^c T^d]
\,.\end{equation}
A priori, there are twice as many color structures as in \eq{ggggg_color} with a relative plus sign instead of a minus sign between the two traces. Under charge conjugation, the operators transform as
\begin{equation}
\C\, O_{\la_1\la_2\la_3\la_4\la_5}^{abcde}\,\vT^{abcde}\, \C = -O_{\la_1\la_2\la_3\la_4\la_5}^{edcba}(\vT^{abcde})
\,.\end{equation}
Therefore, charge conjugation invariance implies for the Wilson coefficients
\begin{equation}
C_{\la_1\la_2\la_3\la_4\la_5}^{abcde} = - C_{\la_1\la_2\la_3\la_4\la_5}^{edcba}
\,,\end{equation}
and hence these additional color structures cannot appear at any order in perturbation theory, either through matching or renormalization group evolution.

The color decomposition of the QCD amplitude into partial amplitudes using the color basis of \eq{ggggg_color} is
\begin{align} \label{eq:ggggg_QCD}
&\cA(g_1 g_2 g_3 g_4g_5)
\nn \\ & \qquad
= \frac{\img}{2T_F} \biggl[\sum_{\si \in S_5/Z_5}\! \tr[a_{\si(1)} a_{\si(2)} a_{\si(3)} a_{\si(4)} a_{\si(5)}]
\nn \\ & \qquad\qquad \times
A\bigl(\si(1),\si(2),\si(3),\si(4), \si(5)\bigr)
\nn\\ & \qquad\quad
+ \sum_{\si \in S_5/(Z_3 \times Z_2)} \tr[a_{\si(1)} a_{\si(2)} a_{\si(3)}]\, \tr[a_{\si(4)} a_{\si(5)}]
\nn\\ & \qquad\qquad\times
B\bigl(\si(1),\si(2),\si(3),\si(4),\si(5)\bigr) \biggr]
\,,\end{align}
from which we obtain the Wilson coefficients 
\begin{align} \label{eq:ggggg_coeffs}
\vC_{+++--}(\lp_1, \dots ,\lp_5)
&= 2\begin{pmatrix}
 A_\fin(1^+,2^+,3^+,4^-,5^-)  \\  A_\fin(1^+,3^+,4^-,5^-,2^+)  \\  A_\fin(1^+,3^+,2^+,5^-,4^-)  \\
 A_\fin(1^+,2^+,3^+,5^-,4^-)  \\  A_\fin(1^+,2^+,4^-,5^-,3^+)  \\  A_\fin(1^+,3^+,2^+,4^-,5^-)  \\
 A_\fin(1^+,4^-,3^+,5^-,2^+)  \\  A_\fin(1^+,4^-,3^+,2^+,5^-)  \\  A_\fin(1^+,5^-,2^+,4^-,3^+)  \\
 A_\fin(1^+,2^+,4^-,3^+,5^-)  \\  A_\fin(1^+,5^-,3^+,2^+,4^-)  \\  A_\fin(1^+,3^+,5^-,2^+,4^-)  \\
 B_\fin(3^+,5^-,4^-,1^+,2^+)  \\  B_\fin(1^+,2^+,5^-,3^+,4^-)  \\  B_\fin(1^+,3^+,4^-,2^+,5^-)  \\
 B_\fin(2^+,5^-,3^+,1^+,4^-)  \\  B_\fin(1^+,4^+,2^-,3^+,5^-)  \\  B_\fin(1^+,3^+,5^-,2^+,4^-)  \\
 B_\fin(2^+,4^-,3^+,1^+,5^-)  \\  B_\fin(1^+,5^-,4^-,2^+,3^+)  \\  B_\fin(1^+,3^+,2^+,4^-,5^-)  \\
 B_\fin(2^+,5^-,4^-,1^+,3^+)
\end{pmatrix}
, \nn \\
\vC_{----\pm}(\lp_1, \ldots, \lp_5)
&= \vC_{++++\mp}(\lp_1, \ldots, \lp_5)\Big|_{\langle..\rangle \leftrightarrow [..]}
 \,,\nn\\
\vC_{---++}(\lp_1, \dots, \lp_5)
&= \vC_{+++--}(\lp_1, \dots, \lp_5)\Big|_{\langle..\rangle \leftrightarrow [..]}
\,.\end{align}
For brevity, we have not written out the coefficients $\vC_{++++-}$ and $\vC_{+++++}$. They have exactly the same structure as $\vC_{+++--}$ with the replacements $4^- \to 4^+$ and $4^-,5^-\to 4^+, 5^+$, respectively, in the arguments of the helicity amplitudes. The remaining Wilson coefficients are given by parity invariance as shown. The overall factor of two comes from combining the two color structures in \eq{ggggg_color}, which are related by charge conjugation. 

At tree level, all the $B$ amplitudes vanish, as do all the amplitudes in $\vC_{++++\pm}$ and $\vC_{----\mp}$. By the parity relations given in \eq{ggggg_coeffs}, only the $A$ amplitudes in $\vC_{+++--}$ are then required for the tree-level matching. Since these amplitudes correspond to single trace color structures, which posses a cyclic symmetry, the required partial amplitudes are invariant under the corresponding cyclic permutations of their arguments. Therefore, at tree level, there are only two independent amplitudes, which we take to be $A_\fin(1^+,2^+,3^+,4^-,5^-)$ and $A_\fin(1^+,2^+,4^-,3^+,5^-)$. These are given in \app{gggggamplitudes}. Simplifications also occur at one loop, since the $B$ amplitudes can be expressed in terms of sums of permutations of the $A$ amplitudes \cite{Bern:1990ux, Bern:1994zx}.

\section{Renormalization Group Evolution}
\label{sec:running}

In this section, we discuss the renormalization group evolution (RGE) of the Wilson coefficients. We start with a general discussion and give the solution of the RGE to all orders in perturbation theory. For completeness, we also explicitly derive the (known) anomalous dimension at one loop. To discuss the RGE, it is convenient to consider the operators $\Op^\dagger$ in \eq{Opm_color}, which are vectors in color space. Lastly, we give explicit results, in a manifestly crossing symmetric form, for the relevant color mixing matrices for the color bases we have used in the previous sections. Since the operators' renormalization is independent of their helicity structure, we drop all helicity labels throughout this section for notational simplicity.

\subsection{General Discussion}
\label{subsec:running_general}

The renormalization of the hard scattering in SCET can either be carried out as operator renormalization, where the relation between bare and renormalized matrix elements is $\vev{\Op^\dagger}^\mathrm{bare}= Z_\xi^{-n_q/2} Z_A^{-n_g/2} \vev{\Op^\dagger}^\mathrm{ren} \widehat Z_{O}$, or with coefficient renormalization where $\vev{\Op^\dagger}^{\mathrm{bare}} \vC^\mathrm{bare} = Z_\xi^{n_q/2} Z_A^{n_g/2} \vev{\Op^\dagger}^\mathrm{bare} \widehat Z_C  \vec C^{\mathrm{ren}}$. The relationship between the two is $\widehat Z_C = \widehat Z_{O}^{-1}$.
Here $Z_\xi$ and $Z_A$ are the wave-function renormalizations of the SCET collinear quark and gluon fields $\xi_n$ and $A_n$, defined in \subsec{scet}, and
\begin{equation}
n_g = n_g^+ + n_g^-
\,,\qquad
n_q = n_q^+ + n_q^-
\end{equation}
are the total number of quark and gluon helicity fields in the operator (recall that there are two quark fields in each of the fermionic helicity currents).
The UV divergences for $\vev{\Op^\dagger}^\mathrm{bare}$ are given in terms of a local product (as opposed to a convolution over label momenta), since we are working at leading power where the operators contain a single field per collinear sector.

Let us consider more explicitly how the renormalization works at one loop. The counterterm Feynman rule at this order is
\begin{equation} \label{eq:SCETctFeyn}
\vev{\Op^\dagger}^\tree \, \Bigl(Z_\xi^{n_q/2}\, Z_A^{n_g/2}\, \hZ_C - 1 \Bigr)
\,.\end{equation}
At one loop, the UV divergences of $\vev{\Op^\dagger}^\mathrm{bare}$ are proportional to the tree-level matrix element as $\vev{\Op^\dagger}^\tree\,\hD$, where $\hD$ is a matrix in color space, which denotes the $1/\epsilon^2$ and $1/\epsilon$ UV divergences (with $\mu$ defined in the $\overline{\mathrm{MS}}$ scheme) of the bare matrix element.  The counterterm has to cancel these UV divergences so
\begin{equation} \label{eq:Z_O}
\vev{\Op^\dagger}^\tree \Bigl(Z_\xi^{n_q/2}\, Z_A^{n_g/2}\, \hZ_C - 1 \Bigr)  = -\vev{\Op^\dagger}^\tree\,\hD
\,,
\end{equation}
which fixes $\widehat Z_C$ at one loop.

Next consider the renormalization group equations, working to all orders in $\alpha_s$. As usual, the $\mu$ independence of the bare operator implies the renormalization group equation for the Wilson coefficient
\begin{equation} \label{eq:RGE}
\mu \frac{\df \vC(\mu)}{\df\mu} = \hga_C(\mu)\,\vC(\mu)
\,,\end{equation}
where the anomalous dimension matrix is defined as
\begin{equation} \label{eq:ga_O}
\hga_C(\mu)
 =  - \hZ_C^{-1}(\mu) \Big[ \frac{\df}{\df\ln\mu}\, \hZ_C(\mu)\Big] 
\,.\end{equation}
The solution of the RGE in \eq{RGE} can be written as
\begin{equation}
\vC(\mu) = \hU(\mu_0,\mu)\, \vC(\mu_0)
\,,\end{equation}
with the evolution matrix
\begin{equation} \label{eq:U_general}
\hU(\mu_0, \mu) = \cP \exp\biggl[\intlim{\ln \mu_0}{\ln \mu}{\ln\mu'}\, \hga_C(\mu')\biggr]
\,.\end{equation}
Here, $\cP$ denotes path ordering along increasing $\mu$, and $\mu > \mu_0$. The path ordering is necessary since $\hga_C(\mu)$ is a matrix in color space.

The anomalous dimension matrix has the general form%
\begin{equation} \label{eq:gaO_general}
\hga_C(\mu)
= \Gamma_\cusp[\alpha_s(\mu)]\,\hDe(\mu^2) + \hga[\alpha_s(\mu)]
\,,\end{equation}
where $\Gamma_\cusp$ is the cusp anomalous dimension and $\hDe(\mu^2)$ is a process-dependent mixing matrix in color space, which does not depend on $\alpha_s$. Its $\mu$ dependence is given by
\begin{equation} \label{eq:hDe_mudep}
\hDe(\mu^2) = \id (n_g C_A + n_q C_F) \ln\Bigl(\frac{\mu_0}{\mu}\Bigr) + \hDe(\mu_0^2)
\,,\end{equation}
which will be demonstrated explicitly at one loop in \subsec{loops}.
We can then perform the integral in \eq{U_general} by using the running of the coupling, $\df\alpha_s(\mu)/\df\ln\mu = \beta(\alpha_s)$, to switch variables from $\ln\mu$ to $\alpha_s$. We find
\begin{align}
\hU(\mu_0,\mu)
&= e^{-(n_g C_A + n_q C_F) K_\Ga(\mu_0,\mu)}
\\\nn & \quad\times
\bar \cP_{\alpha_s} \exp\Bigl[\eta_\Gamma(\mu_0, \mu)\,\hDe(\mu_0^2) + \widehat K_\gamma (\mu_0,\mu)\Bigr]
\,,\end{align}
where $\bar \cP_{\alpha_s}$ now denotes path ordering along decreasing $\alpha_s$, with $\alpha_s(\mu) < \alpha_s(\mu_0)$, and
\begin{align} \label{eq:Kw_def}
K_\Ga (\mu_0, \mu)
&= \intlim{\al_s(\mu_0)}{\al_s(\mu)}{\al_s} \frac{\Gamma_\cusp(\al_s)}{\beta(\al_s)}
   \intlim{\al_s(\mu_0)}{\al_s}{\al_s'} \frac{1}{\beta(\al_s')}
\,,\nn\\
\eta_\Ga(\mu_0, \mu)
&= \intlim{\al_s(\mu_0)}{\al_s(\mu)}{\al_s} \frac{\Gamma_\cusp(\al_s)}{\beta(\al_s)}
\,,\nn\\
\widehat K_\gamma(\mu_0, \mu)
&= \intlim{\al_s(\mu_0)}{\al_s(\mu)}{\al_s} \frac{\hga(\al_s)}{\bt(\al_s)}
\,.\end{align}
Up to two loops, the noncusp piece $\hga(\alpha_s)$ in \eq{gaO_general} is proportional to the identity operator~\cite{MertAybat:2006mz,Aybat:2006wq}
\begin{align}\label{eq:non_cusp_simplify}
\hga(\alpha_s)= \left( n_q \gamma_C^q + n_g \gamma_C^g\right) \id \,.
\end{align}
In this case, the evolution factor simplifies to
\begin{align}
\hU(\mu_0,\mu)
&= e^{-(n_g C_A + n_q C_F) K_\Ga(\mu_0,\mu) + K_\gamma(\mu_0,\mu)}
\nn \\ & \quad \times
\exp\Bigl[\eta_\Gamma(\mu_0, \mu)\,\hDe(\mu_0^2)\Bigr]
\,.\end{align}
Starting at three loops the noncusp anomalous dimension is not color diagonal,
and starts to depend on a conformal cross ratio built from factors of $p_i\cdot p_j$~\cite{Almelid:2015jia}. (For earlier work beyond two loops see~Refs.~\cite{Gardi:2009zv, Gardi:2009qi, Dixon:2009ur, Becher:2009cu, Becher:2009qa, DelDuca:2011ae, Bret:2011xm, DelDuca:2013ara, Caron-Huot:2013fea}.  The result of Ref.~\cite{Almelid:2015jia} implies that the conjectured all-order dipole color structure in Refs.~\cite{Becher:2009cu, Becher:2009qa} is violated.)

The evolution factors $K_\Ga(\mu_0,\mu)$, and $\eta_\Gamma(\mu_0,\mu)$ are universal. Explicit expressions for the integrals in \eq{Kw_def} to NNLL order, together with the required coefficients for $\Gamma_\cusp$ and the $\beta$ function to three loops, are given for reference in \app{RGE_factors}.

\subsection{One-loop Anomalous Dimension}
\label{subsec:loops}

\begin{figure}[t]
\centering
\subfigure[]{\label{fig:coll}
\includegraphics[scale=0.55]{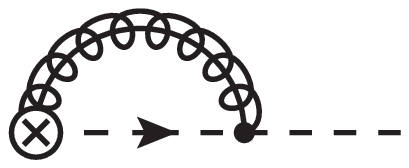}
\hspace{5ex}%
\includegraphics[scale=0.55]{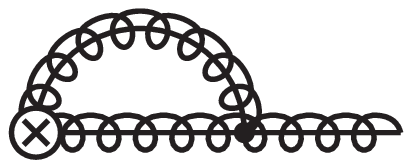}
}
\\
\subfigure[]{\label{fig:soft}%
\includegraphics[scale=0.45]{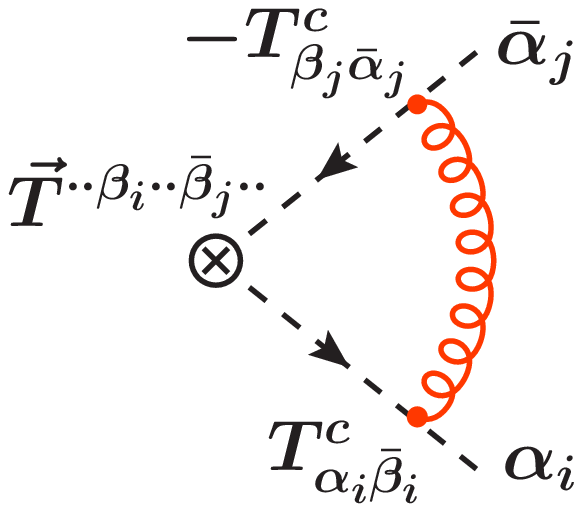}%
\includegraphics[scale=0.45]{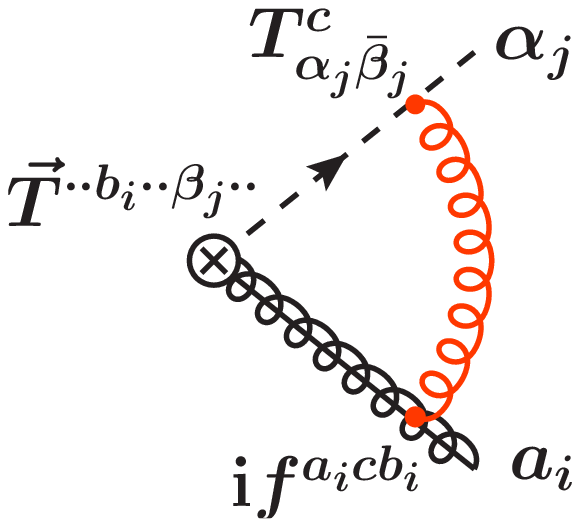}%
\includegraphics[scale=0.45]{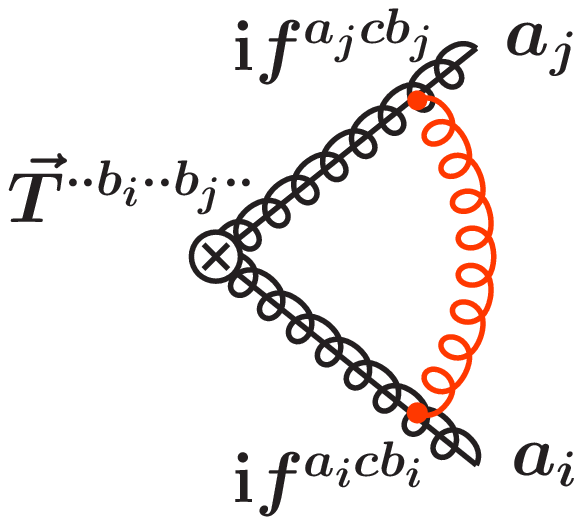}%
}
\caption{(a) Collinear one-loop diagrams. (b) Soft one-loop diagrams connecting two fields $i$ and $j$ in the operator.}
\label{fig:diagrams}
\end{figure}

The anomalous dimension $\hga_C(\mu)$ is process dependent. In this subsection, we derive its general form at one loop.
The anomalous dimension of the operators is determined from the UV divergences in the effective theory. The relevant one-loop diagrams in SCET are shown in \fig{diagrams}. In pure dimensional regularization the UV and IR divergences cancel such that the bare results for the loop diagrams vanish. To extract the UV divergences, we regulate the IR divergences by taking the external particles off shell with $p_i^2 = p_{i\perp}^2 \neq 0$.

Since all fields in the operators correspond to distinct collinear directions, the collinear loop diagrams in \fig{coll} only involve one external line at a time. Different external lines can only interact through the exchange of a soft gluon, shown by the diagrams in \fig{soft}.

When expressing our results, we use the notation [see \eq{Lij_def}]
\begin{align}
L_{{i\perp}} &= \ln \Bigl(-\frac{p_{i\perp}^2}{\mu^2} \Bigr)
\,,\qquad
L_{ij} = \ln \Bigl(-\frac{s_{ij}}{\mu^2} - \img 0\Bigr)
\,,\end{align}
where $s_{ij} = 2p_i\cdot p_j$.

First, we recall the wave function renormalization constants. In Feynman gauge at one loop,
\begin{align} \label{eq:Zi}
Z_\xi &= 1 - \frac{\al_s}{4\pi} \frac{1}{\eps} (C_F + \cdots)
\,,\nn\\
Z_A &= 1 + \frac{\al_s}{4\pi} \frac{1}{\eps} (\beta_0 - 2C_A + \cdots)
\,,\end{align}
where $\beta_0 = 11/3\, C_A - 4/3\, T_F n_f$ is the one-loop beta function coefficient [see \eq{betafunction}],
and $n_f$ is the number of considered quark flavors. Here and below, the ellipses denote possible UV-finite terms, which are irrelevant
for our discussion here. (Using the on-shell scheme for wave function renormalization, the $Z_i$ contain UV-finite pieces,
see \app{IRdiv}.)

The collinear diagrams in \fig{coll} contribute
\begin{align} \label{eq:Ic}
I_c^q = I_c^{\bar{q}}
&= \frac{\al_s C_F}{4\pi} \Bigl(\frac{2}{\eps^2} + \frac{2}{\eps} - \fr{2}{\eps} L_{{i\perp}} + \cdots \Bigr)
\vev{\Op^\dagger}^\tree
\,,\nn\\
I_c^g
&= \frac{\al_s C_A}{4\pi} \Bigl(\frac{2}{\eps^2} + \frac{1}{\eps} - \frac{2}{\eps} L_{{i\perp}}
+ \cdots \Bigr)
\vev{\Op^\dagger}^\tree
\,,\end{align}
where $I_c^i$ denotes the result of the diagram for an external leg of type $i$, either quark or gluon. 

The soft diagrams in \fig{soft} differ from each other only in their color structure. The result of the diagram connecting particles $i$ and $j$ (with $i\neq j$) is given by
\begin{equation} \label{eq:Is}
I_s^{ij}
= \frac{\al_s}{4\pi} \Bigl( \frac{2}{\eps^2} + \frac{2}{\eps} L_{{ij}} - \frac{2}{\eps} L_{{i\perp}}
  - \frac{2}{\eps} L_{{j\perp}}  + \cdots \Bigr)\, \vev{\Op^\dagger}^\tree\, \hatt^c_i\, \hatt^c_j
\,,\end{equation}
where $\hatt^c_i$ and $\hatt^c_j$ are matrices in color space. From \eqs{Ic}{Is} we see explicitly that the operators only mix with respect to the color structure, with no mixing between operators with distinct helicities.

The action of the matrix $\hatt^c_i$ on the color space is to insert a generator acting on the color index of the $i$th particle, i.e.,
\begin{align}
(\vT\, \hatt^c_i)^{\dotsb\alpha_i\dotsb} &= T^c_{\alpha_i\bbeta_i}\,\vT^{\dotsb\beta_i\dotsb}
\,,\nn\\
(\vT\, \hatt^c_i)^{\dotsb\balpha_i\dotsb} &= - \vT^{\dotsb\bbeta_i\dotsb}\,T^c_{\beta_i\balpha_i}
\,,\nn\\
(\vT\, \hatt^c_i)^{\dotsb a_i\dotsb} &= \img f^{a_i c b_i}\, \vT^{\dotsb b_i\dotsb}
\,,\end{align}
for quarks, antiquarks, and gluons, respectively. Our $\hatt^c_i$ is identical to what is usually denoted as $\mathbf{T}_i$ in the notation of Refs.~\cite{Catani:1996jh,Catani:1996vz}.

To give an explicit example, consider $gg\,q\bar{q}$. Then, for quark $i=3$ and antiquark $j=4$ we have
\begin{align}
\Op^\dagger\, \hatt^c_3\, \hatt^c_4
&= O^{a_1 a_2\balpha_3\alpha_4}\,(\vT\, \hatt^c_3\, \hatt^c_4)^{a_1 a_2\alpha_3\balpha_4}
\nn\\
&= O^{a_1 a_2\balpha_3\alpha_4}\, T^c_{\alpha_3 \bbeta_3}\,(-T^c_{\beta_4\balpha_4})
\vT^{a_1 a_2 \beta_3\bbeta_4}
\,,\end{align}
while for gluon $i=1$ and quark $j=3$,
\begin{equation}
\Op^\dagger\, \hatt^c_1\, \hatt^c_3
= O^{a_1 a_2\balpha_3\alpha_4}\, \img f^{a_1 c b_1} T^c_{\alpha_3 \bbeta_3} \vT^{ b_1 a_2 \beta_3\balpha_4}
\,.\end{equation}
Plugging in the explicit basis in \eq{ggqq_color} and using the relations in \app{color}, we can rewrite the resulting color structures above in terms of the basis in \eq{ggqq_color}, which yields 
\begin{align}
\hatt^c_3\, \hatt^c_4
&= - \begin{pmatrix}
   C_F - \tfrac{1}{2}C_A & 0 & 0 \\
   0 & C_F - \tfrac{1}{2}C_A & 0  \\
   T_F & T_F & C_F
\end{pmatrix}
,\nn\\
\hatt^c_1\, \hatt^c_3
&= - \begin{pmatrix}
   \tfrac{1}{2}C_A & 0 & T_F\\
   0 & 0 & -T_F \\
 0 & -T_F & 0
\end{pmatrix}
.\end{align}
The other combinations are computed analogously.

In general, one can easily see that for $i = j$
\begin{equation} \label{eq:titi}
\vT^{ a_1\dotsb\alpha_n}\,\hatt^c_i\, \hatt^c_i = C_i\, \vT^{ a_1\dotsb\alpha_n}
\,,\end{equation}
where $C_i = C_F$ for quarks and $C_i = C_A$ for gluons. By construction, the color basis $\vT^{ a_1\dotsb\alpha_n}$ conserves color, because each index corresponds to an external particle. Since $\hatt_i^c$ measures the color charge of the $i$th particle, color conservation implies
\begin{equation} \label{eq:sumti}
\vT^{a_1\dotsb\alpha_n} \Bigl(\sum_{i=1}^n \hatt_i^c\Bigr) = 0
\,.\end{equation}
As a simple example, consider $gq\bar{q}$ for which $\vT^{ a_1\alpha_2\balpha_3} \equiv T^{a_1}_{\alpha_2\balpha_3}$. In this case, \eq{sumti} gives
\begin{align}
&\img f^{a_1 c b_1} T^{b_1}_{\alpha_2\balpha_3} + T^c_{\alpha_2\bbeta_2} T^{a_1}_{\beta_2\balpha_3} - T^{a_1}_{\alpha_2\bbeta_3} T^c_{\beta_3\balpha_3}
\nn\\ & \quad
= \bigl(\img f^{a_1 c b_1} T^{b_1} + [T^c, T^{a_1}]\bigr)_{\alpha_2\balpha_3} = 0
\,.\end{align}

The total bare one-loop matrix element is given by summing \eq{Ic} for each external particle and \eq{Is} for each pair of distinct particles. The infrared logarithms $L_{{i\perp}}$ have to drop out in the sum of all UV-divergent contributions. To see that this is indeed the case, we can use \eq{titi} to rewrite the collinear contributions. Then, the sum of all $L_{{i\perp}}$ terms is proportional to
\begin{align} \label{eq:Lp2cancel}
&\vev{\Op^\dagger}^\tree \Bigr[\sum_i L_{{i\perp}} \hatt_i^c\,\hatt_i^c + \sum_{i < j}(L_{{i\perp}} + L_{{j\perp}})\, \hatt_i^c\, \hatt_j^c \Bigr]
\nn\\ & \quad
= \vev{\Op^\dagger}^\tree \Bigr(\sum_i L_{{i\perp}} \hatt_i^c\,\hatt_i^c + \sum_{i\neq j} L_{{i\perp}} \hatt_i^c\,\hatt_j^c \Bigr)
\nn\\ & \quad
= \vev{\Op^\dagger}^\tree \Bigl(\sum_i L_{{i\perp}} \hatt_i^c \Bigr)\Bigl(\sum_j\hatt_j^c\Bigr)
= 0
\,,\end{align}
where in the last step we used \eq{sumti}. For the same reason the $1/\eps^2$ poles in the soft diagrams cancel against half of the $1/\eps^2$ poles in the collinear diagrams. The remaining UV-divergent part of the matrix element is given by
\begin{align}\label{eq:Dmatrix}
\vev{\Op^\dagger}^\tree \hD
&= \vev{\Op^\dagger}^\tree\ \frac{\al_s}{4\pi} \biggl[n_g C_A \Bigl(\frac{1}{\eps^2} + \frac{1}{\eps}\Bigr)
\nn\\ &\quad
+ n_q C_F \Bigl(\frac{1}{\eps^2} + \frac{2}{\eps} \Bigr)
- \frac{2}{\eps} \hDe(\mu^2) \biggr]
\,,\end{align}
where the color mixing matrix is given by
\begin{equation} \label{eq:hDe_explicit}
\hDe(\mu^2) = -\sum_{i<j} \hatt_i^c\,\hatt_j^c\, L_{ij}
\,.\end{equation}
Combining this result with the identities in \eqs{titi}{sumti}, we can easily check that the $\mu$ dependence of $\hDe(\mu^2)$ is as in \eq{hDe_mudep}:
\begin{align}
\hDe(\mu^2) - \hDe(\mu_0^2)
&= - 2\sum_{i<j} \hatt_i^c\,\hatt_j^c\,\ln\Bigl(\frac{\mu_0}{\mu}\Bigr)
= \sum_i \hatt_i^c\,\hatt_i^c\,\ln\Bigl(\frac{\mu_0}{\mu}\Bigr)
 \nn\\
&= \id (n_g C_A + n_q C_F) \ln\Bigl(\frac{\mu_0}{\mu}\Bigr)
\,.\end{align}

We can now compute the anomalous dimension of the operators. From \eqs{Z_O}{Dmatrix}, we find at one loop
\begin{equation}\label{z_o}
\hZ_C = \id - \hD - \id \Bigl[\frac{n_g}{2}\,(Z_A-1) + \frac{n_q}{2}\,(Z_\xi-1) \Bigr]
\,,\end{equation}
which using \eq{ga_O} yields the one-loop anomalous dimension
\begin{equation} \label{eq:anomdim}
\hga_C(\mu)
= \frac{\al_s(\mu)}{4\pi} \bigl[4\hDe(\mu^2) - \id(n_g \beta_0 + n_q\,3 C_F) \bigr]
.\end{equation}
The coefficient of $4$ in front of $\hDe(\mu^2)$ is the one-loop cusp anomalous dimension coefficient [see \eq{cusp}]. The remaining terms determine the noncusp $\hga(\alpha_s)$ in \eq{gaO_general} at one loop,%
\begin{equation}
\hga(\alpha_s)
= - \frac{\al_s}{4\pi}\,(n_g \beta_0 + n_q\,3 C_F)\,\id
\,.\end{equation}

\subsection{Mixing Matrices}
\label{subsec:mixing}

In this section, we give explicit expressions for the mixing matrices for the color bases used in \secss{higgs}{vec}{pp}. For simplicity, we only give explicit expressions for up to four partons, but allow for additional colorless particles, such as a Higgs or vector boson. The matrices are straightforward to evaluate using the color relations in \app{color}, but become rather lengthy for more than four partons, due to the large number of allowed color structures, and are more easily evaluated in an automated way (see for example Ref.~\cite{Gerwick:2014gya}). For convenience, we introduce the following shorthand notation for sums and differences of logarithms $L_{ij}$,
\begin{align}
L_{ij \cdot kl \cdot \ldots} &= L_{ij} + L_{kl} + \ldots
\,, \nn \\
L_{ij \cdot \ldots/(kl\cdot \ldots)} &= (L_{ij \cdot \ldots}) -  (L_{kl\cdot\ldots})
\,,\end{align}
with $L_{ij} = \ln(-s_{ij}/\mu^2 -\img0)$ as defined in \eq{Lij_def}.

\begin{widetext}
\subsubsection{Pure Gluon Mixing Matrices}

For $gg$ and $ggg$ in the bases used in \eq{H0_color} and \eqs{H1_color}{Z1g_color}, we have
\begin{align}\label{eq:mix23gluons}
\hDe_{gg}(\mu^2)
= C_A\, L_{12}
\,, \qquad
\hDe_{ggg}(\mu^2)
= \frac{1}{2}C_A\, L_{12 \cdot 13 \cdot 23}
\begin{pmatrix}
 1 & 0 \\
 0 & 1
\end{pmatrix}
.\end{align}
For $gggg$ in the basis used in \eqs{ggggH_color}{gggg_color}, we have
\begin{align} \label{eq:mix4gluons}
&\hDe_{gggg}(\mu^2)
\nn \\ & \quad
= \begin{pmatrix}
\frac{1}{2} C_A L_{12\cdot14\cdot23\cdot34} & 0 & 0 & 2T_F L_{14\cdot23/(13\cdot24)}  & 0 & 2T_F L_{12\cdot34/(13\cdot24)} \\
  0 & \frac{1}{2} C_A L_{12\cdot13\cdot24\cdot34} & 0 & 2T_F L_{13\cdot24/(14\cdot23)} & 2T_F L_{12\cdot34/(14\cdot23)} & 0 \\
  0 & 0 & \frac{1}{2} C_A L_{13\cdot14\cdot23\cdot24} & 0 & 2T_F L_{14\cdot23/(12\cdot34)} & 2T_F L_{13\cdot24/(12\cdot34)} \\
  T_F L_{12\cdot 34/(13 \cdot24)} & T_F L_{12\cdot34/(14\cdot23)} & 0 & C_A L_{12\cdot34} & 0 & 0 \\
  0 & T_F L_{13\cdot24/(14\cdot23)} & T_F L_{13\cdot 24/(12\cdot34)} & 0 & C_A L_{13\cdot 24} & 0 \\
  T_F L_{14\cdot23/(13\cdot24)} & 0 & T_F L_{14\cdot23/(12\cdot34)} & 0 & 0 & C_A L_{14\cdot23}
\end{pmatrix}
.\end{align}
For our color bases formed from multitrace color structures, the structure of the mixing matrices is simple. Since the mixing matrices are determined by single gluon exchange, cyclicity is maintained, and all that can occur in the mixing is that a single trace splits into two or two traces recombine into one. For example, the color structure $\tr[T^a T^b T^c T^d]$ can only mix with
\begin{align}
\tr[T^a T^b T^c T^d]
\,,\qquad
\tr[T^a T^b]\, \tr[T^c T^d]
\,,\qquad \text{and} \qquad
\tr[T^d T^a]\, \tr[T^b T^c]
\,.\end{align}
Therefore, although the mixing matrices quickly get large as the number of color structures grows, their structure remains relatively simple. (An alternative approach to the organization of the anomalous dimensions for a large number of partons has been given in Ref.~\cite{Platzer:2013fha}.) For the dijet case, i.e., in the absence of additional colorless particles, the kinematics simplifies to
\begin{equation} \label{eq:dijetkin}
s = s_{12} = s_{34}
\,,\qquad
t = s_{13} = s_{24}
\,,\qquad
u = s_{14} = s_{23}
\,,\end{equation}
and these matrices were given in Ref.~\cite{Kidonakis:1998nf}, which also gives their eigenvectors and eigenvalues.

\subsubsection{Mixing Matrices Involving $q \bar q$ Pairs}

For $q\bar q$ and $gq\bar q$ in the bases used in \eq{Z0_color} and \eqs{H1_color}{Z1q_color}, we have
\begin{align}\label{eq:mixquarks1}
\hDe_{q\bq}(\mu^2)
= C_F\, L_{12}
\,, \qquad
\hDe_{g\,q\bq}(\mu^2)
= \frac{1}{2}\Big[C_A L_{12 \cdot 13} + (2C_F -C_A) L_{23}\Big]\,.
\end{align}
For $q\bar q q'\bar q'$ in the basis used in \eqss{qqqqH_color}{qqqq_color}{qqqqV_color}, we have
\begin{align} \label{eq:mixing_qqqq}
\hDe_{q\bq\, q\bq}(\mu^2)
= \hDe_{q\bar{q}\, q'\bq'}(\mu^2)
= \begin{pmatrix}
  C_F\, L_{14 \cdot 23} + (C_F - \frac{1}{2}C_A)\, L_{12\cdot 34/(13\cdot24)} & T_F\, L_{14\cdot23/(13\cdot24)} \\
  T_F\, L_{12\cdot34/(13\cdot24)} & C_F\, L_{12\cdot34} + (C_F - \frac{1}{2}C_A)\, L_{14\cdot23/(13\cdot24)}
\end{pmatrix}
.\end{align}
For $gg q\bar q$ in the basis used in \eqss{ggqqH_color}{ggqq_color}{ggqqV_color}, we have 
\begin{align} \label{eq:mixing_ggqq}
\hDe_{gg\,q\bar q}(\mu^2)
&= \begin{pmatrix}
   \frac{1}{2} C_A\, L_{12\cdot13\cdot24} +(C_F - \frac{1}{2} C_A)\, L_{34} & 0 & T_F\, L_{13\cdot24/(14\cdot23)} \\
   0 & \frac{1}{2} C_A\, L_{12\cdot14\cdot23} + (C_F - \frac{1}{2} C_A)\, L_{34} & T_F\, L_{14\cdot23/(13\cdot24)}  \\
   T_F\, L_{12\cdot34/(14\cdot23)} & T_F\, L_{12\cdot34/(13\cdot24)} &  C_A\, L_{12} +  C_F\, L_{34}
   \end{pmatrix}
.\end{align}
\end{widetext}
Again, these simplify in the dijet case, for which they were given along with their eigenvectors and eigenvalues in Ref.~\cite{Kidonakis:1998nf}.

\subsection{Soft Function Evolution}\label{sec:soft_evolution}

In this section, we review the renormalization group evolution of the soft function, focusing on our use of the color basis notation of \subsec{color} for nonorthogonal bases.  We will consider the particular case of the $N$-jettiness event shape \cite{Stewart:2010tn}, which allows for a definition of exclusive $N$-jet production with a factorization theorem of the form of \eq{sigma}.

The color mixing matrices of the previous section are in general complex valued for physical kinematics. For a physical channel, some of the appearing $s_{ij}$ are positive, giving rise to imaginary terms from the logarithms, as in \eq{Lij_def}. Since the cross section is real, these imaginary terms generated by the renormalization group evolution must drop out of the final result. We start by describing the properties of the soft function that ensure that this is the case.

Recall that the hard function $\hH_\kappa$ for a particular partonic channel $\kappa$ has its color indices contracted with those of the soft function. Explicitly,
\begin{align}\label{eq:soft_hard_trace}
  \tr(\hH_\kappa \hS_\kappa)
  &= H_\kappa^{a_1 \cdots \al_n b_1 \cdots \bt_n} S_\kappa^{b_1 \cdots \bt_n a_1 \cdots \al_n}
  \\ \nn
  & =\sum_{\{ \lambda_i \}}
  \Bigl[C_{\lambda_1 \cdot\cdot(\cdot\cdot \lambda_n )}^{b_1 \cdots \bt_n}\Bigr]^*
  S_\kappa^{b_1 \cdots \bt_n a_1 \cdots \al_n} C_{\lambda_1 \cdot\cdot(\cdot\cdot \lambda_n )}^{a_1 \cdots \al_n} \,.
\end{align}
The soft function is defined as a vacuum matrix element of a product of soft Wilson lines $\widehat{Y}$ as
\begin{align}\label{eq:softfunction_def}
\hS_\kappa(M,\{n_i\})= \Big\langle 0 \Big |  {\rm\bar T}\,  \widehat{Y}^{ \dagger}(\{n_i\})\ \delta(M-\hat{M}\,)\ {\rm T}\, \widehat{Y} (\{n_i\}) \Big |  0 \Big \rangle ,
\end{align}
where $\widehat{Y}(\{n_i\})$ is a product of soft Wilson lines in the $n_i$ directions. It is a matrix in color space, and $\widehat{Y}^\dagger$ is its Hermitian conjugate. Here T and ${\rm \bar T}$ denote time ordering and antitime ordering respectively.  The matrices $\widehat{Y}$ and $\widehat{Y}^\dagger$ are multiplied with each other, i.e.\ one of the color indices of the corresponding Wilson lines are contracted, and the external indices correspond to $b_1 \cdots \bt_n$ and $a_1 \cdots \al_n$, respectively.  Thus, for example $\widehat{Y}^\dagger \widehat{Y} = \delta^{a_1 b_1} \cdots \delta^{\alpha_n\beta_n}$.
The dependence of the soft function on the particular measurement, as well as the details of the jet algorithm, are encoded in the measurement function $\hat{M}$, whose precise form is not relevant for the current discussion.

From the definition of the soft function in \eq{softfunction_def} we see that it is Hermitian, namely $(S_\kappa^{b_1 \cdots \bt_n a_1 \cdots \al_n})^* = S_\kappa^{a_1 \cdots \al_n b_1 \cdots \bt_n}$. In abstract notation, this means $\hS_\kappa^\dagger = \hS_\kappa$, which implies that the product $\vC^\dagger \widehat S_\kappa \vC$ appearing in the cross section is real, so imaginary terms that appear in the Wilson coefficients due to renormalization group evolution drop out in the final cross section.

While this argument is trivial in a basis independent form, it is important to emphasize that in a nonorthogonal basis it takes a slightly more complicated form. As discussed in \subsec{color}, in a specific nonorthogonal color basis, \eq{soft_hard_trace} takes the form
$\vC^\dagger\, \hS_\kappa \vC = \vC^{*T} \hT\, \hS_\kappa \vC$ as in \eq{trHS},
where the matrix $\hT$ is defined in \eq{hatT_def}.
Similarly, the matrix representation of $\hS_\kappa$ is not Hermitian with respect to the naive conjugate transpose of its components. Instead, the condition on the reality of the cross section is given by [see \eq{def_daggermatrix}]
\begin{align}
\hS_\kappa = \hS^\dagger_\kappa = \hT^{-1}\, \hS_\kappa^{*T}\, \hT
\,.\end{align}

The invariance of the cross section under the RGE
\begin{align}\label{eq:inv_xs}
\mu \frac{\df}{\df\mu} \sigma_N =0\,,
\end{align}
implies relations between the anomalous dimensions of the SCET functions appearing in the factorization theorem of \eq{sigma}. In particular, it allows the anomalous dimension of the soft function to be determined from the anomalous dimensions of the Wilson coefficients, along with the anomalous dimensions of the beam and jet functions. The anomalous dimensions of the jet and beam functions are proportional to the color-space identity. The anomalous dimensions of the beam and jet functions appearing in the $N$-jettiness factorization theorem are equal to all orders in perturbation theory \cite{Stewart:2010qs} allowing us to use only the jet function anomalous dimension in the following discussion.
Renormalization group consistency then implies that the contributions of the soft function anomalous dimension not proportional to the identity, including the color off-diagonal components, are completely determined by the anomalous dimensions of the Wilson coefficients.

The soft function for $N$-jettiness can be written in the general form of \eq{softfunction_def}, but with an explicit measurement function
\begin{align}
&\hS_\kappa(k_a, k_b, k_1,\ldots, k_N, \{ n_i \})
\\ & \qquad
=\Big\langle 0 \Big |  {\rm\bar T}\, \widehat{Y}^{ \dagger}( \left\{   n_i\right\})\, \prod\limits_i \delta(k_i-\mathcal{\hat T}_i\,)\  {\rm T}\, \widehat{Y} (\{n_i\}) \Big |  0 \Big \rangle\,. \nonumber
\end{align}
Here $\mathcal{\hat T}_i$ picks out the contribution to the $N$-jettiness observable from the momentum region $i$, whose precise definition can be found in Ref.~\cite{Jouttenus:2011wh}. The soft function for $N$-jettiness was first presented to NLO in Ref.~\cite{Jouttenus:2011wh}, and more recently analyzed to NNLO in Ref.~\cite{Boughezal:2015eha}.

The all-orders structure of the renormalization group evolution for the soft function can be derived from \eq{inv_xs}, and is given by \cite{Kelley:2010fn, Jouttenus:2011wh}
\begin{align}\label{eq:soft_evolution}
&\mu \frac{\df}{\df\mu} \hS_\kappa (\{k_i\},\mu)
\nn\\ & \qquad
= \int \biggl[  \prod\limits_i \df k'_i \biggr] \frac{1}{2} \Bigl[   \hga_S (\{ k_i - k'_i \}, \mu)\, \hS_\kappa (\{k_i'\},\mu)
\nn\\ & \qquad\qquad
+  \hS_\kappa (\{k_i'\},\mu)\, \hga_S^\dagger (\{ k_i - k'_i \}, \mu)  \Bigr]
\,.\end{align}
The soft anomalous dimension $\hga_S$, and its conjugate $\hga_S^\dagger$, are given in terms of the anomalous dimension $\gamma_J$ of the jet function and the anomalous dimension of the Wilson coefficients, $\hga_C$ defined in \eqs{RGE}{ga_O}, as
\begin{align}\label{eq:soft_anom}
\hga_S (\{k_i\},\mu)
&= -\id \sum\limits_i Q_i\, \gamma_J^i(Q_i k_i, \mu) \prod\limits_{j\neq i} \delta(k_j)
\nn \\ & \quad
-2 \hga_C^\dagger (\mu) \prod\limits_i \delta(k_i)
\,.\end{align}
(Here, the $Q_i$ are related to the precise $N$-jettiness definition, see Ref.~\cite{Jouttenus:2011wh}.)
The Hermitian conjugates of $\hga_C$ and $\hga_S$ above again refer to the abstract Hermitian conjugate in color space. In a nonorthogonal color basis, they are given in terms of the complex conjugate transpose components according to \eq{def_daggermatrix} as
\begin{equation}
\hga_C^\dagger = \hT^{-1}\, \hga_C^{*T}\, \hT
\,, \qquad
\hga_S^\dagger = \hT^{-1}\, \hga_S^{*T}\, \hT
\,.\end{equation}

%

\section{Conclusions}
\label{sec:conclusions}

In this paper, we have presented a helicity operator approach to SCET. Helicities are naturally defined with respect to the external lightlike reference vectors specifying the jet directions in the effective theory, eliminating the need to consider complicated Lorentz and gamma matrix structures in the operator basis. The helicity operators correspond directly to physical states of definite helicity and color, which when combined with color organization techniques, greatly simplifies the construction of a minimal operator basis. Furthermore, the helicity operators are automatically crossing symmetric, and make manifest parity and charge conjugation symmetries, making it simple to determine relations amongst Wilson coefficients.

We demonstrated the utility of the helicity operator approach by explicitly constructing the basis valid to all orders in perturbation theory for a number of key processes at the LHC involving jets, and then determining the matching coefficients. In particular we considered $pp\to H + 0,1$ jets, $pp\to W/Z/\gamma + 0,1$ jets, and $pp\to 2$ jets at next-to-leading order, and $pp\to H + 2$ jets, $pp\to W/Z/\gamma + 2$ jets, and $pp\to 3$ jets at leading order.  We also discussed the dependence of this matching on the regularization scheme, considering schemes with helicities in $4$ and $d$ dimensions. An important and well-known simplification of the SCET approach is that when dimensional regularization is used for both IR and UV divergences, all loop graphs in the effective theory are scaleless, and thus vanish. As a result, the hard SCET Wilson coefficients in the $\overline{\rm MS}$ scheme, determined from matching QCD to SCET, are given directly by the IR-finite parts of color-ordered helicity amplitudes, defined using \eq{matching_general}. The use of our helicity operator basis therefore makes it simple to combine analytic resummation in SCET with fixed-order calculations of helicity amplitudes.

The all-orders structure for the renormalization group evolution of the helicity operator basis was discussed in detail. At leading power, distinct helicity structures do not mix, with renormalization group evolution causing mixing only in color space. This feature is made manifest at the level of the SCET Lagrangian due to the expansion in the soft and collinear limits. Subtleties associated with the use of nonorthogonal color bases were carefully treated, and expressions for the color sum matrix $\widehat T$ are given for the used color bases for all processes considered in the paper. Explicit results are also given for the one-loop mixing matrices describing the renormalization group evolution in color space for the case of $pp\to \text{up to } 2$ jets with an arbitrary number of uncolored external particles and in a manifestly crossing symmetric form.

Combining the methods of this paper with known expressions for jet, beam, and soft functions for particular exclusive jet cross sections, or jet shapes/observables, should facilitate analytic resummation for a large number of processes for which fixed-order amplitudes are known, or are soon to be calculated.

\begin{acknowledgments}
We thank Lance Dixon for useful discussions. This work was supported in part by the Office of Nuclear Physics of the U.S.
Department of Energy under Contract No. DE-SC0011090,
the DFG Emmy-Noether Grant No. TA 867/1-1, 
the Marie Curie International Incoming Fellowship PIIF-GA-2012-328913 within the 7th European Community Framework Program,
the Simons Foundation Investigator Grant No. 327942,
NSERC of Canada,
and the D-ITP consortium, a program of the Netherlands Organization for Scientific Research (NWO) that is funded by the Dutch Ministry of Education, Culture and Science (OCW).
We thank the Department of Energy's Institute for Nuclear Theory at the University of Washington for its hospitality and partial support during this work. 

\end{acknowledgments}

\appendix

\section{Spinor and Color Identities}
\label{app:useful}

\subsection{Spinor Algebra}
\label{app:helicity}

The overall phase of the spinors $\ket{p\pm}$ is not determined by the Dirac equation, $\pslash\, \ket{p\pm} = 0$, and so can be chosen freely. 
In the Dirac representation,
\begin{equation}
\ga^0 = \begin{pmatrix} 1 & 0 \\ 0 & -1 \end{pmatrix}
\,,\quad
\ga^i = \begin{pmatrix} 0 & \sigma^i \\ -\sigma^i & 0 \end{pmatrix}
\,,\quad
\ga_5 = \begin{pmatrix} 0 & 1 \\ 1 & 0 \end{pmatrix}
\,,\end{equation}
and taking $n_i^\mu = (1,0,0,1)$, we have the standard solutions~\cite{Dixon:1996wi}
\begin{equation} \label{eq:ket_explicit}
\ket{p+} = \frac{1}{\sqrt{2}}
\begin{pmatrix}
   \sqrt{p^-} \\
   \sqrt{p^+} e^{\img \phi_p} \\
   \sqrt{p^-} \\
   \sqrt{p^+} e^{\img \phi_p}
\end{pmatrix}
,\quad
\ket{p-} = \frac{1}{\sqrt{2}}
\begin{pmatrix}
    \sqrt{p^+} e^{-\img\phi_p} \\
    -\sqrt{p^-} \\
    -\sqrt{p^+} e^{-\img\phi_p} \\
    \sqrt{p^-}
\end{pmatrix}
,\end{equation}
where
\begin{equation}
p^\pm = p^0 \mp p^3
\,,\qquad
\exp(\pm \img \phi_p) = \frac{p^1 \pm \img p^2}{\sqrt{p^+ p^-}}
\,.\end{equation}
For negative $p^0$ and $p^\pm$ we use the usual branch of the square root, such that for $p^0 > 0$
\begin{equation}
\ket{(-p)\pm} = \img \ket{p\pm}
\,.\end{equation}
The conjugate spinors, $\bra{p\pm}$, are defined as
\begin{equation}
\bra{p\pm} = \mathrm{sgn}(p^0)\, \overline{\ket{p\pm}}
\,.\end{equation}
The additional minus sign for negative $p^0$ is included to use the same branch of the square root for both types of spinors, i.e., for $p^0 > 0$
\begin{equation}
\bra{(-p)\pm} = - \overline{\ket{(-p)\pm}} = -(-\img) \bra{p\pm} = \img\bra{p\pm}
\,.\end{equation}
In this way all spinor identities are automatically valid for both positive and negative momenta, which makes it easy to use crossing symmetry. The additional signs only appear in relations which involve explicit complex conjugation. The most relevant is
\begin{equation} \label{eq:spin_conj}
\braket{p-}{q+}^* = \mathrm{sgn}(p^0 q^0)\, \braket{q+}{p-}
\,.\end{equation}

The spinor products are denoted by
\begin{equation}
\langle p q \rangle = \braket{p-}{q+}
\,,\qquad
[p q] = \braket{p+}{q-}
\,.\end{equation}
Similarly, for products involving additional gamma matrices, we write
\begin{align}
\bra{p}\gamma^\mu |q] &= \bra{p-} \gamma^\mu \ket{q-}
\,,\quad 
[p|\gamma^\mu\ket{q} = \bra{p+} \gamma^\mu \ket{q+}
\,, \\
\bra{p}k |q] &= \bra{p-} \kslash \ket {q-}
\,,\quad
[p | k \ket{q} = \bra{p+} \kslash \ket {q+}
\,,\\
\bra{p}qk \ket{l} &= \bra{p-} \qslash \kslash \ket {l+}
\,,\quad
[p|qk|l] = \bra{p+} \qslash \kslash \ket {l-}
\,,\end{align}
etc.

Some useful identities, that follow directly from the definition of the spinors, are
\begin{align}
&\ang{pq} = - \ang{qp}
\,,\qquad
[pq] = - [qp]
\,,\\
 &[p|\ga^\mu \ket{p}=\bra{p} \ga^\mu |p] = 2p^\mu
\,.\end{align}
From the completeness relations
\begin{align} \label{eq:comp}
 \ket{p\pm}\bra{p\pm} = \frac{1 \pm \ga_5}{2}\,\pslash
\,, \\
\pslash=|p]\bra{p}+\ket{p}[p|,
\end{align}
one finds
\begin{equation}
\langle pq \rangle [qp] = \frac{1}{2}\,\tr\bigl\{(1 - \ga_5) \pslash \qslash \bigr\} = 2 p\cdot q
\,.\end{equation}
Combining this with \eq{spin_conj}, it follows that
\begin{equation}
 |\ang{pq}| = |[pq]| = \sqrt{|2p \cdot q|}
\,.\end{equation}
The completeness relation is also useful to reduce typical expressions like
\begin{equation}
[p|q\ket{k}= [pq]\langle qk\rangle\,,
\end{equation}
to spinor products.

Charge conjugation invariance of the current, the Fierz identity and the Schouten identity are
\begin{align} \label{eq:Fierzetc}
\bra{p}\gamma^\mu |q]  &= [q|\gamma^\mu \ket{p}
\,,\\
[p|\gamma_\mu \ket{q} [k|\gamma^\mu \ket{l}&= 2[pk]\ang{lq}
\,, \\
\ang{pq} \ang{kl} &= \ang{pk} \ang{ql} + \ang{pl} \ang{kq}
\,.\end{align}
Finally, momentum conservation $\sum_{i=1}^n p_i = 0$ implies
\begin{equation} \label{eq:spinormomcons}
\sum_{i=1}^n [ji] \ang{ik} = 0
\,.\end{equation}

From \eq{ket_explicit}, we see that under parity the spinors transform as
\begin{equation} \label{eq:spinorparity}
\ket{p^\P\pm} = \pm e^{\pm \img\phi_p}\gamma^0\,\ket{p\mp}
\,,\end{equation}
and therefore
\begin{align}
\ang{p^\P q^\P} &= -e^{\img(\phi_p + \phi_q)} [pq] \label{eq:app_parity1}
\,,\\
[p^\P q^\P] &= -e^{-\img(\phi_p + \phi_q)} \ang{pq} \label{eq:app_parity2}
\,.\end{align}

When applying the above result to a helicity amplitude, the phases which appear are determined by the little group scaling (see e.g. Refs.~\cite{Dixon:1996wi, Dixon:2013uaa,Elvang:2013cua} for a review). The little group is the subgroup of the Lorentz transformations that fixes a particular momentum. In terms of the spinor helicity variables, the action of the little group, which preserves the momentum vector $p$, is given by
\begin{equation}
|p\rangle \to z |p\rangle, \qquad [p| \to \frac{1}{z}[p|\,.
\end{equation}
In the case that the particle with momentum $p$ has helicity $h$, the corresponding helicity amplitude scales as $z^{-2h}$ under the little group scaling. This property of the helicity amplitudes then predicts the phases that appear in the amplitude under a parity transformation.

The following completeness relation for the polarization vectors is also useful
\begin{align}
\sum\limits_{\lambda=\pm} \epsilon^\lambda_\mu(p,q) \left (\epsilon^\lambda_\nu(p,q) \right )^*=-g_{\mu \nu}+\frac{p_\mu q_\nu+p_\nu q_\mu}{p\cdot q}.
\end{align}

In SCET the collinear quark fields produce projected spinors 
\begin{align}
\ket{p\pm}_{n}=\frac{\nslash\bnslash}{4}\ket{p\pm}\,.
\end{align}
The projected spinor trivially satisfies the relation
\begin{equation}
\nslash\,\Bigl(\frac{\nslash\bnslash}{4}\ket{p\pm}\Bigr) = 0
\,,\end{equation}
so it is proportional to $\ket{n\pm}$. Working in the basis in \eq{ket_explicit}, we have 
\begin{align}\label{eq:ketn}
\frac{\nslash\bnslash}{4} \ket{p}=& \sqrt{p^0}\left[ \cos\left (\frac{\theta_n}{2}\right )\cos \left(\frac{\theta_p}{2}\right )\right. \nonumber \\
& \left.+e^{i(\phi_p-\phi_n)} \sin \left(\frac{\theta_n}{2}\right )   \sin \left(\frac{\theta_p}{2} \right )   \right ] \ket{n}\,, \nonumber \\
\frac{\nslash\bnslash}{4} |p]=& \sqrt{p^0}\left [ e^{i \left(\phi _p-\phi _n\right)} \cos \left(\frac{\theta _n}{2}\right)  \cos \left(\frac{\theta _p}{2}\right) \right. \nonumber \\
&\left. +\sin \left(\frac{\theta _n}{2}\right) \sin \left(\frac{\theta _p}{2}\right)   \right ] |n]\,.
\end{align}
Here $\theta_n, \phi_n$, and $\theta_p, \phi_p$, are the polar and azimuthal angle of the $n$ and $p$ vectors, respectively.
In particular, we see that choosing $n^\mu=p^\mu/p^0$, which can always be done at leading power since there is a single particle per collinear sector, we have $\phi_p=\phi_n$, $\theta_p=\theta_n$, and the simple relation
\begin{align}
\frac{\nslash\bnslash}{4} \ket{p\pm}=\sqrt{\frac{\bn\cdot p}{2}}\: \ket{n\pm}\,.
\end{align}

\subsection{Color Algebra}
\label{app:color}

The generators $t^a_r$ of a general irreducible representation $r$ of $\mathrm{SU(N)}$ satisfy
\begin{equation}
[t^a_r, t^b_r] = \img f^{abc} \, t_r^c
\,,\quad
t_r^a\, t_r^a = C_r\mathbf{1}
\,,\quad
\tr[t^a_r\,t^b_r] = T_r\,\delta^{ab}
\,,\end{equation}
where $f^{abc}$ are completely antisymmetric, and $C_r$ is the quadratic Casimir of the representation $r$.
The normalization $T_r$ is given by $T_r = C_r d_r/d$, where $d_r$ is the dimension of the representation
and $d$ the dimension of the Lie algebra.

We denote the generators in the fundamental representation by $t_F^a = T^a$, and the overall
normalization is fixed by choosing a specific value for $T_F$. 
The adjoint representation is given by $(t_A^a)_{bc} = -\img f^{abc}$, which implies
\begin{equation}
f^{acd} f^{bcd} = C_A\,\delta^{ab}
\,.\end{equation}
We also define the symmetric structure constants as 
\begin{equation}
d^{abc}=\frac{1}{T_F} \tr[T^a\{ T^b,T^c \}]
\,.
\end{equation}
For the fundamental and adjoint representations we have $d_F = N$, $d_A = d = N^2 - 1$, and so
\begin{equation}
C_F = \frac{N^2 - 1}{2N}
\,,\qquad
C_A = N
\,,
\end{equation}
where we have chosen the standard normalization
\begin{equation}
T_F = \frac{1}{2}
\,.\end{equation}
Throughout the text, and for the amplitudes in the appendices, we have kept $T_F$ arbitrary. This can be done using $C_F = T_F (N^2 - 1)/N$, $C_A = 2T_F N$. The strong coupling constant, $g_s$, can be kept convention independent, by using $g_s \to g_s/\sqrt{2T_F}$.

Some additional useful color identities are
\begin{align}
t^a_r t^b_r t^a_r &= \Bigl(C_r - \frac{C_A}{2}\Bigr)\, t^b_r
\,,\\
T^a T^b T^c T^a &= T_F^2\, \delta^{bc} \mathbf{1} + \Bigl(C_F - \frac{C_A}{2} \Bigr)\, T^b T^c
\,,\end{align}
where the second relation is equivalent to the completeness relation
\begin{equation}
T^a_{\alpha \bbeta}\,T^a_{\gamma \bdelta} = T_F \Bigl(\de_{\alpha \bdelta}\,\de_{\gamma \bbeta} - \frac{1}{N} \de_{\alpha \bbeta}\, \de_{\gamma \bdelta} \Bigr)
\,.\end{equation}
We also have
\begin{align}
T^b\, \img f^{bac}\, T^c &= \frac{C_A}{2}\, T^a
\,,\\
T^c\, \img f^{cad} \img f^{dbe} T^e
&= T_F^2\,\delta^{ab}\,\mathbf{1} + \frac{C_A}{2}\,T^a T^b
\,.\end{align}

\subsection{QCD Color Decompositions}
\label{app:color_decomp}

Here we briefly review a common color decomposition for QCD NLO amplitudes \cite{Berends:1987me, Mangano:1987xk, Mangano:1988kk, Bern:1990ux}. The color bases used for the processes discussed in the text are specific examples of the decompositions given below, and were chosen to facilitate the extraction of the matching coefficients from the amplitudes literature. For a pedagogical introduction to color decompositions in QCD amplitudes see for example Refs.~\cite{Dixon:1996wi, Dixon:2013uaa}.

For an $n$ gluon process, a one-loop color decomposition in terms of fundamental generators $T^a$ is given by
\begin{widetext}
\begin{align} \label{eq:gluon_decomp}
\cA_n(g_1\cdots g_n)
&= g_s^{n-2}  \sum_{\si \in S_n/Z_n}\! \tr[T^{a_{\si(1)}} \cdot \cdot \cdot T^{a_{\si(n)}}]
\Big[ A_{n}^{\text{tree}}\bigl(\si(1), \cdot \cdot \cdot, \si(n)\bigr)
+ g_s^2\, C_A A_{n;1}\bigl(\si(1), \cdot \cdot \cdot, \si(n)\bigr) \Big]
\nn\\ &
+  g_s^n  \sum\limits_{c=3}^{\left \lfloor{n/2}\right \rfloor+1}     \!\!\sum_{\si \in S_n/S_{c-1,n-c+1}}\!\!\! \tr[ T^{a_{\si(1)}}  \cdot \cdot \cdot    T^{a_{\si(c-1)}}] \tr[ T^{a_{\si(c)}}  \cdot \cdot \cdot    T^{a_{\si(n)}}]
A_{n;c}\bigl(\si(1),\cdot \cdot \cdot,\si(n)\bigr)
\,,\end{align}
where $A_{n;1}$, $A_{n}^{\text{tree}}$ are primitive amplitudes, which can be efficiently calculated using unitarity methods, and the $A_{n;c}$ are partial amplitudes which can be written as sums of permutations of the primitive amplitudes. The amplitudes appearing in this decomposition are separately gauge invariant. In this formula, $S_n$ is the permutation group on $n$ elements, and $S_{i,j}$ is the subgroup of $S_{i+j}$ which leaves the given trace structure invariant. At tree level, only the single trace color structure appears.

In the case that additional noncolored particles are also present, an identical decomposition exists, since the color structure is unaffected. For example, for a process involving $n$ gluons and a Higgs particle, the amplitude satisfies the same decomposition as in \eq{gluon_decomp}, but with the partial and primitive amplitudes in \eq{gluon_decomp} simply replaced by $A\bigl(\phi, \si(1),\cdot \cdot \cdot,\si(n)\bigr)$, where $\phi$ denotes the Higgs particle  \cite{Berger:2006sh}.

A similar decomposition exists for processes involving $q \bar q$ pairs. For example, the one-loop decomposition for a process with a $q\bar q$ pair and $n-2$ gluons is given by \cite{Bern:1994fz}
\begin{align} \label{eq:qq_decomp}
\cA_n \bigl( \bar q_1 q_2 g_3\dots g_n  \bigr)
&=g_s^{n-2}\sum_{\si \in S_{n-2}}\!    \bigl( T^{a_{\si(3)}} \cdot \cdot \cdot T^{a_{\si(n)}}   \bigr)_{\alpha \bbeta}
\Big[A_{n}^{\text{tree}}\bigl( 1_{\bar q}, 2_q;\si(3), \cdot \cdot \cdot, \si(n)\bigr) 
+ g_s^2\, C_A A_{n;1}\bigl(1_{\bar q}, 2_q;\si(3), \cdot \cdot \cdot, \si(n)\bigr)\Big]
\nn\\ &
+g_s^n  \sum\limits_{c=3}^{n-3} ~    \!\!\sum_{\si \in S_{n-2}/Z_{c-1}}\!\!\! \tr[ T^{a_{\si(3)}}  \cdot \cdot \cdot    T^{a_{\si(c+1)}}] \bigl ( T^{a_{\si(c+2)}}  \cdot \cdot \cdot    T^{a_{\si(n)}} \bigr )_{\alpha \bbeta}
A_{n;c}\bigl(1_{\bar q}, 2_q;\si(3),\cdot \cdot \cdot,\si(n)\bigr) 
\nn\\ &
+g_s^n \!\! \!\!\!\! \!\!\sum_{\si \in S_{n-2}/Z_{n-3}}\!\!\! \tr[ T^{a_{\si(3)}}  \cdot \cdot \cdot    T^{a_{\si(n-1)}}] \bigl (  T^{a_{\si(n)}} \bigr )_{\alpha \bbeta}
A_{n;n-2}\bigl(1_{\bar q}, 2_q;\si(3),\cdot \cdot \cdot,\si(n)\bigr) 
\nn\\ &
+g_s^n \!\! \!\!\!\! \!\!  \sum_{\si \in S_{n-2}/Z_{n-2}}\!\!\! \tr[ T^{a_{\si(3)}}  \cdot \cdot \cdot    T^{a_{\si(n)}}]\delta_{\alpha \bbeta}
A_{n;n-1}\bigl(1_{\bar q}, 2_q;\si(3),\cdot \cdot \cdot,\si(n)\bigr) 
\,.\end{align}
This decomposition is easily extended to the case of additional $q\bar q$ pairs.
As with the gluon case, the same color decomposition also applies if additional uncolored particles are included in the amplitude.

For more than five particles, the one-loop color decompositions given above do not give a complete basis of color structures beyond one loop, since color structures with more than two traces can appear. A complete basis of color structures is required for the SCET basis to guarantee a consistent RGE. A convenient basis of color structures for one-loop matching is then given by extending the one-loop decomposition to involve all higher trace structures.

\section{Helicity Amplitudes for Higgs $+$ Jets}
\label{app:Hamplitudes}

In this appendix we give explicit results for the hard matching coefficients for $H+0,1,2$ jets. We only explicitly consider gluon-fusion processes, where the Higgs couples to two gluons through a top-quark loop, and additional jets correspond to additional gluons, or quark antiquark pairs. When matching onto SCET we perform a one-step matching and directly match full QCD onto SCET, as was done for $H+0$ jets in Ref.~\cite{Berger:2010xi}. Most QCD results are obtained in the limit of infinite top quark mass, by first integrating out the top quark and matching onto an effective $ggH$ interaction,
\begin{equation} \label{eq:HiggsLeff}
\cL_\hard = \frac{C_1}{12\pi v} H G_{\mu\nu}^a G^{\mu\nu\,a}
\,,\end{equation}
which is then used to compute the QCD amplitudes. Here $v = (\sqrt{2}G_F)^{-1/2} = 246\GeV$. From the point of view of the one-step matching from QCD onto SCET, using \eq{HiggsLeff} is just a convenient way to compute the full QCD amplitude in the $m_t\to \infty$ limit. In particular, the $\alpha_s$ corrections to $C_1$ in \eq{HiggsLeff} are included in the amplitudes below, and therefore also in the SCET Wilson coefficients. In this way, if higher-order corrections in $1/m_t$ or the exact $m_t$ dependence for a specific amplitude are known, they can easily be included in the QCD amplitudes and the corresponding SCET Wilson coefficients. We illustrate this for the case of $H+0$ jets below.

We separate the QCD amplitudes into their IR-divergent and IR-finite parts
\begin{align}
A &= A_\div + A_\fin\,,
\nonumber \\
B &= B_\div + B_\fin\,,
\end{align}
where $A_\fin$, $B_\fin$ enter the matching coefficients in \sec{higgs}. For simplicity, we drop the subscript ``$\fin$'' for those amplitudes that have no divergent parts, i.e. for $A_\div = 0$ we have $A_\fin \equiv A$. For the logarithms we use the notation
\begin{align}
L_{ij} &= \ln\Bigl(-\frac{s_{ij}}{\mu^2} - \img 0 \Bigr)
\,,\nn\qquad
L_{ij/H} = \ln\Bigl(-\frac{s_{ij}}{\mu^2} - \img 0 \Bigr) - \ln\Bigl(-\frac{m_H^2}{\mu^2} - \img 0 \Bigr)
\,.\end{align}

\subsection{\boldmath $H+0$ Jets}
\label{app:H0amplitudes}

We expand the amplitudes in powers of $\alpha_s(\mu)$ as
\begin{equation} \label{eq:H_expand}
A = \frac{2T_F\alpha_s(\mu)}{3\pi v}\, \sum_{n=0}^\infty A^{(n)} \Bigl(\frac{\alpha_s(\mu)}{4\pi}\Bigr)^n
\,.\end{equation}
The amplitudes with opposite helicity gluons vanish to all orders because of angular momentum conservation,
\begin{align}
A(1^\pm, 2^\mp; 3_H)
&= 0,
\end{align}
corresponding to the fact that the helicity operators for these helicity configurations were not included in the basis of \eq{ggH_basis}.
The lowest order helicity amplitudes including the full $m_t$ dependence are given by
\begin{align}
A^\zero(1^+, 2^+; 3_H)
&= \frac{s_{12}}{2}\frac{[12]}{\ang{12}}\, F^\zero\Bigl(\frac{s_{12}}{4m_t^2}\Bigr)= \frac{s_{12}}{2}\, F^\zero\Bigl(\frac{s_{12}}{4m_t^2}\Bigr)\, e^{\img\Phi_{++ H}}
\,,\nn\\
A^\zero(1^-, 2^-; 3_H)
&=\frac{s_{12}}{2}\frac{\ang{12}}{[12]}\, F^\zero\Bigl(\frac{s_{12}}{4m_t^2}\Bigr)= \frac{s_{12}}{2}\, F^\zero\Bigl(\frac{s_{12}}{4m_t^2}\Bigr)\, e^{\img\Phi_{-- H}}
\,,
\end{align}
where the function $F^{(0)}(z)$ is defined as
\begin{align}
F^\zero(z)
&= \frac{3}{2z} - \frac{3}{2z}\Bigl\lvert 1 - \frac{1}{z}\Bigr\rvert
\begin{cases}
\arcsin^2(\sqrt{z})\, , & 0 < z \leq 1 \,,\\
\ln^2[-\img(\sqrt{z} + \sqrt{z-1})] \,, \quad & z > 1
\,.\end{cases}
\end{align}
For simplicity, we have extracted the (irrelevant) overall phases 
\begin{equation}
e^{\img\Phi_{++H}} = \frac{[12]}{\ang{12}}
\,,\qquad
e^{\img\Phi_{--H}} = \frac{\ang{12}}{[12]}
\,.\end{equation}
Since the two helicity amplitudes for $ggH$ cannot interfere and are equal to each other by parity up to an overall phase, their higher-order corrections are the same as for the spin-summed $ggH$ form factor.
The divergent part of the NLO amplitudes is given by
\begin{equation}
A_\div^\one(1^\pm, 2^\pm; 3_H)
= A^\zero(1^\pm, 2^\pm; 3_H) \biggl[
-\frac{2}{\eps^2}\, C_A + \frac{1}{\eps}\,(2C_A\, L_{12} - \beta_0) \biggr]
\,.\end{equation}
The IR-finite parts entering the matching coefficients in \eq{ggH_coeffs} at NLO are~\cite{Berger:2010xi}
\begin{align}
A_\fin^\one(1^\pm, 2^\pm; 3_H)
&= A^\zero(1^\pm, 2^\pm; 3_H) \biggl[ C_A \Bigl(-L_{12}^2 + \frac{\pi^2}{6} \Bigr)
+ F^\one\Bigl(\frac{s_{12}}{4m_t^2}\Bigr) \biggr]
\,,\nn\\
F^\one(z) &=
C_A\Bigl(5 - \frac{38}{45}\, z - \frac{1289}{4725}\, z^2 - \frac{155}{1134}\, z^3
- \frac{5385047}{65488500}\, z^4\Bigr)
\nn\\ & \quad
+ C_F \Bigl(-3 + \frac{307}{90}\, z + \frac{25813}{18900}\, z^2 + \frac{3055907}{3969000}\, z^3
+ \frac{659504801}{1309770000}\, z^4 \Bigr) + \ord{z^5}
\,.\end{align}
The full analytic expression for $F^\one(z)$ is very long, so we only give the result expanded in $z$. Since the additional $m_t$ dependence coming from $F^\one(z)$ is small and the expansion converges quickly, the expanded result is fully sufficient for on-shell studies of Higgs production. The IR-finite parts at NNLO are~\cite{Berger:2010xi}
\begin{align}
A_\fin^\two(1^\pm, 2^\pm; 3_H)
&= A^\zero(1^\pm, 2^\pm; 3_H) \biggl\{
\frac{1}{2} C_A^2 L_{12}^4 + \frac{1}{3} C_A \beta_0 L_{12}^3 +
  C_A\Bigl[C_A\Bigl(-\frac{4}{3} + \frac{\pi^2}{6} \Bigr)  - \frac{5}{3}\beta_0
  - F^\one\Bigl(\frac{s_{12}}{4m_t^2}\Bigr) \Bigr] L_{12}^2
\nn \\ & \quad
+ \Bigl[C_A^2\Bigl(\frac{59}{9} - 2\zeta_3 \Bigr) + C_A \beta_0 \Bigl(\frac{19}{9}-\frac{\pi^2}{3}\Bigr)
- \beta_0 F^\one\Bigl(\frac{s_{12}}{4m_t^2}\Bigr) \Bigr] L_{12}
+ F^\two\Bigl(\frac{s_{12}}{4m_t^2}\Bigr) \biggr\}
\,,\nn\\
F^\two(z) &=
\bigl(7 C_A^2 + 11 C_A C_F - 6 C_F \beta_0 \bigr) \ln(-4z-\img 0)
+ C_A^2 \Bigl(-\frac{419}{27} + \frac{7\pi^2}{6} + \frac{\pi^4}{72} - 44 \zeta_3 \Bigr)
\nn \\ & \quad
+  C_A C_F \Bigl(-\frac{217}{2} - \frac{\pi^2}{2} + 44\zeta_3 \Bigr)
+ C_A \beta_0 \Bigl(\frac{2255}{108} + \frac{5\pi^2}{12} + \frac{23\zeta_3}{3} \Bigr)
- \frac{5}{6} C_A T_F
\nn \\ & \quad
+ \frac{27}{2} C_F^2
+ C_F \beta_0 \Bigl(\frac{41}{2} - 12 \zeta_3\Bigr)
-\frac{4}{3} C_F T_F
+ \ord{z}
\,.\end{align}
Here we only give the leading terms in the $m_t\to\infty$ limit. The first few higher-order terms in $z$ in $F^\two(z)$ can be obtained from the results of Refs.~\cite{Harlander:2009bw, Pak:2009bx}.

\subsection{\boldmath $H+1$ Jet}
\label{app:H1amplitudes}

The amplitudes for $H+1$ jet were calculated in Ref.~\cite{Schmidt:1997wr} in the $m_t\to\infty$ limit. Reference~\cite{Schmidt:1997wr} uses $T_F=1$ and $g_s T^a/\sqrt{2}$ for the $q\bar{q}g$ coupling. Thus, we can convert to our conventions by replacing $T^a\to \sqrt{2} T^a$, and identifying $1/N = C_A - 2C_F$ and $N = C_A$ in the results of Ref.~\cite{Schmidt:1997wr}. We expand the amplitudes in powers of $\alpha_s(\mu)$ as
\begin{equation}
A = \frac{2T_F\alpha_s(\mu)}{3\pi v}\,g_s(\mu)
\sum_{n=0}^\infty A^{(n)} \Bigl(\frac{\alpha_s(\mu)}{4\pi}\Bigr)^n
\,.\end{equation}

\subsubsection{$gggH$}

The tree-level amplitudes entering the matching coefficient $\vC_{++\pm}$ in \eq{gggH_coeffs} are
\begin{align}
A^\zero(1^+,2^+,3^+;4_H)
&= \frac{1}{\sqrt{2}}\, \frac{m_H^4}{\ang{12}\ang{23}\ang{31}}
= \frac{m_H^4}{\sqrt{2|s_{12} s_{13} s_{23}|}}\,e^{\img\Phi_{+++H}}
\,,\nn\\
A^\zero(1^+,2^+,3^-;4_H)
&= \frac{1}{\sqrt{2}}\, \frac{[12]^3}{[13][23]}
= \frac{s_{12}^2}{\sqrt{2|s_{12} s_{13} s_{23}|}}\,e^{\img\Phi_{++-H}}
\,,\end{align}
where we have extracted the (irrelevant) overall phases
\begin{equation}
e^{\img\Phi_{+++H}} = \frac{\sqrt{|s_{12}|}}{\ang{12}} \frac{\sqrt{|s_{13}|}}{\ang{31}} \frac{\sqrt{|s_{23}|}}{\ang{23}}
\,,\qquad
e^{\img\Phi_{++-H}} = \frac{[12]}{\ang{12}} \frac{\sqrt{|s_{12}|}}{\ang{12}} \frac{\sqrt{|s_{13}|}}{[13]} \frac{\sqrt{|s_{23}|}}{[23]}
\,.\end{equation}
The divergent parts of the one-loop amplitudes are
\begin{equation}
A_\div^\one(1^+,2^+,3^\pm;4_H)
= A^\zero(1^+,2^+,3^\pm,4_H) \biggl\{
-\frac{3}{\eps^2}\, C_A + \frac{1}{\eps}\,\Bigl[C_A\, (L_{12} + L_{13} + L_{23} )
- \frac{3}{2} \beta_0 \Bigr] \biggr\}
\,.\end{equation}
The finite parts of the $gggH$ amplitudes, which enter the matching coefficient $\vC_{++\pm}$ at one loop are
\begin{align}
A_\fin^\one(1^+,2^+,3^+;4_H)
&= A^\zero(1^+,2^+,3^+;4_H) \biggl\{f(s_{12}, s_{13}, s_{23}, m_H^2, \mu)
+ \frac{1}{3}(C_A - 2 T_F n_f)\,\frac{s_{12} s_{13}+ s_{12} s_{23}+ s_{13} s_{23}}{m_H^4}
\biggr\}
\,,\nn\\
A_\fin^\one(1^+,2^+,3^-;4_H)
&= A^\zero(1^+,2^+,3^-;4_H)
\biggl\{f(s_{12}, s_{13}, s_{23}, m_H^2, \mu)
+ \frac{1}{3}(C_A - 2 T_F n_f)\,\frac{s_{13} s_{23}}{s_{12}^2}
\biggr\}
\,,
\end{align}
where we have extracted the common function
\begin{align}
f(s_{12}, s_{13}, s_{23}, m_H^2, \mu)
&= -C_A \biggl[\frac{1}{2} (L_{12}^2+L_{13}^2+L_{23}^2)
 + L_{12/H} L_{13/H} + L_{12/H} L_{23/H} + L_{13/H} L_{23/H}
\nn \\ & \quad
+ 2\Li_2\Bigl(1-\frac{s_{12}}{m_H^2}\Bigr) + 2\Li_2\Bigl(1-\frac{s_{13}}{m_H^2}\Bigr)
+ 2\Li_2\Bigl(1-\frac{s_{23}}{m_H^2}\Bigr) - 5 - \frac{3\pi^2}{4} \biggr] - 3 C_F
\,.\end{align}

\subsubsection{$gq\bq H$}

The tree-level amplitudes entering the matching coefficient $\vC_{\pm(+)}$ in \eq{gqqH_coeffs} are
\begin{align}
A^\zero(1^+;2_q^+,3_\bq^-;4_H)
&= -\frac{1}{\sqrt{2}}\,\frac{[12]^2}{[23]}
= \frac{s_{12}}{\sqrt{2|s_{23}|}}\, e^{\img\Phi_{+(+)H}}
\,,\nn \\
A^\zero(1^-;2_q^+,3_\bq^-;4_H)
&= - \frac{1}{\sqrt{2}}\,\frac{\ang{13}^2}{\ang{23}}
= \frac{s_{13}}{\sqrt{2|s_{23}|}}\, e^{\img\Phi_{-(+)H}}
\,,\end{align}
where the (irrelevant) overall phases are given by
\begin{equation}
e^{\img\Phi_{+(+)H}} = \frac{[12]}{\ang{12}} \frac{\sqrt{|s_{23}|}}{[23]}
\,,\qquad
e^{\img\Phi_{-(+)H}} = \frac{\ang{13}}{[13]} \frac{\sqrt{|s_{23}|}}{\ang{23}}
\,.\end{equation}
The divergent parts of the one-loop amplitudes are
\begin{equation}
A_\div^\one(1^\pm;2_q^+,3^-_\bq; 4_H)
= A^\zero(1^\pm;2_q^+,3^-_\bq; 4_H) \biggl\{
-\frac{1}{\eps^2} (C_A + 2C_F) + \frac{1}{\eps} \Bigl[C_A(L_{12} + L_{13} - L_{23}) + C_F(2L_{23} - 3)
- \frac{\beta_0}{2} \Bigr] \biggr\}
\,.\end{equation}
The finite parts of the $gq\bq H$ amplitudes, which enter the matching coefficient $\vC_{\pm(+)}$ at one loop are
\begin{align}
A_\fin^\one(1^+;2_q^+,3_\bq^-;4_H)
&= A^\zero(1^+;2_q^+,3_\bq^-;4_H) \biggl\{
   g(s_{12}, s_{13}, s_{23}, m_H^2, \mu) + (C_F - C_A)\, \frac{s_{23}}{s_{12}} \biggr\}
\,,\nn\\
A_\fin^\one(1^-;2_q^+,3_\bq^-;4_H)
&= A^\zero(1^-;2_q^+,3_\bq^-;4_H) \biggl\{
   g(s_{12}, s_{13}, s_{23}, m_H^2, \mu) + (C_F - C_A)\, \frac{s_{23}}{s_{13}} \biggr\}
\,,
\end{align}
where we have extracted the common function
\begin{align}
g(s_{12}, s_{13}, s_{23}, m_H^2, \mu)
&= C_A\biggl[-\frac{1}{2}(L_{12}^2 + L_{13}^2 - L_{23}^2) + L_{12/H} L_{13/H}
  - (L_{12/H} + L_{13/H}) L_{23/H}
  -2\Li_2\Bigl(1-\frac{s_{23}}{m_H^2}\Bigr)
\nn \\ & \quad
  + \frac{22}{3} +\frac{\pi^2}{4} \biggr]
+ C_F\biggl[ -L_{23}^2 + 3L_{23} - 2 L_{12/H} L_{13/H}
  - 2\Li_2\Bigl(1-\frac{s_{12}}{m_H^2}\Bigr) - 2\Li_2\Bigl(1-\frac{s_{13}}{m_H^2}\Bigr)
\nn \\ & \quad
  - 11 + \frac{\pi^2}{2} \biggr]
+ \beta_0 \Bigl(-L_{23} + \frac{5}{3}\Bigr)
\,.\end{align}

\subsection{\boldmath $H+2$ Jets}
\label{app:H2amplitudes}

The full set of tree-level helicity amplitudes for $H+2$ jets in the $m_t\to\infty$ limit were calculated in Ref.~\cite{Kauffman:1996ix}, and all amplitudes below are taken from there. We expand the amplitudes $A$, $B$, in the decomposition of \eq{qqQQH_QCD}, \eq{ggqqH_color}, and \eq{ggggH_color}, as
\begin{align}
A &= \frac{2T_F\alpha_s(\mu)}{3\pi v}\,[g_s(\mu)]^2\,
\sum_{n=0}^\infty A^{(n)} \Bigl(\frac{\alpha_s(\mu)}{4\pi}\Bigr)^n
\,, \nonumber \\
B&= \frac{2T_F\alpha_s(\mu)}{3\pi v}\,[g_s(\mu)]^2\,
\sum_{n=0}^\infty B^{(n)} \Bigl(\frac{\alpha_s(\mu)}{4\pi}\Bigr)^n
\,.\end{align}
For simplicity, we only give explicit results for the tree-level amplitudes in this appendix. To reduce the length of expressions, we use the kinematic variables $s_{ijk}$ defined by
\begin{equation}
s_{ijk}=(p_i+p_j+p_k)^2
= s_{ij} + s_{ik} + s_{jk}
\,.\end{equation}
The $H+2$ jets process is nonplanar, which means that we cannot remove all the relative phases in the amplitudes. It is therefore most convenient to keep all expressions in spinor helicity notation. We will explicitly demonstrate an example of the phases which appear in \eqs{nonplanar_phase}{nonplanar_phase2}.

\subsubsection{$q\bq\, q'\bq' H$ and $q\bq\, q\bq H$}\label{app:qqqqH}

The tree-level amplitudes entering the Wilson coefficients $\vC_{(+;\pm)}$ and $\vC_{(+\pm)}$ in \eqs{qqQQH_coeffs}{qqqqH_coeffs} are
\begin{align}
A^\zero(1_q^+,2_\bq^-;3_{q'}^+,4_{\bq'}^-;5_H)
= - B^\zero(1_q^+,2_\bq^-;3_{q'}^+,4_{\bq'}^-;5_H)
&= \frac{1}{2}\biggl[\frac{\ang{24}^2}{\ang{12}\ang{34}} + \frac{[13]^2}{[12][34]}\biggr]
\,,\nn\\
A^\zero(1_q^+,2_\bq^-;3_{q'}^-,4_{\bq'}^+;5_H)
= -B^\zero(1_q^+,2_\bq^-;3_{q'}^-,4_{\bq'}^+;5_H)
&= - \frac{1}{2}\biggl[\frac{\ang{23}^2}{\ang{12}\ang{34}} + \frac{[14]^2}{[12][34]}\biggr]
\,,\end{align}

\subsubsection{$ggq\bar q H$}\label{app:ggqqH}

The tree-level amplitudes entering the Wilson coefficients $\vC_{+-(+)}$, $\vC_{++(+)}$, and $\vC_{--(+)}$ in \eq{ggqqH_coeffs} are
\begin{align}
A^\zero(1^+,2^-;3_q^+,4_{\bar q}^-;5_H)
&= \frac{\ang{24}^3}{\ang{12}\ang{14}\ang{34}}- \frac{[13]^3}{[12][23][34]}
\,,\nn \\
A^\zero(2^-,1^+;3_q^+,4_{\bar q}^-;5_H)
&= \frac{[13]^2[14]}{[12][24][34]}-\frac{\ang{23}\ang{24}^2}{\ang{12}\ang{13}\ang{34}}
\,,\nn \\
A^\zero(1^+,2^+;3_q^+,4_{\bar q}^-;5_H)
&= -\frac{[1|2+3|4\rangle^2[23]}{s_{234}\ang{24}} \Bigl(\frac{1}{s_{23}} + \frac{1}{s_{34}}\Bigr)
   + \frac{[2|1+3|4\rangle^2[13]}{s_{134}s_{34}\ang{14}}
   - \frac{[3|1+2|4\rangle^2}{\ang{12}\ang{14}\ang{24}[34]}
\,,\nn \\
A^\zero(1^-,2^-;3_q^+,4_{\bar q}^-;5_H)
&= \frac{\langle2|1+4|3]^2\ang{14}}{s_{134}[13]} \Big(\frac{1}{s_{14}} + \frac{1}{s_{34}}\Big)
- \frac{\langle1|2+4|3]^2 \ang{24}}{s_{234}s_{34}[23]}
+ \frac{\langle 4|2+1|3]^2}{[12][13][23]\ang{34}}
\,.\end{align}
In these expressions we have eliminated the Higgs momentum, $p_5$, using momentum conservation, so that all momenta appearing in the above expressions are lightlike. We have also used an extended spinor-helicity sandwich, defined by $[ i|j+k|l\rangle=[ i|j|l\rangle+[ i|k|l\rangle$ to simplify notation.

All the $B$ amplitudes vanish at tree level,
\begin{align}
B^\zero(1^+,2^-;3_q^+,4_{\bar q}^-;5_H)
= B^\zero(1^+,2^+;3_q^+,4_{\bar q}^-;5_H)
= B^\zero(1^-,2^-;3_q^+,4_{\bar q}^-;5_H)
= 0
\,.\end{align}

\subsubsection{$ggggH$}\label{app:ggggH}

The tree-level amplitudes entering the Wilson coefficients $\vC_{++--}$, $\vC_{+++-}$, and $\vC_{++++}$ in \eq{ggggH_coeffs} are
\begin{align}
A^\zero(1^+,2^+,3^+,4^+;5_H)
&= \frac{-2 M_H^4}{\ang{12}\ang{23}\ang{34}\ang{41}}
 \,, \nn\\
A^\zero(1^+,2^+,3^+,4^-;5_H)
 &=2 \bigg[\frac{[1|2+3|4\rangle^2[23]^2}{s_{234}s_{23}s_{34}}
+\frac{[2|1+3|4\rangle^2[13]^2}{s_{134}s_{14}s_{34}}
+\frac{[3|1+2|4\rangle^2[12]^2}{s_{124}s_{12}s_{14}}
\nn \\ & \quad
+\frac{[13]}{[41]\ang{12}\ang{23}[34]}\bigg(    \frac{s_{12}[1|2+3|4\rangle}{\ang{34}}
+ \frac{s_{23}[3|1+2|4\rangle}{\ang{41}}+[13]s_{123}\bigg) \bigg]
 \,,\nn \\
A^\zero(1^+,2^+,3^-,4^-;5_H)
&= 2 \bigg[\frac{[12]^4}{[12][23][34][41]}+\frac{\ang{34}^4}{\ang{12}\ang{23}\ang{34}\ang{41}}\bigg]\,,
\nn \\
A^\zero(1^+,4^-,2^+,3^-;5_H)
&= 2 \bigg[\frac{[12]^4}{[13][14][23][24]}+\frac{\ang{34}^4}{\ang{13}\ang{14}\ang{23}\ang{24}} \bigg]
\,. 
\end{align}

To illustrate the relative phases that appear in these amplitudes, we can rewrite the amplitude $A^\zero(1^+,2^+,3^-,4^-;5_H)$ in terms of the Lorentz invariants $s_{ij}$
\begin{align} \label{eq:nonplanar_phase}
A^\zero(1^+,2^+,3^-,4^-;5_H)
&= 2 e^{\img \Phi_{++--H}}\, \bigg[\frac{s_{12}^2}{\sqrt{|s_{12}s_{23}s_{34}s_{14}|}} + e^{\img \varphi} \frac{s_{34}^2}{\sqrt{|s_{12}s_{23}s_{34}s_{14}|}}  \bigg]\,,
\end{align}
with
\begin{align} \label{eq:nonplanar_phase2}
 \varphi &= -2 \bt\, \mathrm{arg} \bigg\{\frac{\img \sqrt{s_{23}}[-s_{12} s_{34} + s_{13}s_{24} + s_{14} s_{23} -\img( \sqrt{\al}+2\sqrt{s_{13}} \sqrt{s_{23}} s_{14})]}{-s_{12} s_{34} (\sqrt{s_{13}} - \img \sqrt{s_{23}}) + (s_{13} s_{24} - s_{14} s_{23} + \img \sqrt{\al})(\sqrt{s_{13}}+ \img \sqrt{s_{23}})}\bigg\}
 \,,\nn \\
 \al &= 16 (\eps_{\mu \nu \rho \si} p_1^\mu p_2^\nu p_3^\rho p_4^\si)^2
 = 4 s_{13} s_{14} s_{23} s_{24} - (s_{12} s_{34} - s_{13}s_{24} - s_{14}s_{23})^2 \geq 0
 \,,\nn \\
 \bt &= \mathrm{sgn}(\eps_{\mu \nu \rho \si} p_1^\mu p_2^\nu p_3^\rho p_4^\si)
 \,.
\end{align}
The branch cut of the square root is given by the usual prescription,
$\sqrt{s_{ij}} \equiv \sqrt{s_{ij}+\img 0} = \img \sqrt{|s_{ij}|}$ if $s_{ij}<0$. Our convention for the antisymmetric Levi-Civita tensor is $\eps_{0123} = -1$. For this process we can choose a frame where all but one of the momenta $p_1$ through $p_4$ lie in a plane (with $p_5$ determined by momentum conservation).  The phase $\varphi$ is needed to determine the momentum of the nonplanar momentum and the sign $\bt$ resolves which side of the plane this particle is on, which is not captured by the $s_{ij}$ (because they are symmetric with respect to a reflection about the plane). We note the simplicity of the spinor-helicity expression as compared with the explicit expression for the phases.

\section{Helicity Amplitudes for Vector Boson $+$ Jets}
\label{app:Zamplitudes}

In this appendix we give all required partial amplitudes for the vector boson $+$ jets processes discussed in \sec{vec}. For each of the amplitudes $A_{q,v,a}$, $B_{q,v,a}$ defined in \sec{vec}, we split the amplitude into its IR-divergent and IR-finite parts,
\begin{equation}
X = X_{\div} + X_{\fin}
\,,\end{equation}
where $X$ stands for any of $A_{q,v,a}$ and $B_{q,v,a}$. For the logarithms we use the notation
\begin{equation}
L_{ij} = \ln \Bigl(-\frac{s_{ij}}{\mu^2} - \img 0\Bigr)
\,,\qquad
L_{ij/kl} = L_{ij} - L_{kl}
= \ln \Bigl(-\frac{s_{ij}}{\mu^2} - \img 0\Bigr) - \ln \Bigl(-\frac{s_{kl}}{\mu^2} - \img 0\Bigr)
\,.\end{equation}

\subsection{\boldmath $V+0$ Jets}
\label{app:V0amplitudes}

In this section we give the amplitudes $A_{q,v,a}$ for $V+0$ jets. For each partonic channel, we expand the amplitudes as
\begin{equation}
X = \sum_{n=0}^\infty X^{(n)} \Bigl(\frac{\alpha_s(\mu)}{4\pi}\Bigr)^n
\,.\end{equation}
where $X$ stands for any of $A_{q,v,a} $.
The tree-level and one-loop helicity amplitudes entering the matching coefficient in \eq{qqV_coeffs} are given by
\begin{align}
A_q^\zero(1_q^+,2_{\bar q}^-;3_\ell^+,4_{\bar \ell}^-) &= -2\img\, \frac{[13]\ang{24}}{s_{12}}
 \,, \nn\\
A_{q,\div}^\one(1_q^+,2_{\bar q}^-;3_\ell^+,4_{\bar \ell}^-)
&= A_q^\zero(1_q^+,2_{\bar q}^-;3_\ell^+,4_{\bar \ell}^-)\, C_F \biggl[
-\frac{2}{\eps^2} + \frac{1}{\eps}\,\big(2L_{12} - 3\big) \biggr]\, ,
\nn \\
A_{q,\fin}^\one(1_q^+,2_{\bar q}^-;3_\ell^+,4_{\bar \ell}^-)
&= A_q^\zero(1_q^+,2_{\bar q}^-;3_\ell^+,4_{\bar \ell}^-)\, C_F \biggl[ -L_{12}^2 + 3 L_{12} - 8 + \frac{\pi^2}{6} \biggr]\, ,
\nn \\
A_v^{(0)}&=A_v^{(1)}=A_a^{(0)}=A_a^{(1)}=0\,.
\end{align}

\subsection{\boldmath $V+1$ Jet}
\label{app:V1amplitudes}

In this section we give the amplitudes $A_{q,v,a}$ for $V+1$ jets. Each amplitude is expanded as
\begin{equation}
X = g_s(\mu) \sum_{n=0}^\infty X^{(n)} \Bigl(\frac{\alpha_s(\mu)}{4\pi}\Bigr)^n
\end{equation}
where $X$ stands for any of $A_{q,v,a} $.
The tree-level and one-loop helicity amplitudes for $V+1$ jets were calculated in Refs.~\cite{Giele:1991vf, Arnold:1988dp, Korner:1990sj, Bern:1997sc}. We use the results given in Ref.~\cite{Bern:1997sc}, which uses $T_F=1$ and $g_s T^a/\sqrt{2}$ for the $q\bar{q}g$ coupling. We can thus convert to our conventions by replacing $T^a\to \sqrt{2} T^a$, and identifying $1/N = C_A - 2C_F$ and $N = C_A$. The one-loop amplitudes are given in the FDH scheme in Ref.~\cite{Bern:1997sc}, which we convert to the HV scheme using \eqs{DR_HV_UV}{DR_HV_IR}.

The tree-level amplitudes entering the matching coefficient $\vC_{x+(+;+)}$ in \eq{gqqV_coeffs} is given by
\begin{align}
A_q^\zero(1^+; 2_q^+,3_{\bar q}^-;4_\ell^+,5_{\bar \ell}^-)
&= -2\sqrt{2}  \frac{\ang{35}^2}{\ang{12}\ang{13}\ang{45}}
\nn \\
A_v^\zero &=  A_{a}^\zero=0
\,.\end{align}
The divergent part of the one-loop helicity amplitude is given by
\begin{align}
A_{q,\div}^\one(1^+\!; 2_q^+,3_{\bar q}^-;4_\ell^+,5_{\bar \ell}^-)
&= A_q^\zero(1^+\!; 2_q^+,3_{\bar q}^-;4_\ell^+,5_{\bar \ell}^-)
\nn \\ & \quad \times
 \biggl\{\! -\frac{1}{\eps^2}(C_A + 2C_F)  + \frac{1}{\eps}\Bigl[C_A(L_{12}+L_{13}-L_{23}) + C_F(2L_{23} - 3)- \frac{\bt_0}{2} \Bigr] \biggr\}
.\end{align}
The finite parts entering the matching coefficients at one loop are
\begin{align}
A_{q,\fin}^\one(1^+; 2_q^+,3_{\bar q}^-;4_\ell^+,5_{\bar \ell}^-)
&= A_q^\zero(1^+,2_q^+,3_{\bar q}^-;4_\ell^+,5_{\bar \ell}^-)
 \nn \\ & \quad\times
 \biggl\{ \frac{C_A}{2} \biggl(-L_{12}^2-L_{13}^2 + 3 L_{13} - 7 + \frac{\pi^2}{3} \biggr)
 + \Bigl(C_F - \frac{C_A}{2}\Bigr) \Bigl(-L_{23}^2+ 3 L_{45} - 8 + \frac{\pi^2}{6} \Bigr)
 \nn \\ & \quad
 + C_A \biggl[
   - \text{Ls}_{-1}\Bigl(\frac{s_{12}}{s_{45}},\frac{s_{13}}{s_{45}}\Bigr)
   + \frac{\langle 3|24|5\rangle}{\ang{35} s_{45}} \text{L}_0 \Bigl(\frac{s_{13}}{s_{45}} \Bigr)
   - \frac{1}{2} \frac{\langle 3|24|5\rangle^2}{\ang{35}^2 s_{45}^2} \text{L}_1 \Bigl(\frac{s_{13}}{s_{45}}
  \Bigr)  \biggr]
 \nn \\ & \quad
 + (C_A-2C_F) \biggl[
   \text{Ls}_{-1}\Bigl(\frac{s_{12}}{s_{45}},\frac{s_{23}}{s_{45}}\Bigr)
   + \biggl(\frac{\ang{25}^2\ang{13}^2}{\ang{12}^2\ang{35}^2}
   - \frac{\ang{15}\ang{23} + \ang{13}\ang{25}}{\ang{12}\ang{35}} \biggr)
      \text{Ls}_{-1}\Bigl(\frac{s_{13}}{s_{45}},\frac{s_{23}}{s_{45}}\Bigr)
\nn \\ & \qquad
   + \frac{2[12]\ang{25}\ang{13}}{\ang{35}s_{45}} \text{L}_0 \Bigl( \frac{s_{13}}{s_{45}} \Bigr)
   +\frac{\ang{25}^2   [2|1|3\rangle\ang{13}}{\ang{12}\ang{35}^2 s_{13}} \text{L}_0 \Bigl(\frac{s_{45}}{s_{13}}\Bigr)
   - \frac{ \langle 3|21|5 \rangle   \ang{25} \ang{13}}{\ang{12}\ang{35}^2 s_{23}} \text{L}_0 \Bigl(\frac{s_{45}}{s_{23}}\Bigr)
\nn \\ & \qquad
   - \frac{\ang{13}^2 [1|2|5\rangle^2}{2\ang{35}^2s_{13}^2} \text{L}_{1} \Bigl(\frac{s_{45}}{s_{13}} \Bigr)
   - \frac{ [1|2|3\rangle [4|1|5\rangle \ang{45}\ang{13}}{\ang{35}^2 s_{23}^2} \text{L}_1\Bigl(\frac{s_{45}}{s_{23}}\Bigr)
\nn \\ & \qquad
   + \frac{[14]([12][34]+[13][24])\ang{13}\ang{45}}{2\ang{35}^2[13][23][45]} \biggr] \biggr\}
\,, \nn\\
A_{a}^\one(1^+; 2_q^+,3_{\bar q}^-;4_\ell^+,5_{\bar \ell}^-)
&= 4\sqrt{2} T_F\, [12][14]\ang{35} \biggl[ \frac{1}{s_{45}^2} \text{L}_1\Big(\frac{s_{23}}{s_{45}}\Big) - \frac{1}{12 s_{45} m_t^2}\biggr]
\,, \nn \\
A_v^{(1)} &=0
\,.\end{align}
The contributions from virtual top quark loops are calculated in an expansion in $1/m_t$ to order $1/m_t^2$ in Ref.~\cite{Bern:1997sc}, hence the divergent behavior of $A_{a}^\one$ as $m_t \to 0$. To reduce the length of the expressions, we have used the commonly defined functions
\begin{align}
 \text{L}_0(r) = \frac{\ln r}{1-r}
 \,, \qquad
 \text{L}_1(r) = \frac{\text{L}_0(r) +1}{1-r}
 \,, \qquad
 \text{Ls}_{-1}(r_1,r_2) = \Li_2(1-r_1) + \Li_2(1-r_2) + \ln r_1\, \ln r_2 - \frac{\pi^2}{6}
 \,.
\end{align}
The proper branch cut of logarithms follows from the prescriptions $s_{ij} \to s_{ij} + \img 0$.
The proper branch cut of the dilogarithm follows from that of the logarithm through the identity
\begin{equation}
  \text{Im}\bigr [\Li_2(1-r) \bigr ]= -\ln(1-r)\ \text{Im}\bigr [ \ln r \bigr ]
\,.\end{equation}

\subsection{\boldmath $V+2$ Jets}
\label{app:V2amplitudes}

In this section, we give the amplitudes $A_{q,v,a}$, $B_{q,v,a}$ for $V+2$ jets. Each amplitude is expanded as
\begin{equation}\label{eq:X_series}
X = [g_s(\mu)]^2 \sum_{n=0}^\infty X^{(n)} \Bigl(\frac{\alpha_s(\mu)}{4\pi}\Bigr)^n
\end{equation}
where $X$ stands for any of $A_{q,v,a}$ or $B_{q,v,a}$. We also define the kinematic variables $s_{ijk}$ as
\begin{equation}
s_{ijk}=(p_i+p_j+p_k)^2=s_{ij}+s_{ik}+s_{jk}
\,.\end{equation}
The one-loop helicity amplitudes for  $q' \bar q' q\bar q\, V$ and $q \bar q\, q\bar q\, V$ were calculated in Ref.~\cite{Bern:1996ka}.  The one-loop helicity amplitudes for $gg\,q\bar q\,V$ were calculated in Ref.~\cite{Bern:1997sc}, which also gives compact expressions for the four-quark amplitudes, which we use here. The contributions from virtual top quark loops are calculated in an expansion in $1/m_t$ to order $1/m_t^2$ in Ref.~\cite{Bern:1997sc}.

Reference~\cite{Bern:1997sc} uses $T_F=1$ and $g_s T^a/\sqrt{2}$ for the $q\bar{q}g$ coupling. We can thus convert to our conventions by replacing $T^a\to \sqrt{2} T^a$, and identifying $1/N = C_A - 2C_F$ and $N = C_A$. The one-loop amplitudes are given in the FDH scheme in Ref.~\cite{Bern:1997sc}, which we convert to the HV scheme using \eqs{DR_HV_UV}{DR_HV_IR}.

\subsubsection{$q' \bar q' q\bar q\, V$ and $q \bar q\, q\bar q\, V$ }
\label{app:V2qqqq}
 
The tree-level amplitudes for $q' \bar q' q\bar q\, V$ and $q \bar q\, q\bar q\, V$ entering the Wilson coefficients in \eqs{qqQQV_coeffs}{qqqqV_coeffs} are given by
\begin{align}\label{eq:ap_qqqqV_tree}
A_q^\zero(1_{q'}^+,2_{\bar q'}^-;3_q^+,4_{\bar q}^-;5_\ell^+,6_{\bar \ell}^-)
&= - B_q^\zero(1_{q'}^+,2_{\bar q'}^-;3_q^+,4_{\bar q}^-;5_\ell^+,6_{\bar \ell}^-)
\nn \\
&= \frac{2}{s_{12}s_{56}}
  \biggl[ \frac{[13] \ang{46} (\ang{12}[15] - \ang{23}[35])}{s_{123}}
  + \frac{\ang{24}[35] ([12]\ang{26} + [14]\ang{46})}{s_{124}} \biggr]
\,, \nn \\
A_q^\zero(1_{q'}^-,2_{\bar q'}^+;3_q^+,4_{\bar q}^-;5_\ell^+,6_{\bar \ell}^-)
&= - B_q^\zero(1_{q'}^-,2_{\bar q'}^+;3_q^+,4_{\bar q}^-;5_\ell^+,6_{\bar \ell}^-)
\nn \\
&= \frac{2}{s_{12}s_{56}}
   \biggl[ \frac{[23] \ang{46}(\ang{12}[25]+\ang{13}[35])}{s_{123}}
   + \frac{\ang{14}[35]([12]\ang{16}-[24]\ang{46})}{s_{124}} \biggr]
\,, \nn \\
A_v^\zero &= A_a^\zero = B_v^\zero = B_a^\zero = 0
\,.\end{align}

Due to the length of the one-loop $q'\bar q' q \bar q\, V$ amplitudes, we only show how to translate the decomposition of the amplitude in Ref.~\cite{Bern:1997sc} to our notation. The one-loop amplitudes are given in terms of the bare partial amplitudes $A_{i;j}(3_q,2_{\bar Q},1_Q,4_{\bar q})$ of Ref.~\cite{Bern:1997sc} as
\begin{align}\label{eq:convert_to_BDK_qqqq}
A_{q}^\one(1_{q'},2_{\bar q'};3_q,4_{\bar q};6_\ell^+,5_{\bar \ell}^-)
&= -\img\, 32\pi^2\,  N\, A_{6;1}(3_q,2_{\bar Q},1_Q,4_{\bar q})
   - \Bigl(\frac{\beta_0}{\eps}+2C_F-\frac{1}{3}C_A\Bigr) A_{q}^\zero(1_{q'},2_{\bar q'};3_q,4_{\bar q};6_\ell^+,5_{\bar \ell}^-)
\,, \nn \\
B_{q}^\one(1_{q'},2_{\bar q'};3_q,4_{\bar q};6_\ell^+,5_{\bar \ell}^-)
&= -\img\, 32\pi^2\, N  \, A_{6;2}(3_q,2_{\bar Q},1_Q,4_{\bar q})
   - \Bigl(\frac{\beta_0}{\eps}+2C_F-\frac{1}{3}C_A \Bigr) B_{q}^\zero(1_{q'},2_{\bar q'};3_q,4_{\bar q};6_\ell^+,5_{\bar \ell}^-)
\,, \nn \\
A_{a}^\one(1_{q'},2_{\bar q'};3_q,4_{\bar q};6_\ell^+,5_{\bar \ell}^-)
&= - B_{a}^\one(1_{q'},2_{\bar q'};3_q,4_{\bar q};6_\ell^+,5_{\bar \ell}^-)
= -\img\, 32\pi^2  \, A_{6;3}(3_q,2_{\bar Q},1_Q,4_{\bar q})
\,, \nn \\
A_{v}^\one &= B_{v}^\one = 0
\,.\end{align}
The overall factor $-\img\, 32\pi^2$ is due to our different normalization conventions. We have not included helicity labels, as these relations are true for all helicity combinations. Note that the partial amplitudes $A_{i;j}$ do not include labels for the lepton momenta, which are implicitly taken as $6_\ell^+$, $5_{\bar \ell}^-$. The terms in the first two lines proportional to $A_q^\zero$ and $B_q^\zero$ come from the UV renormalization and switching from FDH to HV.

\subsubsection{$gg\,q\bar q\, V$}
\label{app:V2ggqq}

The tree-level amplitudes for $gg\,q\bar q\, V$ entering the matching coefficients in \eq{ggqqV_coeffs} are given by
\begin{align}\label{eq:ap_ggqqV_tree}
  A_q^\zero(1^+,2^+;3_q^+,4_{\bar q}^-;5_\ell^+,6_{\bar \ell}^-)
 &= -4 \frac{\ang{46}^2}{\ang{12}\ang{13}\ang{24}\ang{56}}
 \,, \nn \\
  A_q^\zero(1^+,2^-;3_q^+,4_{\bar q}^-;5_\ell^+,6_{\bar \ell}^-)
 &= \frac{4}{s_{12}s_{56}} \bigg[\frac{[13] \ang{23} \ang{46} (\ang{23}[35]-\ang{12}[15])}{\ang{13} s_{123}}+
 \frac{\ang{24}[35] [14] ([12]\ang{26} + [14]\ang{46})}{[24]s_{124}}
 \nn \\ & \quad
 +
 \frac{([12]\ang{26}+[14]\ang{46})(\ang{23}[35]-\ang{12}[15])}{\ang{13}[24]} \bigg]
 \,, \nn \\
 A_q^\zero(1^-,2^+;3_q^+,4_{\bar q}^-;5_\ell^+,6_{\bar \ell}^-)
 &=\frac{4}{s_{12}s_{56}} \bigg [ \frac{[23]^2 \ang{46}(\ang{12}[25] +\ang{13}[35]) }{ [13] s_{123}  }
    + \frac{\ang{14}^2[35]([12]\ang{16} - [24]\ang{46})}{\ang{24} s_{124}} \nn \\
 &\quad +\frac{[23] \ang{14} [35] \ang{46}}{[13] \ang{24}}     \bigg]
 \,, \nn \\
A_v^\zero &= A_a^\zero= B_q^\zero = B_v^\zero = B_a^\zero = 0
 \,.
\end{align}

Due to the length of the one-loop $gg\,q\bar q\, V$ amplitudes, we again only show how to translate the decomposition of the amplitude in Ref.~\cite{Bern:1997sc} to our notation. The one-loop amplitudes are given in terms of the bare partial amplitudes $A_{i;j}(3_q,1,2,4_{\bar q})$, $A_{i;j}^\mathrm{v}(3_q,4_{\bar q}; 1,2)$, and $A_{i;j}^\mathrm{ax}(3_q,4_{\bar q};1,2)$ of Ref.~\cite{Bern:1997sc} as
\begin{align}\label{eq:convert_to_BDK_ggqq}
A_{q}^\one(1,2;3_q,4_{\bar q};6_\ell^+,5_{\bar \ell}^-)
&= -\img\, 64\pi^2\, N\, A_{6;1}(3_q,1,2,4_{\bar q})
   - \Bigl( \frac{\beta_0}{\eps}+C_F\Bigr)\, A_{q}^\zero(1,2;3_q,4_{\bar q};6_\ell^+,5_{\bar \ell}^-)
\,, \nn \\
B_{q}^\one(1,2;3_q,4_{\bar q};6_\ell^+,5_{\bar \ell}^-)
&= -\img\, 64\pi^2\, A_{6;3}(3_q, 4_{\bar q}; 1,2)
\,, \nn \\
A_{v}^\one(1,2;3_q,4_{\bar q};6_\ell^+,5_{\bar \ell}^-)
&= -\img\, 64\pi^2\, A_{6;4}^\mathrm{v}(3_q,4_{\bar q}; 1,2)
\,, \nn \\
B_{v}^\one(1,2;3_q,4_{\bar q};6_\ell^+,5_{\bar \ell}^-)
&=+\img\, 64\pi^2\, \frac{2}{N}\,  A_{6;4}^\mathrm{v}(3_q, 4_{\bar q}; 1,2)
\,, \nn \\
A_{a}^\one(1,2;3_q,4_{\bar q};6_\ell^+,5_{\bar \ell}^-)
&= -\img\, 64\pi^2\, A_{6;4}^\mathrm{ax}(3_q,4_{\bar q}; 1,2)
\,, \nn \\
B_{a}^\one(1,2;3_q,4_{\bar q};6_\ell^+,5_{\bar \ell}^-)
&= -\img\, 64\pi^2\, \frac{1}{N }\,
\bigl[A_{6;5}^\mathrm{ax}(3_q, 4_{\bar q}; 1,2)- A_{6;4}^\mathrm{ax}(3_q, 4_{\bar q}; 1,2) - A_{6;4}^\mathrm{ax}(3_q, 4_{\bar q}; 2,1) \bigr]
\,.\end{align}
The overall factor $-\img\, 64\pi^2$ is due to our different normalization conventions. We have not included helicity labels, as these relations are true for all helicity combinations. Note that the partial amplitudes $A_{i;j}$ do not include labels for the lepton momenta, which are implicitly taken as $6_\ell^+$, $5_{\bar \ell}^-$. The term in the first line proportional to $A_q^\zero$ comes from the UV renormalization and switching from FDH to HV.

\section{\boldmath Helicity Amplitudes for $pp\to$ Jets}
\label{app:helicityamplitudes}

\subsection{\boldmath $pp \to 2$ Jets}\label{app:pp2jets_app}

In this appendix we give explicit expressions for all partial amplitudes that are required in Eqs.~\eqref{eq:qqQQ_coeffs}, \eqref{eq:qqqq_coeffs}, \eqref{eq:ggqq_coeffs}, and \eqref{eq:gggg_coeffs}, for the various partonic channels of the $pp \to 2$ jets process. 
Since this process is planar, we can write all amplitudes for a given set of helicities with a common overall phase extracted, which is determined by the phases of the external particles. In this way, we do not need to worry about relative phases between the Wilson coefficients for different color structures when they mix under renormalization. The cross section does not depend on this overall phase. This simplifies the numerical implementation considerably for this process, as it avoids having to implement the complex spinor algebra. To extract the overall phase from the amplitudes, the following relations for the relative phases between the spinor products are useful:
\begin{equation} \label{eq:phases}
\frac{\ang{12}}{[34]} = \frac{\ang{34}}{[12]} = \frac{\ang{14}}{[23]} = \frac{\ang{23}}{[14]}
= - \frac{\ang{13}}{[24]} = - \frac{\ang{24}}{[13]}
\,.\end{equation}
These relations follow from \eq{spinormomcons} with $n = 4$.

We split the partial amplitudes into their IR-divergent and IR-finite parts,
\begin{equation}
A = A_\div + A_\fin
\,,\qquad
B = B_\div + B_\fin
\,,\end{equation}
where the IR-finite parts enter the matching coefficients. We expand the amplitudes and Wilson coefficients in powers of $\alpha_s(\mu)$ as
\begin{equation} \label{eq:asexp}
X = [g_s(\mu)]^2 \sum_{n=0}^\infty X^{(n)} \Bigl(\frac{\alpha_s(\mu)}{4\pi}\Bigr)^n
\,,\end{equation}
where $X$ stands for any of $A_{\div,\fin}$, $B_{\div,\fin}$, and $X^\zero$ and $X^\one$ are the tree-level and one-loop contributions, respectively. For simplicity, we drop the subscript ``$\fin$'' for those amplitudes that have no divergent parts, e.g., for the tree-level amplitudes $A^\zero_\div = 0$ and $A^\zero_\fin \equiv A^\zero$. For the logarithms we use the notation
\begin{equation}
L_{ij} = \ln \Bigl(-\frac{s_{ij}}{\mu^2} - \img 0\Bigr)
\,,\qquad
L_{ij/kl} = L_{ij} - L_{kl}
= \ln \Bigl(-\frac{s_{ij}}{\mu^2} - \img 0\Bigr) - \ln \Bigl(-\frac{s_{kl}}{\mu^2} - \img 0\Bigr)
\,.\end{equation}

\subsubsection{$q\bq \,q'\bq'$ and $q\bq\, q\bq$ }
\label{app:qqqqamplitudes}

Here we list all partial amplitudes up to one loop entering the Wilson coefficients in \eqs{qqQQ_coeffs}{qqqq_coeffs}. The one-loop helicity amplitudes for $q\bq \,q'\bq'$ and $q\bq\, q\bq$ were first calculated in Ref.~\cite{Kunszt:1993sd}, and the two-loop helicity amplitudes were computed in Refs.~\cite{Glover:2004si,Freitas:2004tk}. We find agreement between the one-loop results of Refs.~\cite{Glover:2004si} and \cite{Freitas:2004tk}, from which we take our results.\footnote{Note that there is a minor disagreement here with the earlier calculation in Ref.~\cite{Kunszt:1993sd}, presumably due to typos. Specifically, in Ref.~\cite{Kunszt:1993sd} the factors $\big( \log^2 \frac{s_{14}}{s_{12} }+\pi^2\big)$ and $\big( \log^2 \frac{s_{14}}{s_{13} }+\pi^2\big)$ in Eqs.~(5.10) and (5.12) respectively, must be swapped to achieve agreement with the results of Refs.~\cite{Glover:2004si,Freitas:2004tk,Kelley:2010fn}.   Reference~\cite{Freitas:2004tk} also has a minor typo, having a flipped overall sign for the IR-divergent terms.}  Our one-loop matching coefficients agree with the calculation of Ref.~\cite{Kelley:2010fn}.

The tree-level amplitudes are
\begin{align}
A^\zero(1_q^+,2_\bq^-; 3_{q'}^+,4_{\bq'}^-)
&=-B^\zero(1_q^+,2_\bq^-; 3_{q'}^+,4_{\bq'}^-)= -\frac{\ang{24}[13]}{s_{12}} = \frac{s_{13}}{s_{12}}\,e^{\img\Phi_{(+;+)}}\,,
\nn\\
A^\zero(1_q^+,2_\bq^-; 3_{q'}^-,4_{\bq'}^+)
&=-B^\zero(1_q^+,2_\bq^-; 3_{q'}^-,4_{\bq'}^+)= -\frac{\ang{23}[14]}{s_{12}} = \frac{s_{14}}{s_{12}}\,e^{\img\Phi_{(+;-)}}
\,,\end{align}
where the phases are given by
\begin{equation}
e^{\img\Phi_{(+;+)}} = \frac{\ang{24}}{\ang{13}}
\,, \qquad
e^{\img\Phi_{(+;-)}} = \frac{\ang{23}}{\ang{14}}
\,.\end{equation}
We have chosen to express all the one-loop amplitudes in terms of $A^\zero(1_q^+,2_\bq^-; 3_{q'}^+,4_{\bq'}^-)$ and $A^\zero(1_q^+,2_\bq^-; 3_{q'}^-,4_{\bq'}^+)$. The divergent parts of the one-loop amplitudes are
\begin{align}
A_\div^\one(1_q^+,2_\bq^-; 3_{q'}^+,4_{\bq'}^-)
&= A^\zero(1_q^+,2_\bq^-; 3_{q'}^+,4_{\bq'}^-) \biggl\{
-\frac{4}{\eps^2}\,C_F  + \frac{2}{\eps}
\bigl[C_F (2L_{12} - 4L_{13/14} - 3) + C_A (L_{13/14} - L_{12/13} )
\bigr] \biggr\}
\,,\nn\\
B_\div^\one(1_q^+,2_\bq^-; 3_{q'}^+,4_{\bq'}^-)
&= A^\zero(1_q^+,2_\bq^-; 3_{q'}^+,4_{\bq'}^-)\, \left \{  \frac{4}{\eps^2} C_F - \frac{2}{\eps}
\bigl[C_F (2L_{12} - 2L_{13/14} - 3) + C_A (L_{13/14} - L_{12/13} )
\bigr]  \right \}
\,,\nn\\[1ex]
A_\div^\one(1_q^+,2_\bq^-; 3_{q'}^-,4_{\bq'}^+)
&= A^\zero(1_q^+,2_\bq^-; 3_{q'}^-,4_{\bq'}^+) \biggl\{
-\frac{4}{\eps^2}\,C_F  + \frac{2}{\eps}
\bigl[C_F (2L_{12} - 4L_{13/14} - 3) + C_A (L_{13/14} - L_{12/13} )
\bigr] \biggr\}
\,,\nn\\
B_\div^\one(1_q^+,2_\bq^-;3_{q'}^-,4_{\bq'}^+)
&= A^\zero(1_q^+,2_\bq^-;3_{q'}^-,4_{\bq'}^+)\, \left \{  \frac{4}{\eps^2} C_F - \frac{2}{\eps}
\bigl[C_F (2L_{12} - 2L_{13/14} - 3) + C_A (L_{13/14} - L_{12/13} )
\bigr]  \right \}
\,.\end{align}
The finite parts entering the Wilson coefficients are
\begin{align}
A_\fin^\one(1_q^+,2_\bq^-; 3_{q'}^+,4_{\bq'}^-)
&= A^\zero(1_q^+,2_\bq^-; 3_{q'}^+,4_{\bq'}^-) \bigl[
f(s_{12}, s_{13}, s_{14}, \mu) + (4C_F-C_A)\, g(s_{12}, s_{13}, s_{14}) \bigr]
\,, \nn\\
B_\fin^\one(1_q^+,2_\bq^-; 3_{q'}^+,4_{\bq'}^-)
&= A^\zero(1_q^+,2_\bq^-; 3_{q'}^+,4_{\bq'}^-) \, \bigl[
4C_F\, L_{12}L_{13/14} - f(s_{12}, s_{13}, s_{14}, \mu) + (C_A - 2C_F)\, g(s_{12}, s_{13}, s_{14}) \bigr]
\,,\nn\\[1ex]
A_\fin^\one(1_q^+,2_\bq^-;3_{q'}^-,4_{\bq'}^+)
&= A^\zero(1_q^+,2_\bq^-;3_{q'}^-,4_{\bq'}^+) \bigl[
f(s_{12}, s_{13}, s_{14}, \mu) + 2(C_A-2C_F)\, g(s_{12}, s_{14}, s_{13}) \bigr]
\,,\nn\\
B_\fin^\one(1_q^+,2_\bq^-;3_{q'}^-,4_{\bq'}^+)
&= A^\zero(1_q^+,2_\bq^-;3_{q'}^-,4_{\bq'}^+) \,\big[
4C_F\, L_{12} L_{13/14} - f(s_{12}, s_{13}, s_{14}, \mu) + 2(C_F-C_A)\, g(s_{12}, s_{14}, s_{13})\bigr]
\,,\nn\\
f(s_{12}, s_{13}, s_{14}, \mu)
&= C_F \Bigl[-2 L_{12}^2 + 2 L_{12} (3 + 4 L_{13/14}) - 16 + \frac{\pi^2}{3} \Bigr]
+ C_A \Bigl(2 L_{12}(L_{12/13} - L_{13/14}) + \frac{10}{3} + \pi^2 \Bigr)
\nn\\ & \quad
- \beta_0 \Bigl(L_{12} - \frac{5}{3} \Bigr)
\,,\nn\\
g(s_{12}, s_{13}, s_{14})
&= \frac{s_{12}}{s_{13}} \biggl[\frac{1}{2} \Bigl(1-\frac{s_{14}}{s_{13}}\Bigr) \Bigl(L_{12/14}^2+\pi^2\Bigr) + L_{12/14}\biggr]
\,.\end{align}

\subsubsection{$gg\,q\bar{q}$}
\label{app:ggqqamplitudes}

The one-loop helicity amplitudes for $gg\,q\bar{q}$ were first calculated in Ref.~\cite{Kunszt:1993sd}, and the two-loop helicity amplitudes were computed in Refs.~\cite{Bern:2003ck,Glover:2003cm}. We take our results from Ref.~\cite{Bern:2003ck}, converted to our conventions.%
\footnote{We find a slight disagreement with the earlier results of Ref.~\cite{Kunszt:1993sd} for their subleading color amplitude in Eq.~(5.24). This amplitude appears to have typos since it does not have the correct IR structure, as determined by the general formula~\cite{Catani:1998bh} or by the SCET result in \eq{IR_1loop}. Comparing with the matching calculation of Ref.~\cite{Kelley:2010fn}, we find a typo in the $\pi^2$ term in $W_4$ in Eq.~(54), which should have $3\pi^2 u^2/(2 ts )\to -3\pi^2 u/(4t)$.}

Here we list all partial amplitudes up to one loop entering the Wilson coefficients in \eq{ggqq_coeffs}. We start with the partial amplitudes where the gluons have opposite helicity, which are the only ones having a nonzero tree-level contribution. The tree-level amplitudes are given by
\begin{align}
A^\zero(1^+,2^-;3_q^+,4_\bq^-)
&= -2\,\frac{\ang{23}\ang{24}^3}{\ang{12}\ang{24}\ang{43}\ang{31}}
= 2\,\frac{\sqrt{\abs{s_{13}\,s_{14}}}}{s_{12}}\,e^{\img\Phi_{+-(+)}}
\,,\nn\\
A^\zero(2^-,1^+;3_q^+,4_\bq^-)
&= -2\,\frac{\ang{23}\ang{24}^3}{\ang{21}\ang{14}\ang{43}\ang{32}}
= 2\,\frac{s_{13}\sqrt{\abs{s_{13}\,s_{14}}}}{s_{12}\,s_{14}}\,e^{\img\Phi_{+-(+)}}
\,,\nn\\
B^\zero(1^+,2^-;3_q^+,4_\bq^-) &= 0
\,.\end{align}
In the second step we extracted a common overall phase from the amplitudes, which is given by
\begin{equation}
e^{\img\Phi_{+-(+)}}
= \frac{\ang{24}}{[24]}\,\frac{[13][14]}{\sqrt{\abs{s_{13}\,s_{14}}}}
\,.\end{equation}
The divergent parts of the corresponding one-loop amplitudes are
\begin{align}
A_\div^\one(1^+,2^-;3_q^+,4_\bq^-)
&= A^\zero(1^+,2^-;3_q^+,4_\bq^-) \biggl[
-\frac{2}{\eps^2}\, (C_A + C_F) + \frac{1}{\eps}\,(2C_F\, L_{12} + 2C_A\, L_{13} - 3C_F - \beta_0) \biggr]
\,,\nn\\
A_\div^\one(2^-,1^+;3_q^+,4_\bq^-)
&= A^\zero(2^-,1^+;3_q^+,4_\bq^-) \biggl[
-\frac{2}{\eps^2}\, (C_A + C_F) + \frac{1}{\eps}\,(2C_F\, L_{12} + 2C_A\, L_{14} - 3C_F - \beta_0) \biggr]
\,,\nn\\
B_\div^\one(1^+,2^-;3_q^+,4_\bq^-)
&= A^\zero(1^+,2^-;3_q^+,4_\bq^-)\, \frac{1}{\eps}\,4T_F
\Bigl( L_{12/14} + \frac{s_{13}}{s_{14}}\, L_{12/13} \Bigr)
\,.\end{align}
The corresponding finite parts entering the Wilson coefficient $\vC_{+-(+)}$ at one loop are
\begin{align}
A_\fin^\one(1^+,2^-;3_q^+,4_\bq^-)
&= A^\zero(1^+,2^-;3_q^+,4_\bq^-) \biggl\{
C_A \Bigl(-L_{13}^2 + L_{12/13}^2 + 1 + \frac{7 \pi^2}{6} \Bigr)
+ C_F \Bigl(-L_{12}^2 + 3 L_{12} - 8 + \frac{\pi^2}{6} \Bigr)
\nn\\ & \quad
+ (C_A-C_F) \frac{s_{12}}{s_{14}}\, (L^2_{12/13} + \pi^ 2)
\biggr\}
\,,\nn\\
A_\fin^\one(2^-,1^+;3_q^+,4_\bq^-)
&= A^\zero(2^-,1^+;3_q^+,4_\bq^-) \biggl\{
\frac{C_A}{2}\Bigl(-2L_{14}^2 + L_{12/14}^2 - 3\,L_{12/14} + 1 + \frac{4\pi^2}{3} \Bigr)
\nn\\ & \quad
+ C_F\Bigl(-L_{12}^2 + 3 L_{12} - 8 + \frac{\pi^2}{6} \Bigr)
- \frac{C_A}{2}\,\frac{s_{14}}{s_{13}} \biggl[ \Bigl(1 - \frac{s_{14}}{s_{13}}\, L_{12/14} \Bigr)^2  + L_{12/14} + \frac{s_{14}^2}{s_{13}^2}\,\pi^2 \biggr]
\nn\\ &\quad
+ \Bigl(\frac{C_A}{2} - C_F\Bigr)\,\frac{s_{12}}{s_{13}} \biggl[
\Bigl(1 + \frac{s_{12}}{s_{13}}\, L_{12/14} \Bigr)^2 - L_{12/14} + \frac{s_{12}^2}{s_{13}^2}\,\pi^2
\biggr] \biggr\}
\,,\nn\\
B_\fin^\one(1^+,2^-;3_q^+,4_\bq^-)
&= A^\zero(1^+,2^-;3_q^+,4_\bq^-)\, 4T_F
\biggl[-L_{12} L_{13/14} + \frac{s_{12}}{s_{14}}\,L_{14} L_{12/13}
- \frac{3}{4}\,\frac{s_{12}}{s_{13}}\,\big(L_{12/14}^2 + \pi^2\bigr) \biggr]
\,.\end{align}

The partial amplitudes where both gluons have the same helicity vanish at tree level,
\begin{align}
A^\zero(1^+, 2^+; 3_q^+, 4_\bq^-) &= B^\zero(1^+, 2^+; 3_q^+, 4_\bq^-) = 0
\,,\nn\\
A^\zero(1^-, 2^-; 3_q^+, 4_\bq^-) &= B^\zero(1^-, 2^-; 3_q^+, 4_\bq^-) = 0
\,.\end{align}
The corresponding one-loop amplitudes entering the Wilson coefficients $\vC_{++(+)}$ and $\vC_{--(+)}$ are IR finite. They are
\begin{align}
A^\one(1^+, 2^+;3_q^+,4_\bq^-)
&= 2\sqrt{\abs{s_{13}\,s_{14}}}\, e^{\img\Phi_{++(+)}}
\Bigl[(C_A - C_F) \frac{1}{s_{13}} + \frac{1}{3}(C_A - 2T_F\,n_f) \frac{1}{s_{12}} \Bigr]
\,,\nn\\
A^\one(2^+, 1^+;3_q^+,4_\bq^-)
&= -2\sqrt{\abs{s_{13}\,s_{14}}}\, e^{\img\Phi_{++(+)}}
\Bigl[(C_A - C_F)\frac{1}{s_{14}} + \frac{1}{3}(C_A - 2T_F\,n_f) \frac{1}{s_{12}} \Bigr]
\,,\nn\\
B^\one(1^+, 2^+;3_q^+,4_\bq^-) &= 0
\,,\end{align}
and
\begin{align}
A^\one(1^-, 2^-;3_q^+, 4_\bq^-)
&= 2\sqrt{\abs{s_{13}\,s_{14}}}\,e^{\img\Phi_{--(+)}}
\Bigl[(C_A - C_F)\frac{1}{s_{13}} + \frac{1}{3}(C_A - 2T_F\,n_f) \frac{1}{s_{12}} \Bigr]
\,,\nn\\
A^\one(2^-, 1^-;3_q^+, 4_\bq^-)
&= -2\sqrt{\abs{s_{13}\,s_{14}}}\,e^{\img\Phi_{--(+)}}
\Bigl[(C_A - C_F)\frac{1}{s_{14}} + \frac{1}{3}(C_A - 2T_F\,n_f) \frac{1}{s_{12}} \Bigr]
\,,\nn\\
B^\one(1^-, 2^-;3_q^+,4_\bq^-) &= 0
\,,\end{align}
with the overall phases
\begin{equation}
e^{\img\Phi_{++(+)}}
= \frac{[12]}{\ang{12}}\,\frac{[13]\ang{14}}{\sqrt{\abs{s_{13}\,s_{14}}}}
\,,\qquad
e^{\img\Phi_{--(+)}}
= \frac{\ang{12}}{[12]}\,\frac{[13]\ang{14}}{\sqrt{\abs{s_{13}\,s_{14}}}}
\,.\end{equation}

\subsubsection{$gggg$}
\label{app:ggggamplitudes}

The one-loop helicity amplitudes for $gggg$ were first calculated in Ref.~\cite{Kunszt:1993sd}, and the two-loop amplitudes were computed in Refs.~\cite{Glover:2001af,Bern:2002tk}. The results given here are taken from Ref.~\cite{Bern:2002tk}, and converted to our conventions. We also find complete agreement with the expressions given in Ref.~\cite{Kunszt:1993sd}.\footnote{We have also compared with the matching calculation of Ref.~\cite{Kelley:2010fn}, which has a minor typo. In particular, in ${\cal F}(s,t,u)$ in Eq.~(61) the $n_f$ terms must be dropped and $\beta_0$ set to $11 C_A/3$. Also as noted in Ref.~\cite{Broggio:2014hoa}, the last column of Table 5 in Ref.~\cite{Kelley:2010fn} applies to helicities $7,8$, while the second-to-last column applies to helicities $9$--$16$.}

The amplitudes inherit the cyclic symmetry of the traces, which means that many of the amplitudes appearing in \eq{gggg_coeffs} are related, for example
\begin{equation} \label{eq:gggg_cyclic}
A(1^+,3^-,4^-,2^+) = A(2^+,1^+,3^-,4^-)
\,.\end{equation}
For the convenience of the reader, we will explicitly give all amplitudes needed in \eq{gggg_coeffs}. We start with the partial amplitudes with two positive-helicity and two negative-helicity gluons, which are the only nonvanishing amplitudes at tree level. We have
\begin{align} \label{eq:gggg_zero}
A^\zero(1^+,2^+,3^-,4^-)
&= 4\frac{\ang{34}^4}{\ang{12}\ang{23}\ang{34}\ang{41}}
= 4\frac{s_{12}}{s_{14}}\,e^{\img\Phi_{++--}}
\,,\nn\\
A^\zero(1^+,3^-,4^-,2^+)
&= 4\frac{\ang{34}^4}{\ang{13}\ang{34}\ang{42}\ang{21}}
= 4\frac{s_{12}}{s_{13}}\,e^{\img\Phi_{++--}}
\,,\nn\\
A^\zero(1^+,4^-,2^+,3^-)
&= 4\frac{\ang{34}^4}{\ang{14}\ang{42}\ang{23}\ang{31}}
= 4\frac{s_{12}^2}{s_{13}\,s_{14}}\,e^{\img\Phi_{++--}}
\,,\end{align}
with the common overall phase
\begin{equation}
e^{\img\Phi_{++--}}
= -\frac{[12]}{\ang{12}}\,\frac{\ang{34}}{[34]}
\,.\end{equation}
The corresponding $B^\zero$ all vanish,
\begin{equation}
B^\zero(1^+,2^+,3^-,4^-)
= B^\zero(1^+,3^-,4^-,2^+) = B^\zero(1^+,4^-,2^+,3^-) = 0
\,.\end{equation}
At one loop the $B^\one$ amplitudes can be expressed in terms of the $n_f$-independent part of the $A^\one$,
\begin{align}
B^\one(1^+,2^+,3^-,4^-)
= \frac{2T_F}{C_A}\,2\bigl[A^\one(1^+,2^+,3^-,4^-) + A^\one(1^+,3^-,4^-,2^+) + A^\one(1^+,4^-,2^+,3^-) \bigr] \Bigl\lvert_{n_f = 0}
\,.\end{align}
The same relation also holds for the other helicity assignments. Using the cyclic symmetries of the amplitudes, it follows that the last three entries in the Wilson coefficients in \eq{gggg_coeffs} at one loop are all equal to each other and are given by $2T_F/C_A$ times the sum of the first three entries at $n_f = 0$. The divergent parts of the one-loop amplitudes are
\begin{align}
A_\div^\one(1^+,2^+,3^-,4^-)
&= A^\zero(1^+,2^+,3^-,4^-) \biggl[
-\frac{4}{\eps^2}\,C_A  + \frac{2}{\eps}\,(C_A\,L_{12} + C_A\, L_{14} - \beta_0) \biggr]
\,,\nn\\
A_\div^\one(1^+,3^-,4^-,2^+)
&= A^\zero(1^+,3^-,4^-,2^+) \biggl[
-\frac{4}{\eps^2}\,C_A  + \frac{2}{\eps}\,(C_A\,L_{12} + C_A\, L_{13} - \beta_0) \biggr]
\,,\nn\\
A_\div^\one(1^+,4^-,2^+,3^-)
&= A^\zero(1^+,4^-,2^+,3^-) \biggl[
- \frac{4}{\eps^2}\,C_A  + \frac{2}{\eps}\,(C_A\,L_{13} + C_A\, L_{14} - \beta_0) \biggr]
\,,\nn\\
B_\div^\one(1^+,2^+,3^-,4^-) &=A^\zero(1^+,2^+,3^-,4^-)\biggl[ \frac{8T_F}{\epsilon} \left (     L_{12/13}+\frac{s_{14}}{s_{13}}L_{12/14}   \right )  \biggr]  \,, \nn \\
B_\div^\one(1^+,3^-,4^-,2^+)&= A^\zero(1^+,2^+,3^-,4^-) \biggl[\frac{8T_F}{\epsilon}    \left (   \frac{s_{14}}{s_{13}}  L_{13/14}+\frac{s_{12}}{s_{13}}L_{13/12}   \right )        \biggr] \,, \nn \\
B_\div^\one(1^+,4^-,2^+,3^-)&= A^\zero(1^+,2^+,3^-,4^-) \biggl[\frac{8T_F}{\epsilon}      \left (     L_{14/13}+\frac{s_{12}}{s_{13}}L_{14/12}   \right )       \biggr]
\,.\end{align}
The finite parts entering the Wilson coefficient $\vC_{++--}$ at one loop are
\begin{align} \label{eq:gggg_one}
A_\fin^\one(1^+,2^+,3^-,4^-)
&= A^\zero(1^+,2^+,3^-,4^-) \biggl[
C_A\Bigl(- 2L_{12} L_{14} - \frac{4}{3} + \frac{4\pi^2}{3} \Bigr)
+ \beta_0\Bigl(L_{14} - \frac{5}{3} \Bigr) \biggr]
\,,\nn\\
A_\fin^\one(1^+,3^-,4^-,2^+)
&= A^\zero(1^+,3^-,4^-,2^+) \biggl[
C_A\Bigl(-2L_{12} L_{13} - \frac{4}{3} + \frac{4\pi^2}{3} \Bigr)
+ \beta_0\Bigl(L_{13} - \frac{5}{3} \Bigr) \biggr]
\,,\nn\\
A_\fin^\one(1^+,4^-,2^+,3^-)&=A^\zero(1^+,4^-,2^+,3^-)   \biggl\{ C_A \left (-2L_{14}L_{13} +\frac{4}{3} \pi^2 -\frac{4}{3}  \right ) -\beta_0 \left (\frac{5}{3}+\frac{s_{13}}{s_{12}}L_{14}+\frac{s_{14}}{s_{12}}L_{13}  \right )
\nn \\
&-(C_A-2T_F n_f) \frac{s_{13}s_{14}}{s_{12}^2} \biggl[ 1+\left ( \frac{s_{13}}{s_{12}}-\frac{s_{14}}{s_{12}}\right )L_{13/14} 
+\left (2- \frac{s_{13}s_{14}}{s_{12}^2}\right ) \left (L^2_{13/14}+\pi^2  \right)  \biggr] 
\nn \\
& -3T_F n_f \frac{s_{13}s_{14}}{s_{12}^2} \left( L^2_{13/14}   +\pi^2 \right)    \biggr\}
\,,\nn\\
B_\fin^\one(1^+,2^+,3^-,4^-)
&= B_\fin^\one(1^+,3^-,4^-,2^+) = B_\fin^\one(1^+,4^-,2^+,3^-)
\nn\\
&= -4T_F A^{(0)}(1^+,2^+,3^-,4^-)\, \biggl[
\frac{s_{14}}{s_{13}}\, 2 L_{13}\, L_{12/14}
+  2 L_{14}\, L_{12/13}
+\frac{s_{14}}{s_{12}} +  \frac{s_{14}}{s_{12}}  \Bigl(\frac{s_{13}}{s_{12}}-\frac{s_{14}}{s_{12}}\Bigr) L_{13/14}
\nn\\ & \quad
+ \frac{s_{14}}{s_{12}} \Bigl(2 - \frac{s_{13}\,s_{14}}{s_{12}^2}\Bigr)\bigl(L_{13/14}^2 + \pi^2\bigr)
\biggr]
\,.\end{align}
Due to \eq{gggg_cyclic}, the first two amplitudes in \eq{gggg_zero}, as well as the first two in \eq{gggg_one}, can be obtained from each other by interchanging $1^+\leftrightarrow 2^+$ which corresponds to $s_{13} \leftrightarrow s_{14}$ without an effect on the overall phase.

The amplitudes with only one or no gluon with negative helicity vanish at tree level,
\begin{align}
A^\zero(1^+, 2^+, 3^+, 4^\pm)
&= A^\zero(1^+, 3^+, 4^\pm, 2^+)
= A^\zero(1^+, 4^\pm, 2^+, 3^+) = 0
\,,\nn\\
B^\zero(1^+, 2^+, 3^+, 4^\pm)
&= B^\zero(1^+, 3^+, 4^\pm, 2^+)
= B^\zero(1^+, 4^\pm, 2^+, 3^+) = 0
\,.\end{align}
The corresponding one-loop amplitudes are infrared finite. Those entering $\vC_{+++-}$ are given by
\begin{align}
A^\one(1^+,2^+,3^+,4^-)
&= 4\frac{[13]^2}{[41]\ang{12}\ang{23}[34]} \frac{1}{3}(C_A - 2T_F\,n_f)\,
   (s_{14} + s_{34} )=4\,e^{\img\Phi_{+++-}}\, \frac{1}{3}(C_A - 2T_F\,n_f)\,
   \Bigl(\frac{s_{13}}{s_{12}} + \frac{s_{13}}{s_{14}}\Bigr)
\,,\nn\\
A^\one(1^+,3^+,4^-,2^+)
&= 4\frac{[23]^2}{[42]\ang{21}\ang{13}[34]} \frac{1}{3}(C_A - 2T_F\,n_f)\,
   (s_{13} + s_{12} )= 4\,e^{\img\Phi_{+++-}}\, \frac{1}{3}(C_A - 2T_F\,n_f)\,
   \Bigl(\frac{s_{14}}{s_{12}} + \frac{s_{14}}{s_{13}}\Bigr)
\,,\nn\\
A^\one(1^+,4^-,2^+,3^+)
&= 4\frac{[21]^2}{[42]\ang{23}\ang{31}[14]} \frac{1}{3}(C_A - 2T_F\,n_f)\,
   (s_{13} + s_{14} )= 4\,e^{\img\Phi_{+++-}}\, \frac{1}{3}(C_A - 2T_F\,n_f)\,
   \Bigl(\frac{s_{12}}{s_{14}} + \frac{s_{12}}{s_{13}}\Bigr)
\,,\nn\\
B^\one(1^+,2^+,3^+,4^-)
&= B^\one(1^+,3^+,4^-,2^+) = B^\one(1^+,4^-,2^+,3^+)
= -16T_F\,e^{\img\Phi_{+++-}}
\,,\end{align}
and those for $\vC_{++++}$ are
\begin{align}
A^\one(1^+,2^+,3^+,4^+)
&= A^\one(1^+,3^+,4^+,2^+) = A^\one(1^+,4^+,2^+,3^+)
= 4\,e^{\img\Phi_{++++}}\,\frac{1}{3}(C_A - 2T_F\,n_f)
\,,\nn\\
B^\one(1^+,2^+,3^+,4^+)
&= B^\one(1^+,3^+,4^+,2^+) = B^\one(1^+,4^+,2^+,3^+)
= 16T_F\,e^{\img\Phi_{++++}}
\,,\end{align}
where for convenience we have extracted the overall phases
\begin{equation}
e^{\img\Phi_{+++-}}
= \frac{[12]}{\ang{12}}\,\frac{[13]}{\ang{13}}\,\frac{\ang{14}}{[14]}
\,,\qquad
e^{\img\Phi_{++++}}
= -\frac{[12]}{\ang{12}}\,\frac{[34]}{\ang{34}}
\,.\end{equation}

\subsection{\boldmath $pp \to 3$ Jets}\label{app:pp3jets_app}
In this appendix we give explicit expressions for all partial amplitudes that are required in Eqs.~\eqref{eq:gqqQQ_coeffs}, \eqref{eq:gqqqq_coeffs}, \eqref{eq:gggqq_coeffs}, and \eqref{eq:ggggg_coeffs}, for the various partonic channels for the $pp \to 3$ jets process. The one-loop amplitudes for these processes were calculated in Refs.~\cite{Kunszt:1994tq,Bern:1994fz,Bern:1993mq}, respectively. These papers use $T_F=1$ and $g_s T^a/\sqrt{2}$ for the $q\bar{q}g$ coupling. Thus, we can convert to our conventions by replacing $T^a\to \sqrt{2} T^a$, and identifying $1/N = C_A - 2C_F$ and $N = C_A$. Below we restrict ourselves to giving explicit expressions for the tree-level amplitudes, since the one-loop expressions are fairly lengthy.

For each partonic channel, we expand the amplitude as
\begin{equation} \label{eq:3jdecomp}
X = [g_s (\mu)]^3 \sum_{n=0}^\infty X^{(n)} \Bigl(\frac{\alpha_s(\mu)}{4\pi}\Bigr)^n
\,,\end{equation}
where $X$ stands for any of $A_{\div,\fin}$, $B_{\div,\fin}$.

\subsubsection{$gq\bq \,q'\bq'$ and $g\,q\bq\, q\bq$}
\label{app:gqqqqamplitudes}

The tree-level amplitudes entering the Wilson coefficients in \eqs{gqqQQ_coeffs}{gqqqq_coeffs} are given by
\begin{align}
A^\zero(1^+; 2_q^+,3_\bq^-; 4_{q'}^+,5_{\bq'}^-)
&= \sqrt{2} \frac{\ang{25}\ang{35}^2}{\ang{12}\ang{15}\ang{23}\ang{45}}
\,,&
A^\zero(1^+; 4_{q'}^+,5_{\bq'}^-; 2_q^+,3_\bq^-)
&= -\sqrt{2}\frac{\ang{35}^2\ang{34}}{\ang{13}\ang{14}\ang{23}\ang{45}}
\,,\nn\\
A^\zero(1^+; 2_q^+,3_\bq^-; 4_{q'}^-,5_{\bq'}^+)
&= -\sqrt{2}\frac{\ang{25}\ang{34}^2}{\ang{12}\ang{15}\ang{23}\ang{45}}
\,,&
A^\zero(1^+; 4_{q'}^-,5_{\bq'}^+; 2_q^+,3_\bq^-)
&= \sqrt{2}\frac{\ang{34}^3}{\ang{13}\ang{14}\ang{23}\ang{45}}
\,,\nn\\
B^\zero(1^+; 2_q^+,3_\bq^-; 4_{q'}^+,5_{\bq'}^-)
&= -\sqrt{2}\frac{\ang{23}\ang{35}^2}{\ang{12}\ang{13}\ang{23}\ang{45}}
\,,&
B^\zero(1^+; 4_{q'}^+,5_{\bq'}^-;2_q^+,3_\bq^-)
&= -\sqrt{2}\frac{\ang{35}^2\ang{45}}{\ang{14}\ang{15}\ang{23}\ang{45}}
\,,\nn\\
B^\zero(1^+; 2_q^+,3_\bq^-; 4_{q'}^-,5_{\bq'}^+)
&= \sqrt{2}\frac{\ang{23}\ang{34}^2}{\ang{12}\ang{13}\ang{23}\ang{45}}
\,,&
B^\zero(1^+; 4_{q'}^-,5_{\bq'}^+; 2_q^+,3_\bq^-)
&= \sqrt{2}\frac{\ang{34}^2\ang{45}}{\ang{14}\ang{15}\ang{23}\ang{45}}
\,.\end{align}
Of these helicity amplitudes only 4 are independent. The one-loop amplitudes were computed in Ref.~\cite{Kunszt:1994tq}.

\subsubsection{$ggg\,q\bar{q}$}
\label{app:gggqqamplitudes}

The three independent tree-level partial amplitudes which enter the Wilson coefficients in \eq{gggqq_coeffs} are given by
\begin{align}
A^\zero(1^+,2^+,3^-;4_q^+,5_\bq^-)
&= 2\sqrt{2} \frac{\ang{34}\ang{35}^2}{\ang{12}\ang{14}\ang{23}\ang{45}}
\,,\nn\\
A^\zero(2^+,3^-,1^+;4_q^+,5_\bq^-)
&= -2\sqrt{2}  \frac{\ang{34} \ang{35}^3}{\ang{13}\ang{15}\ang{23}\ang{24}\ang{45}}
\,,\nn\\
A^\zero(3^-,1^+,2^+;4_q^+,5_\bq^-)
&= -2\sqrt{2}  \frac{\ang{35}^3}{\ang{12}\ang{13}\ang{25}\ang{45}}
\,,\nn\\
B^\zero &= C^\zero = 0
\,.\end{align}
At tree level, the partial amplitudes for the other color structures vanish, $B^\zero = C^\zero = 0$. The one-loop amplitudes were computed in Ref.~\cite{Bern:1994fz}.

\subsubsection{$ggggg$}
\label{app:gggggamplitudes}

The two independent partial amplitudes that enter the Wilson coefficients in \eq{ggggg_coeffs} are given by the Parke-Taylor formula~\cite{Parke:1986gb}
\begin{align} \label{eq:ggggg_zero}
A^\zero(1^+,2^+,3^+,4^-,5^-)
&= 4\sqrt{2} \frac{\ang{45}^4}{\ang{12}\ang{23}\ang{34}\ang{45}\ang{51}}
\,,\nn\\
A^\zero(1^+,2^+,4^-,3^+,5^-)
&= 4\sqrt{2} \frac{\ang{45}^4}{\ang{12}\ang{15}\ang{24}\ang{34}\ang{35}}
\,,\nn\\
B^\zero &= 0
\,.\end{align}
All other amplitudes can be obtained by cyclic permutations. The double-trace color structure does not appear at tree level, so $B^\zero = 0$. The one-loop amplitudes were calculated in Ref.~\cite{Bern:1993mq}.

\section{RGE Ingredients}
\label{app:RGE_factors}

In this appendix, we collect explicit results required for the running of the hard matching coefficients required to NNLL order.
We expand the $\beta$ function and cusp anomalous dimension in powers of $\alpha_s$ as
\begin{align} \label{eq:betafunction}
\beta(\alpha_s) =
- 2 \alpha_s \sum_{n=0}^\infty \beta_n\Bigl(\frac{\alpha_s}{4\pi}\Bigr)^{n+1}
\,,\qquad
\Gamma_{\rm cusp}(\alpha_s) =
\sum_{n=0}^\infty \Gamma_n \Bigl(\frac{\alpha_s}{4\pi}\Bigr)^{n+1}
\,.\end{align}
Up to three-loop order in the $\overline {\rm MS}$ scheme, the coefficients of the $\beta$ function are~\cite{Tarasov:1980au, Larin:1993tp}
\begin{align} \label{eq:cusp}
\beta_0 &= \frac{11}{3}\,C_A -\frac{4}{3}\,T_F\,n_f
\,, \qquad
\beta_1 = \frac{34}{3}\,C_A^2  - \Bigl(\frac{20}{3}\,C_A\, + 4 C_F\Bigr)\, T_F\,n_f
\,, \nn\\
\beta_2 &=
\frac{2857}{54}\,C_A^3 + \Bigl(C_F^2 - \frac{205}{18}\,C_F C_A
 - \frac{1415}{54}\,C_A^2 \Bigr)\, 2T_F\,n_f
 + \Bigl(\frac{11}{9}\, C_F + \frac{79}{54}\, C_A \Bigr)\, 4T_F^2\,n_f^2\,,
\end{align}
and for the cusp anomalous dimension they are~\cite{Korchemsky:1987wg, Moch:2004pa}
\begin{align}
& \qquad \qquad \qquad \qquad \qquad \qquad \qquad  \Gamma_0 = 4
\,, \qquad
\Gamma_1 = \Bigl( \frac{268}{9} -\frac{4\pi^2}{3} \Bigr)\,C_A  -
   \frac{80}{9}\,T_F\, n_f
\,,\nn\\
&\Gamma_2 =
\Bigl(\frac{490}{3} -\frac{536 \pi^2}{27} + \frac{44 \pi ^4}{45}
  + \frac{88 \zeta_3}{3}\Bigr)C_A^2
  + \Bigl(\frac{80 \pi^2}{27} - \frac{836}{27} - \frac{112 \zeta_3}{3} \Bigr)C_A\, 2T_F\,n_f
  + \Bigl(32 \zeta_3 - \frac{110}{3}\Bigr) C_F\, 2T_F\,n_f
  - \frac{64}{27}\,T_F^2\, n_f^2
\,.\end{align}
Note that here $\Gamma_\cusp$ does not include an overall color factor; it differs from the usual $q\bar{q}$ case by a factor of $C_F$.

For the noncusp anomalous dimension of the Wilson coefficient, which is color diagonal to two loops, we write
\begin{align}
\hga(\alpha_s)= [ n_q \gamma_C^q(\alpha_s) + n_g \gamma_C^g(\alpha_s)]  \id + \ord{\al_s^3}
\,,\end{align}
as in \eq{non_cusp_simplify}. The quark and gluon noncusp anomalous dimensions, 
\begin{equation}
\gamma_C^q(\alpha_s) =\Bigl( \frac{\alpha_s}{4 \pi}\Bigr) \gamma_{C \, 0}^{q} + \Bigl( \frac{\alpha_s}{4 \pi}\Bigr)^2 \gamma_{C \, 1}^{q}
\,,\qquad
\gamma_C^{g}(\alpha_s) =\Bigl( \frac{\alpha_s}{4 \pi}\Bigr) \gamma_{C \, 0}^{g} +\Bigl( \frac{\alpha_s}{4 \pi}\Bigr)^2 \gamma_{C \, 1}^{g}
\,,\end{equation}
have the following coefficients
\begin{align}
\gamma_{C \, 0}^{q}
&=-3C_F
\,, \nn \\
\gamma_{C \, 1}^{q}
&= -C_F \biggl[  \biggl( \frac{41}{9}-26 \zeta_3  \biggr)C_A +\biggl( \frac{3}{2}-2\pi^2+24 \zeta_3 \biggr) C_F +\biggl( \frac{65}{18}+\frac{\pi^2}{2}   \biggr)  \beta_0  \biggr]\,, \nn \\
\gamma_{C \, 0}^{g} &=-\beta_0
\,, \nn \\
\gamma_{C \, 1}^{g} &= \biggl( -\frac{59}{9}+2\zeta_3 \biggr) C_A^2 + \biggl(-\frac{19}{9}+\frac{\pi^2}{6} \biggr) C_A \beta_0 -\beta_1
\,.\end{align}

The evolution kernels required for the resummation were defined in \eq{Kw_def} by the integrals 
\begin{align}
K_\Ga(\mu_0, \mu)
&= \intlim{\al_s(\mu_0)}{\al_s(\mu)}{\al_s} \frac{\Gamma_\cusp(\al_s)}{\beta(\al_s)}
   \intlim{\al_s(\mu_0)}{\al_s}{\al_s'} \frac{1}{\beta(\al_s')}
\,,\nn\\
\eta_\Ga(\mu_0, \mu)
&= \intlim{\al_s(\mu_0)}{\al_s(\mu)}{\al_s} \frac{\Gamma_\cusp(\al_s)}{\beta(\al_s)}
\,, \nn \\
\widehat K_\gamma(\mu_0, \mu)
&= \intlim{\al_s(\mu_0)}{\al_s(\mu)}{\al_s} \frac{\hga(\al_s)}{\bt(\al_s)}
\,.
\end{align}
Up to two loops, we can simplify the noncusp evolution kernel as
\begin{align}
\widehat K_\gamma(\mu_0, \mu)= \left( n_q  K^q_\gamma(\mu_0, \mu)+ n_g  K^g_\gamma(\mu_0, \mu)  \right)  \id\,.
\end{align}
Explicit results to NNLL order are given by
\begin{align}
K_\Ga(\mu_0, \mu) &= -\frac{\Gamma_0}{4\beta_0^2}\,
\biggl\{ \frac{4\pi}{\alpha_s(\mu_0)}\, \Bigl(1 - \frac{1}{r} - \ln r\Bigr)
   + \biggl(\frac{\Gamma_1 }{\Gamma_0 } - \frac{\beta_1}{\beta_0}\biggr) (1-r+\ln r)
   + \frac{\beta_1}{2\beta_0} \ln^2 r
\nn\\ & \quad
+ \frac{\alpha_s(\mu_0)}{4\pi}\, \biggl[
  \biggl(\frac{\beta_1^2}{\beta_0^2} - \frac{\beta_2}{\beta_0} \biggr) \Bigl(\frac{1 - r^2}{2} + \ln r\Bigr)
  + \biggl(\frac{\beta_1\Gamma_1 }{\beta_0 \Gamma_0 } - \frac{\beta_1^2}{\beta_0^2} \biggr) (1- r+ r\ln r)
  - \biggl(\frac{\Gamma_2 }{\Gamma_0} - \frac{\beta_1\Gamma_1}{\beta_0\Gamma_0} \biggr) \frac{(1- r)^2}{2}
     \biggr] \biggr\}
\,, \nn\\
\eta_\Gamma(\mu_0, \mu) &=
 - \frac{\Gamma_0}{2\beta_0}\, \biggl[ \ln r
 + \frac{\alpha_s(\mu_0)}{4\pi}\, \biggl(\frac{\Gamma_1 }{\Gamma_0 }
 - \frac{\beta_1}{\beta_0}\biggr)(r-1)
 + \frac{\alpha_s^2(\mu_0)}{16\pi^2} \biggl(
    \frac{\Gamma_2 }{\Gamma_0 } - \frac{\beta_1\Gamma_1 }{\beta_0 \Gamma_0 }
      + \frac{\beta_1^2}{\beta_0^2} -\frac{\beta_2}{\beta_0} \biggr) \frac{r^2-1}{2}
    \biggr]\nonumber \,, \\
    K^{q}_\gamma(\mu_0, \mu)
&= - \frac{\gamma^{q}_{C\,0}}{2\beta_0}\, \biggl[ \ln r
 + \frac{\alpha_s(\mu_0)}{4\pi}\, \biggl(\frac{\gamma^{q}_{C\,1} }{\gamma^{q}_{C\,0} }
 - \frac{\beta_1}{\beta_0}\biggr)(r-1)
    \biggr]\,, \nonumber \\
     K^{g}_\gamma(\mu_0, \mu)
&= - \frac{\gamma^{g}_{C\,0}}{2\beta_0}\, \biggl[ \ln r
 + \frac{\alpha_s(\mu_0)}{4\pi}\, \biggl(\frac{\gamma^{g}_{C\,1} }{\gamma^{g}_{C\,0} }
 - \frac{\beta_1}{\beta_0}\biggr)(r-1)
    \biggr]   
\,,\end{align}
with $r = \alpha_s(\mu)/\alpha_s(\mu_0)$. The running coupling in the above equations is given by the three-loop expression
\begin{equation}
\frac{1}{\alpha_s(\mu)} = \frac{X}{\alpha_s(\mu_0)}
  +\frac{\beta_1}{4\pi\beta_0}  \ln X
  + \frac{\alpha_s(\mu_0)}{16\pi^2} \biggr[
  \frac{\beta_2}{\beta_0} \Bigl(1-\frac{1}{X}\Bigr)
  + \frac{\beta_1^2}{\beta_0^2} \Bigl( \frac{\ln X}{X} +\frac{1}{X} -1\Bigr) \biggl]
\,,\end{equation}
with $X\equiv 1+\alpha_s(\mu_0)\beta_0 \ln(\mu/\mu_0)/(2\pi)$. 
\end{widetext}

\section{Color Sum Matrices}
\label{app:treesoft}

For each specific process considered in the text we decomposed the Wilson coefficients in a color basis as
\begin{equation} 
C_{+\cdot\cdot(\cdot\cdot-)}^{a_1\dotsb\alpha_n}
= \sum_k C_{+\cdot\cdot(\cdot\cdot-)}^k T_k^{a_1\dotsb\alpha_n}
\equiv \vT^{ a_1\dotsb\alpha_n} \vC_{+\cdot\cdot(\cdot\cdot-)}
\,,\end{equation}
where $\vT^{ a_1\dotsb\alpha_n} $ is a row vector of color structures which form a complete basis of the allowed color structures for the particular process. Since convenient color bases are generically not orthogonal, the scalar product between Wilson coefficients is nontrivial. The $\vec C^\dagger $ is given by
\begin{align}
 \vec C^\dagger = \left[C^{ a_1\dotsb\alpha_n} \right]^* \vT^{a_1\dotsb\alpha_n}
= \vC^{*T}\, \hT
\,,\end{align}
where
\begin{equation}
\hT = \sum_{a_1,\ldots,\alpha_n} (\vT^{a_1\dotsb\alpha_n})^\dagger \vT^{a_1\dotsb\alpha_n}
\end{equation}
is the matrix of color sums.

In this appendix we give explicit expressions for $\hT$ for all the processes in this paper, both for general $SU(N)$,
as well as a numerical result for the specific case of $N=3$. 
For simplicity, in this section we restrict ourselves to the normalization convention $T_F=1/2$, and $C_A=N$, and write the results for general $SU(N)$ in terms of only $C_A$ and $C_F$.

For $q\bar{q}$ and $gg$ in the basis in \eq{H0_color}, we have
\begin{equation}
\hT_{q\bar{q}} = C_A = 3
\,, \qquad
\hT_{gg} = 2C_A C_F = 8
\,.\end{equation}
For $g\,q\bar{q}$ and $ggg$ in the basis \eq{H1_color}, we have
\begin{align}
\hT_{g\,q\bar{q}} &= C_A C_F = 4
\,, \nn \\
\hT_{ggg} &= 2C_F
\begin{pmatrix}
   C_A^2 & 0 \\
   0 & C_A^2-4
\end{pmatrix}
= \frac{8}{3}
\begin{pmatrix}
   9 & 0 \\
   0 & 5
\end{pmatrix}
\,.\end{align}
For $q\bq\,q\bq$ and $q\bq\,q'\bq'$ in the basis \eq{qqqq_color}, we have
\begin{equation}
\hT_{q\bq\,q\bq} = \hT_{q\bq\,q'\bq'}
= \begin{pmatrix}
   C_A^2 & C_A \\
   C_A & C_A^2
\end{pmatrix}
= \begin{pmatrix}
   9 & 3 \\
   3 & 9
\end{pmatrix}
.\end{equation}
For $gg\,q\bq$ in the basis \eq{ggqq_color}, we have
\begin{align}
\hT_{gg\,q\bar q} &= \frac{C_A C_F}{2} \begin{pmatrix}
   2C_F & 2C_F - C_A & 1 \\
   2C_F - C_A & 2C_F & 1 \\
   1 & 1 & C_A
\end{pmatrix}
\nn \\ &
= \frac{2}{3}
\begin{pmatrix}
  8 &\! -1 & 3 \\
 \! -1 & 8 & 3 \\
  3 & 3 & 9
\end{pmatrix}
,\end{align}
and for $gggg$ in the basis \eq{gggg_color}, we have
\begin{align}
\hT_{gggg} &= \frac{C_A C_F}{4}
\begin{pmatrix}
  a & b & b & c & d & c \\
  b & a & b & c & c & d \\
  b & b & a & d & c & c \\
  c & c & d & e & f & f \\
  d & c & c & f & e & f \\
  c & d & c & f & f & e
\end{pmatrix},
\end{align}
where
\begin{align}
a &= C_A^2 - \frac{9}{2} C_A C_F + 6 C_F^2 + \frac{1}{4} = \frac{23}{12}
\,, \nn \\
b &= C_A^2 - 5C_A C_F + 6C_F^2 = - \frac{1}{3}
\,,\nn \\
c &= C_F  = \frac{4}{3}
\,, \qquad
d =\frac{ (2C_F - C_A)}{2} = -\frac{1}{6}
\,, \nn \\
e &= C_F C_A = 4
\,, \qquad
f=\frac{1}{2}
\,.\end{align}
For $g\,q\bq\,q\bq$ and $g\,q\bq\,q'\bq'$ in the basis \eq{gqqqq_color} we have
\begin{align}
\hT_{g\,q\bar q\,q\bar q} &= C_A C_F \begin{pmatrix}
 C_A & 0 & 1 & 1 \\
 0 & C_A & 1 & 1 \\
 1 & 1 & C_A & 0 \\
 1 & 1 & 0 & C_A
\end{pmatrix}
\nn \\ & 
= 4
\begin{pmatrix}
 3 & 0 & 1 & 1 \\
 0 & 3 & 1 & 1 \\
 1 & 1 & 3 & 0 \\
 1 & 1 & 0 & 3
\end{pmatrix}
.\end{align}
For $ggg\, q\bq$ in the basis \eq{gggqq_color} we have
\begin{equation}
\hT_{ggg\, q\bq} = 
\frac{C_F}{4}
\begin{pmatrix}
  a & b & b & c & d & d & e & f & f & i & j \\
  b & a & b & d & c & d & f & e & f & i & j \\
  b & b & a & d & d & c & f & f & e & i & j \\
  c & d & d & a & b & b & e & f & f & j & i \\
  d & c & d & b & a & b & f & e & f & j & i \\
  d & d & c & b & b & a & f & f & e & j & i \\
  e & f & f & e & f & f & g & h & h & 0 & 0 \\
  f & e & f & f & e & f & h & g & h & 0 & 0 \\
  f & f & e & f & f & e & h & h & g & 0 & 0 \\
  i & i & i & j & j & j & 0 & 0 & 0 & i & j \\
  j & j & j & i & i & i & 0 & 0 & 0 & j & i
\end{pmatrix}
,\end{equation}
where
\begin{align}
a &= 4C_A C_F^2 = \frac{64}{3}
\,, \qquad
b = C_A - 2C_F = \frac{1}{3}
\,, \nn \\
c &= (C_A^2+1)(C_A-2C_F) = \frac{10}{3}
\,, \qquad
d = -2C_F = -\frac{8}{3}
\,, \nn \\
e &= -1
\,, \qquad
f = 2 C_A C_F = 8
\,,\qquad
g = 2C_A^2 C_F = 24
\,,\nn\\
h & = C_A = 3
\,,\qquad
i = C_A^2-2 = 7
\,, \qquad
j = -2
\,.\end{align}
For $ggggg$ in the basis \eq{ggggg_color} we have
\begin{equation}
\hT_{ggggg} = \frac{C_F}{32}
\begin{pmatrix}
   \hX_1 & \hX_2 \\
   \hX_2^T & \hX_3 \\
\end{pmatrix}
,\end{equation}
where
\begin{align}
\hX_1 &=
\begin{pmatrix}
 a &\! -b &\! -c &\! -b & c &\! -b &\! -c & b &\! -c &\! -b &\! -c & 0 \\
\! -b & a & b &\! -c & b & c &\! -b &\! -c & b &\! -c & 0 & c \\
\! -c & b & a &\! -b &\! -c &\! -b &\! -c &\! -b &\! -c & 0 & c &\! -b \\
\! -b &\! -c &\! -b & a & b &\! -c & b &\! -c & 0 & c & b & c \\
 c & b &\! -c & b & a &\! -b &\! -c & 0 & c &\! -b & c & b \\
\! -b & c &\! -b &\! -c &\! -b & a & 0 & c & b & c &\! -b & c \\
\! -c &\! -b &\! -c & b &\! -c & 0 & a &\! -b & c &\! -b &\! -c &\! -b \\
 b &\! -c &\! -b &\! -c & 0 & c &\! -b & a & b & c & b &\! -c \\
\! -c & b &\! -c & 0 & c & b & c & b & a &\! -b & c &\! -b \\
\! -b &\! -c & 0 & c &\! -b & c &\! -b & c &\! -b & a & b & c \\
\! -c & 0 & c & b & c &\! -b &\! -c & b & c & b & a &\! -b \\
 0 & c &\! -b & c & b & c &\! -b &\! -c &\! -b & c &\! -b & a
\end{pmatrix}
, \nn \\
\hX_2 &=
\begin{pmatrix}
\! -d & d &\! -e & e & e &\! -e &\! -d &\! -d &\! -d & e \\
\! -d &\! -d & d & e &\! -e &\! -e & e & e & d &\! -d \\
\! -e &\! -e & d & d & e &\! -e &\! -e & d & d & d \\
 d &\! -e &\! -e &\! -d &\! -d &\! -e & e & d &\! -d &\! -e \\
\! -d &\! -e & e &\! -e &\! -d &\! -d &\! -e & e &\! -d &\! -d \\ 
 e &\! -e &\! -e &\! -e & e & d & d &\! -d & d &\! -d \\
 d &\! -d &\! -d &\! -d & d &\! -e &\! -e & e &\! -e & e \\
\! -e & d &\! -d & d &\! -e &\! -e & d &\! -d &\! -e &\! -e \\
\! -e &\! -d &\! -d & e & e &\! -d & d &\! -e & e &\! -d \\
 d & d & e & e &\! -d & d & d & e & e & e \\
 e & e &\! -e & d &\! -d &\! -d & d & d &\! -e & e \\
\! -e & e & d &\! -d &\! -d & d &\! -e &\! -e &\! -e &\! -d
\end{pmatrix}
, \nn \\
\hX_3 &=
\begin{pmatrix}
 f & 0 &\! -g &\! -g & 0 & g & g & g & 0 & g \\
 0 & f & 0 & g &\! -g & g & 0 &\! -g &\! -g & g \\
\! -g & 0 & f & 0 &\! -g & g &\! -g & g & g & 0 \\
\! -g & g & 0 & f & 0 &\! -g & g & 0 & g & g \\
 0 &\! -g &\! -g & 0 & f & 0 &\! -g &\! -g & g & g \\
 g & g & g &\! -g & 0 & f & 0 &\! -g & g & 0 \\
 g & 0 &\! -g & g &\! -g & 0 & f & 0 & g &\! -g \\
 g &\! -g & g & 0 &\! -g &\! -g & 0 & f & 0 & g \\
 0 &\! -g & g & g & g & g & g & 0 & f & 0 \\
 g & g & 0 & g & g & 0 &\! -g & g & 0 & f
\end{pmatrix}
,
\end{align}
and 
\begin{align}
a &= C_A^4-4C_A^2+10 = 55
\,,\qquad
b = 2C_A^2-4 =  14
\,, \nn \\
c &= 2
\,, \qquad
d = 2C_A^2 C_F =24
\,, \qquad
e = C_A = 3
\,,\nn \\
f &= 2C_A^3 C_F = 72
\,,\qquad
g = C_A^2 = 9
\,.\end{align}

\section{IR Divergences}
\label{app:IRdiv}

In this appendix, we explicitly check that the IR divergences of QCD are reproduced by SCET. This ensures that they drop out in the one-loop matching, and that the resulting Wilson coefficients are IR finite. They also provide a very useful cross check when converting from the different conventions used in the literature to ours.

The one-loop matching equation relating the SCET operators and their Wilson coefficients to the QCD amplitude is
\begin{equation}
\vev{\Op^\dagger}^\zero \vec{C}^\one  + \vev{\Op^\dagger}^\one \vec{C}^\zero  = -\img \cA^\one
\,.
\end{equation}
First we determine the residues of the propagators entering the LSZ reduction formula. Regulating both UV and IR divergences in dimensional regularization, all bare loop integrals in SCET are scaleless and vanish, i.e.\ the UV and IR divergences cancel. In particular, for the self-energy diagrams, we have
\begin{equation}
    \Sigma = \Sigma_\UV + \Sigma_\IR = 0
\,.\end{equation}
The UV divergences $\Sigma_\UV$ plus possible additional UV finite terms $\Sigma_x$ (as dictated by the renormalization scheme) determine the wave function renormalization $Z_\xi$. The remainder $\Sigma_\IR-\Sigma_x$ enters the residue $R_\xi$
\begin{align}
Z_\xi^{-1} &= 1 - \frac{\df(\Sigma_\UV+\Sigma_x)}{\df\pslash}\biggr|_{\pslash=0}
\,,\nn \\
R_\xi^{-1} &= 1 - \frac{\df(\Sigma_\IR-\Sigma_x)}{\df\pslash}\biggr|_{\pslash=0}
\,.\end{align}
At one loop in pure dimensional regularization, we then have $R_\xi = Z_\xi^{-1}$, and similarly for gluons $R_A = Z_A^{-1}$. In the on-shell scheme $\Sigma_x=\Sigma_\IR$, so with pure dimensional regularization $Z_\xi = R_\xi = Z_A = R_A = 1$.

Since all loop diagrams contributing to $\vev{\Op^\dagger}^\one$ vanish, the only nonzero contributions come from the counterterm in \eq{Z_O} and the one-loop residues. At one loop we find
\vspace{9mm}

\begin{widetext}
\begin{align}\label{eq:IR_1loop}
 \vev{\Op^\dagger}^\one  \vC^{(0)}
    &= \vev{\Op^\dagger}^{(0)} \Bigl[ \bigl(Z_\xi^{n_q/2}\, Z_A^{n_g/2}\, \hZ_C - 1 \bigr)
+ \big(R_\xi^{n_q/2} R_A^{n_g/2} - 1\big)  \Bigr] \vC^{(0)}
    = \vev{\Op^\dagger}^{(0)} (\hZ_C-1) \vC^{(0)}
\nn \\
    &= \vev{\Op^\dagger}^{(0)}\, \frac{\al_s}{4\pi} \biggl[-\frac{1}{\eps^2}\,(n_g C_A + n_q C_F)
+ \frac{1}{\eps}\Bigl( -\frac{1}{2}n_g \beta_0 - \frac{3}{2} n_q C_F + 2\hDe(\mu^2)\Bigr) \biggr] 
\vC^{(0)}
\,,\end{align}
where we used the explicit expression for $\hZ_C$ derived in \subsec{loops}. One can easily check that this exactly reproduces the IR-divergent parts of the QCD amplitudes. For example, for $gg\,q\bar{q}$, we have
\begin{equation}
\biggl[-\frac{1}{\eps^2}(2 C_A + 2 C_F) + \frac{1}{\eps}\Bigl( -\beta_0 -3 C_F + 2\hDe_{gg\,q\bar{q}}(\mu^2)\Bigr) \biggr]
\vC_{+-(+)}^{(0)} (p_1,p_2;p_3,p_4)
    = \begin{pmatrix}
    A_\div^\one(1^+,2^-;3_q^+,4_{\bar q}^-) \\
    A_\div^\one(2^-,1^+;3_q^+,4_{\bar q}^-) \\
    B_\div^\one(1^+,2^-;3_q^+,4_{\bar q}^-) \\
  \end{pmatrix}
\,.\end{equation}
Hence, the IR divergences in  $\vev{\Op^\dagger}^\one \vC^{(0)}$ and $\cA^\one$ cancel each other and do not enter in $\vC^\one$, as must be the case.
\end{widetext}


\bibliography{helicityOps}

\end{document}